\documentclass[acmsmall,manuscript]{acmart}

\AtBeginDocument{%
  \providecommand\BibTeX{{%
    \normalfont B\kern-0.5em{\scshape i\kern-0.25em b}\kern-0.8em\TeX}}}

\setcopyright{acmcopyright}
\copyrightyear{2018}
\acmYear{2018}
\acmDOI{10.1145/1122445.1122456}

\usepackage{array}
\usepackage{epigraph}
\usepackage{etoolbox}

\newcommand{\italicsmallquote}[1]{\small\textit{#1}}
\AtBeginEnvironment{quote}{\italicsmallquote}

\newcommand{\software}{RoamResearch Discourse Graph extension}
\newcommand{\sysshort}{\texttt{DiscourseGraph} extension}

\newcommand{\model}{discourse graph}
\newcommand{\modelp}{discourse graphs}

\begin{document}

\title{Steps Towards an Infrastructure for Scholarly Synthesis}

\author{Joel Chan}
\email{joelchan@umd.edu}
\affiliation{%
  \institution{University of Maryland}
}

\author{Matthew Akamatsu}
\email{akamatsm@uw.edu}
\affiliation{%
  \institution{University of Washington}
}

\author{David Vargas}
\email{dvargas92495@gmail.com}
\affiliation{%
  \institution{Vellum AI}
}

\author{Lukas Kawerau}
\email{lukas@kawerau.org}
\affiliation{%
  \institution{Empiria Insight}
}

\author{Michael Gartner}
\email{michael@cybrarian.net}
\affiliation{%
  \institution{Cybrarian Services}
}

\renewcommand{\shortauthors}{Chan et al.}

\begin{abstract}
  Sharing, reusing, and synthesizing knowledge is central to the research process, both individually, and with others. These core functions are not supported by our formal scholarly publishing infrastructure: instead of the smooth functioning of functional infrastructure, researchers resort to laborious "hacks" and workarounds to "mine" publications for what they need, and struggle to efficiently share the resulting information with others. 
Information scientists have proposed an alternative infrastructure based on the more appropriately granular 
model of a discourse graph of claims, and evidence, along with key rhetorical relationships between them. However, despite significant technical progress on standards and platforms, the predominant infrastructure remains steadfastly document-based. 
Drawing from infrastructure studies, we locate the current infrastructural bottlenecks in the lack of local systems that integrate discourse-centric models to augment synthesis work, from which an infrastructure for synthesis can be grown.
Through 3 years of research through design and field deployment in a distributed community of hypertext notebook users, we 
elaborate a design vision of what can and should be built in order to grow a discourse-centric synthesis infrastructure: 
a thriving ``installed base'' of researchers authoring local, shareable discourse graphs to improve synthesis work, enhance primary research and research training, and augment collaborative research. 
We discuss how this design vision --- and our empirical work --- contributes steps towards a new infrastructure for synthesis, and 
increases HCI's capacity to advance collective intelligence and solve infrastructure-level problems. 
\end{abstract}




\maketitle

\section{Introduction}
\label{sec:Intro}

To advance the state of knowledge, researchers must \textit{synthesize} what is currently known and unknown about problems. Effective synthesis generates new knowledge, integrating relevant theories, concepts, claims, and evidence into novel conceptual wholes \cite{strikeTypesSynthesisTheir1983,blakeCollaborativeInformationSynthesis2006}. Synthesis may be supported by and manifested in various forms (e.g., a theory, a systematic or integrative literature review, or a model of a design space). The advanced understanding from synthesis can be a powerful force multiplier for choosing effective studies and operationalizations \cite{vanrooijTheoryTestHow2021,mcelreathReplicationCommunicationPopulation2015,scheelWhyHypothesisTesters2020}. Unfortunately, synthesis work is frequently experienced as arduous \cite{ervinMotivatingAuthorsUpdate2008,knightEnslavedTrappedData2019,granelloPromotingCognitiveComplexity2001}, documented as extremely cost-intensive \cite{shojaniaHowQuicklySystematic2007,petrosino1999lead,ervinMotivatingAuthorsUpdate2008,borahAnalysisTimeWorkers2017,michelsonSignificantCostSystematic2019}, and frequently absent in published research outputs, such as literature reviews in doctoral dissertations \cite{lovittsMakingImplicitExplicit2007,holbrookInvestigatingPhDThesis2004,booteScholarsResearchersCentrality2005} and published papers \cite{alton-leeTroubleshooterChecklistProspective1998,alvessonGeneratingResearchQuestions2011,vanrooijTheoryTestHow2021,mcphetresDecadeTheoryReflected2020,bhurkeUsingSystematicReviews2015,flemingCochraneNonCochraneSystematic2013}. \textbf{What, if anything, can be done to make effective synthesis more commonplace? And what solutions might HCI contribute to complement the predominant focus on institution- and incentive-level interventions from a metaresearch perspective?}

Information scientists have offered a compelling diagnosis of the problem of synthesis: there is a fundamental mismatch between researchers' information needs for synthesis work, and the document-centric data model of our predominant scholarly communication infrastructure. 
To do effective synthesis work smoothly, researchers need ready access to information artifacts at the level of claims, theories or empirical results \cite{blakeCollaborativeInformationSynthesis2006,ribaupierreExtractingDiscourseElements2017,shumScholOntoOntologybasedDigital2000}. These information artifacts are needed to grapple with the key questions that constitute the work of synthesis. For example, what empirical \textit{evidence} supports/opposes the major \textit{claims} under consideration?  What are some alternative \textit{claims} that are also consistent with the major \textit{evidence} for our key claims or hypotheses? How might we weave together systems of discursively related \textit{claims} into a coherent theory that can explain the key phenomena we seek to understand? What are the respective \textit{evidence} bases (and that support theory X vs. theory Y? And what key points of (in)consistency in the \textit{evidence} base might point to areas of productive controversy \cite{strikeTypesSynthesisTheir1983}? 

To meet these information needs, information scientists propose that researchers need an infrastructure based on a \textit{discourse-oriented} data model for the representation of scientific knowledge 
\cite{clarkMicropublicationsSemanticModel2014,shumScholOntoOntologybasedDigital2000,waardProteinsFairytalesDirections2010,brushSEPIOSemanticModel2016,kuhnBroadeningScopeNanopublications2013}. One prominent example is the micropublication model \cite{clarkMicropublicationsSemanticModel2014}, which models scientific knowledge as a \textbf{discourse graph} of claims, contextualized by relevant context, such as authorship, evidence, and methods. 

However, despite their significant potential for accelerating synthesis, \modelp{} have not yet gained widespread acceptance \cite{kuhnBroadeningScopeNanopublications2013}; instead, the predominant scholarly communication infrastructures remain steadfastly coarse-grained and document-based. 
For example, digital libraries and search engines such as Google Scholar, PubMed, and Semantic Scholar, primarily index topics and documents, not the claims, theories or results that researchers need for synthesis.
Citation databases largely do not show \textit{why} scholarly works cite each other: we cannot trace lines of evidence, or how theories are used and applied in different settings, or quickly assemble collections of consistent or contradictory claims or results: Scite.ai's distinction between supportive vs. contradictory citations, and Semantic Scholar's distinctions between "background", "methods", and "results" citations, take steps towards this deeper understanding, but are still operating on the paper as a fundamental unit. 

Understanding the mismatch between researchers' information needs and the basic data model of our current scholarly communication infrastructure clarifies why synthesis work is hard. Lacking a true infrastructure for synthesis, researchers make do with an assemblage of workarounds and hacks to find and synthesize relevant literature: they cobble across published syntheses; piece together leads on papers and authors and verbal statements of key ideas from knowledgeable colleagues; and engage in laborious citation tracing, manual reference checks, and extensive error-prone keyword search and screening of abstracts and full-text. 
From this emerging collection, researchers then construct \textit{ad-hoc} databases of claims and data, strewn across annotations in PDFs, notes and partial drafts in Google Docs and spreadsheets \cite{bosmanInnovationsScholarlyCommunication2016,qianOpeningBlackBox2020,willisDocumentsDistributedScientific2014,hoangOpportunitiesComputerSupport2018}. This bespoke database and system 
may work well enough for individual projects, but is rarely systematically transferred across projects and people, even within the same research group, leaving further related projects or people to repeat the arduous process, or worse: settle for subpar synthesis, or focus on diminishing low-hanging fruit instead of carefully and creatively conceptualizing new research directions. 

In this paper, we build on this work to frame the problem of synthesis as a problem of \textbf{infrastructure development}. 
We conceptualize \textit{infrastructure} here in the specific sense from infrastructure studies \cite{edwards2007understanding}. We recognize that infrastructure is fundamentally \textit{sociotechnical} \cite{leeHumanInfrastructureCyberinfrastructure2006}: it consists of technical components intertwined with, and sustained/constrained by, social phenomena and structures (e.g., culture, practices, institutions). 
We also recognize that infrastructures cannot be designed or constructed \textit{a priori}, but instead, are \textbf{grown} and accreted over past infrastructures \cite{aanestadInformationInfrastructuresChallenge2017,edwards2009introduction,starStepsEcologyInfrastructure1996}. Edwards et al elaborate on this intuition by drawing three key implications from historical models of infrastructural development from history and sociology \cite{hughesNetworksPowerElectrification1983}: ``First, true infrastructures only begin to form when locally constructed, centrally controlled \textit{systems} are linked into \textit{networks} and \textit{internetworks} governed by distributed control and coordination processes. Second, infrastructure formation typically starts with \textit{technology transfer} from one location or domain to another; adapting a system to new conditions introduces technical variations as well as social, cultural, organization, legal, and financial adjustment. Third, infrastructures are consolidated by means of \textit{gateways} that permit the linking of heterogeneous systems into networks'' (p. 8; emphasis in original). 

Mapping this formulation of the infrastructure development problem to our setting, we see that existing work on the \model{} data model has laid a foundation for \textit{gateways} that can weave networks of networks into larger scale infrastructure for synthesis. The remaining problem is an HCI problem: how to develop \textit{local systems} that enhance synthesis, but leave seeds for development into infrastructure: (socio)technical means of \textit{networking} these systems in a distributed fashion (e.g., via peer-to-peer networking), and \textit{technology transfer} across different contexts. This formulation of the remaining open problem aligns with researchers' diagnosis of the current main blocker to progress towards a discourse-centric synthesis infrastructure: the lack of sociotechnical means --- such as people to do the work of creating discourse graphs, and tools for them to do this work --- for creating and sustaining this infrastructure \cite{kuhnBroadeningScopeNanopublications2013,kuhnGenuineSemanticPublishing2017,renearStrategicReadingOntologies2009,griffithCIViCCommunityKnowledgebase2017}. This paper therefore explores the following core research question: \textbf{What might it look like to integrate the seeds of a discourse-centric infrastructure into local scientific practices, and the collaborative and institutional structures of research?}

We approached this question using \textbf{Research through Design} (RTD) \cite{zimmermanResearchDesignHCI2014}:   
we \textit{designed} and built a software extension for a hypertext notebook --- used by thousands of researchers for notetaking, lab management, and writing --- that could structure and share research knowledge in a discourse-centric data model. 
We then \textit{deployed} our software prototype in this user community over the course of 2.5 years, iterated on the design in response to authentic usage. Finally, we \textit{reflected} on conceptual and empirical findings from the deployment, including direct interactions and observations with 48 researchers in the user community. 

Our research through design work resulted in 
concrete design patterns, prototypes, and empirical evidence of authentic usage that elaborated a design vision of what can and should be built --- at least at the core/base --- in order to grow a discourse-centric synthesis infrastructure: 
a thriving ``installed base'' \cite{hansethTechnologyTraitorEmergent1998,starStepsEcologyInfrastructure1996} of researchers authoring local discourse graphs to improve synthesis work (\S\ref{sec:improving-synthesis}), enhance primary research and research training (\S\ref{sec:improving-primary-research}), and augment collaborative research (\S\ref{sec:augmenting-collaboration}). The work of discourse graph authoring is technically enabled by local systems of incrementally formalizable (\S\ref{sec:incremental-formalization} and locally extensible \S\ref{sec:extending-grammar}) discourse graph authoring mechanisms. These local systems can be transferred across a diverse ecosystem of personal and lab hypertext notebooks 
(\S\ref{sec:tool-transfer}), 
that can then be networked in a peer-to-peer fashion using the discourse graph model as a shared data exchange protocol. 

This paper's primary contribution is therefore \textbf{a design vision for --- and concrete steps towards --- a new infrastructure for synthesis}.  
larger infrastructure for effective scholarly synthesis. 
Because research synthesis shares many characteristics with --- and interacts with --- other forms of complex (and collective) creative work, such as sensemaking (e.g., intelligence analysis), design innovation, and collective deliberation and governance, our research also 
increases HCI's capacity to advance collective intelligence.

\section{Related Work}
\label{sec:RW}
\subsection{Knowledge infrastructures in science}
\label{sec:rw-science-infrastructures}

Our focus on a discourse-centric infrastructure for synthesis here complements existing infrastructure studies on knowledge infrastructure, which tend to focus primarily on \textit{data} (collection, sharing, and reuse) \cite{edwardsVastMachineComputer2010,edwardsScienceFrictionData2011,borgman2013knowledge,rollandTrustReliabilityReusing2013,dagiralMakingKnowledgeBoundary2016}, not on discourse and synthesis of granular concepts and claims. For example, Edwards et al \cite{edwardsVastMachineComputer2010} described the large-scale distributed research infrastructure for climate studies and modeling, and Mosconi et al \cite{mosconiThreeGapsOpening2019} described tensions between data sharing mandates and large-scale infrastructures on the one hand, and interdisciplinary researchers' data management and sharing practices on the ground. 

Additionally, since a key analytic strategy of infrastructural inversion looks for moments of maintenance, upgrade, and breakdown \cite{ribesSociotechnicalStudiesCyberinfrastructure2010}, much of this research identifies where infrastructure is \textit{not} working  \cite{mosconiThreeGapsOpening2019}. While this perspective is valuable for understanding what \textit{is}, it is limited for understanding what \textit{could be} (our focus). Our work here aligns more with constructively oriented work like Bowker's \cite{bowkerBiodiversityDatadiversity2000} synthesis of insights from science studies to propose design principles for constructing a global biodiversity database, and Young et al's \cite{youngInfrastructuringCrossDisciplinarySynthetic2017}, ambitious design arc from an infrastructural investigation of land change scientists' needs for synthesis.


\subsection{New Conceptual Models for Synthesis Infrastructures}
\label{sec:rw-conceptual-models}

The proposal for a discourse-centric infrastructure that we explore here is situated in a larger array of related proposals 
--- including ontologies, semantically rich data models, and other metadata and linked data standards --- for new modes of knowledge representation, sharing, and transfer \cite{renear_strategic_2009,kuhnGenuineSemanticPublishing2017,waardProteinsFairytalesDirections2010}. 
In this paper, we use the term "\textbf{discourse graph}" to refer to a suite of information models \cite{ciccareseSWANBiomedicalDiscourse2008,clarkMicropublicationsSemanticModel2014,brushSEPIOSemanticModel2016,shumModelingNaturalisticArgumentation2006,shumScholOntoOntologybasedDigital2000,grothAnatomyNanopublication2010,mccrickardMakingClaimsKnowledge2012,dewaardHypothesesEvidenceRelationships2009} that share a common underlying model for representing scientific discourse: one that distills traditional forms of publication down into more granular, formalized knowledge \textit{claims}, linked to supporting evidence and \textit{context} through a network or \textit{graph} model. Notable inspirations for this our work here include McCrickard's claims model for representing design knowledge \cite{mccrickardMakingClaimsKnowledge2012, mccrickardAchievingBothCreativity2013}; the \texttt{micropublications} model of claims and evidence \cite{clarkMicropublicationsSemanticModel2014}, \texttt{SEPIO} ontology for representing assertions and evidence lines \cite{brushSEPIOSemanticModel2016}; and the \texttt{nanopublications} model \cite{grothAnatomyNanopublication2010} that aims to bridge less formal natural-language scientific assertions to more formalized machine-readable assertions \cite{kuhnBroadeningScopeNanopublications2013} that can integrate into more sophisticated ``semantic publishing'' infrastructures \cite{kuhnGenuineSemanticPublishing2017}.

We choose the term \model{} to refer to this family of data models partly because of the conceptual roots of these information models in argumentation theory, such as Toulmin's model of argumentation \cite{toulminUsesArgument2003}, and to distinguish them from ontologies and other concept-centric forms of knowledge graphs: with discourse graphs, the emphasis is on representing and relating knowledge claims (rather than concepts) as the central unit, and emphasizing linking and relating these claims (rather than categorizing or filing them).

To understand the potential of this data model as a foundation for an infrastructure for synthesis, consider a researcher who wants to understand what interventions might be most promising for mitigating online harassment. To synthesize a formulation of this complex interdisciplinary problem that can advance the state of the art, she needs material that can help her work through detailed answers to a range of questions. For example, 
which theories have the most empirical support in this particular setting? 
What are the key phenomena to keep in mind when designing an intervention? 
What intervention patterns (whether technical, behavioral, or institutional) have been proposed that are both a) judged on theoretical and circumstantial grounds as likely to be effective in this setting, and b) lacking in direct evidence for efficacy? 

The answers to these questions cannot be found in the common representations of scientific knowledge in current tools, such as titles and abstracts of research papers, groupings of papers by area, or citation or authorship networks.
Instead, the answers lie at the more granular level of theoretical and empirical \textbf{claims} or statements. For example, \textit{``viewers in a Twitch chat engaged in less bad behaviors after a user was banned by a moderator for bad behavior''} 
\cite{seeringShapingProAntiSocial2017}
, and \textit{``users of subreddits quarantined for hate speech subsequently did not reduce levels of hate speech''}
\cite{chandrasekharanQuarantinedExaminingEffects2022} 
are empirical claims that interrelate in complex ways, each supporting claims and theories that may be in tension with each other. 
Granular representations of claims is crucial not just for precisely \textit{finding} relevant information to inform the synthesis, but also for \textit{constructing} more appropriately complex arguments and theories, by connecting statements in logical and discursive relationships.
Beyond operating at the claim level, our researcher will also need to work through a range of contextual, methodological details about empirical \textbf{evidence}. 
For example, to compare, judge, and develop claims (e.g., what has been established with sufficient certainty, where the frontier might be), she would need to know, for example, which empirical evidence came from which measures (e.g., self-report, behavioral measures), and the extent to which results have been replicated across a variety of settings (e.g., different research groups, years, platforms, scales). 

\begin{figure}
    \centering
    \includegraphics[width=\linewidth]{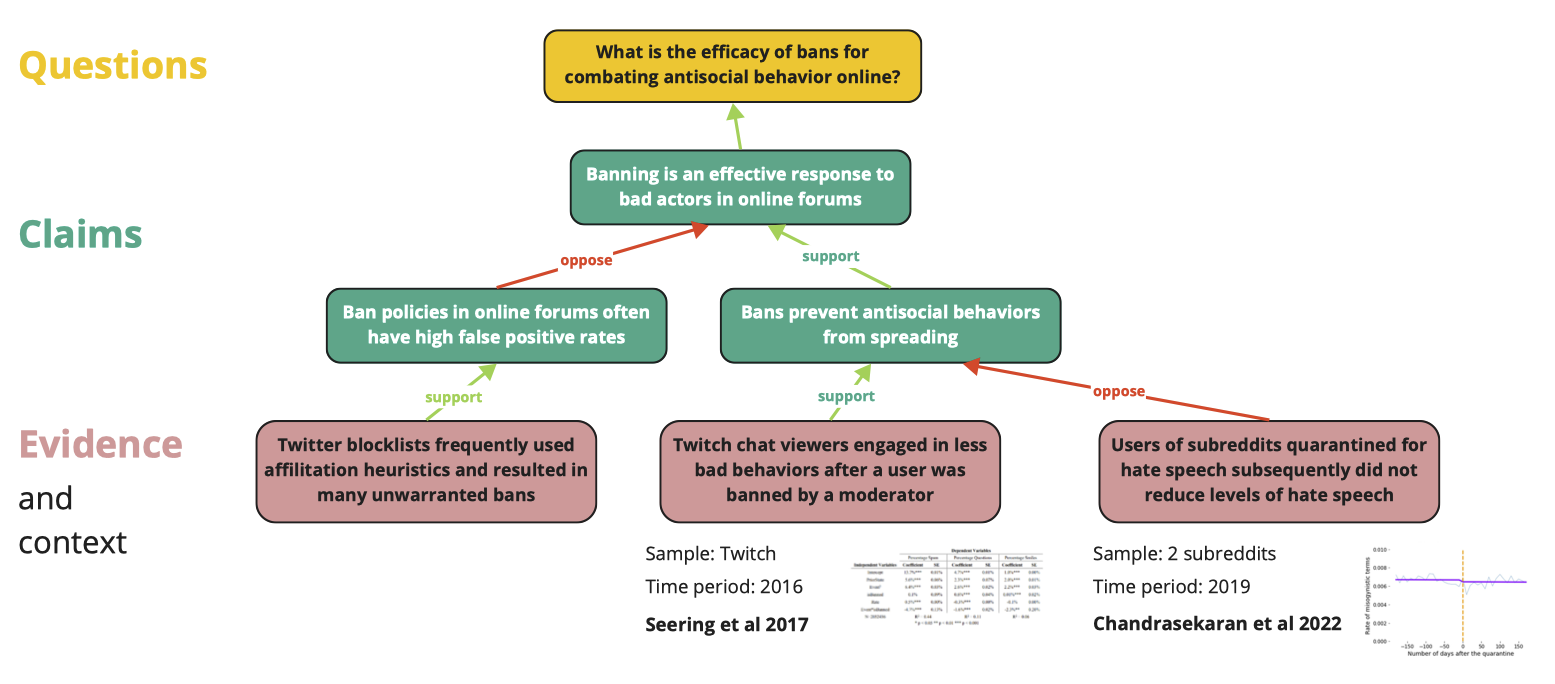}
    \caption{Example \model{} (with claims and associated context) for theories and findings on effects of bans on bad actors in online forums.}
    \label{fig:example-discourse-graph}
\end{figure}

Figure \ref{fig:example-discourse-graph} illustrates a \model{} that might support such an analysis. In this \model{}, the primary unit of analysis of claims matches the information needs of synthesis. Contextual entities and information, such as methodological details and metadata, are explicitly attached to evidence in the \model{}, enabling direct analysis of claims with their evidentiary context, and critical engagement, integration, or reinterpretation of individual findings. And discourse units in the \model{} can have many-to-many relationships to support composition of more complex arguments and theories. If appropriately stable domain-specific vocabulary or ontologies are available, a set of claims may also be modeled as defining a subgraph of a more granular knowledge or causal graph, supporting more sophisticated theoretical synthesis that is still connected to empirical evidence.  

There have been attempts to integrate \modelp{} into scientific communities of practice: for example, the ScholOnto model \cite{shumModelingNaturalisticArgumentation2006} was supporting remote PhD mentoring and distributed collaborations; SWAN \cite{ciccareseSWANBiomedicalDiscourse2008} was integrated into the successful Alzforum online research community for Alzheimers research \cite{clarkAlzforumSWANPresent2007}; and the micropublication model was integrated into the Domeo scientist network \cite{clarkMicropublicationsSemanticModel2014}. However, for a variety of reasons, these nascent infrastructures are no longer active, and the impact of these deployments has not been empirically evaluated. This may partly be due to changes in funding and infrastructure for these research software, leading to deprecation 
of technical infrastructure. Other efforts might not have made it past the experimental prototype stage for similar reasons (lack of funding, incentives). However, the pivot of some information models into more educational or general-purpose applications \cite{liddoContestedCollectiveIntelligence2012a} is suggestive of the open problems around realizing a discourse-centric infrastructure in scientific practice. As we note above, there is some consensus that these problems center on the "human" side of the infrastructure such as tools and incentives for authors, and sociotechnical means of networking and sharing \modelp{} in practice  \cite{kuhnGenuineSemanticPublishing2017,waardProteinsFairytalesDirections2010}. Our work joins and complements existing nascent explorations of different ways to foster sustainable contributions to synthesis infrastructures, such as controlled natural language wikis \cite{kuhnAceWikiNaturalExpressive2009}, markup integrations in word processors \cite{urenSensemakingToolsUnderstanding2006,grozaSALTWeavingClaim2007}, and annotation schemes in publication submission portals \cite{bucurNanopublicationBasedSemanticPublishing2022}.

\subsection{Existing and emerging tools and workflows for synthesis work}
\label{sec:rw-synthesis-tools}

Our work is grounded in fundamental studies of the tools and workflows of scholars \cite{palmerScholarlyInformationPractices2009}, 
which have yielded valuable descriptions of the information needs, pain points, and regularities (and differences) in scholarly workflows and tools. Examples include Ellis's classic study of the information-seeking patterns of academic researchers \cite{ellisModelingInformationSeekingPatterns1993}, O'Hara's diary study of PhD students' use of library documents and reading practices \cite{oharaStudentReadersUse1998}, Ribaupierre et al's description of scholars' strategic reading practices \cite{ribaupierreExtractingDiscourseElements2017}, and Blake and Pratt's detailed description and model of the information behaviors of scientists conducting systematic reviews \cite{blakeCollaborativeInformationSynthesis2006}. 
On the practice side, surveys of tool usage by scientists suggest that document-centric workflows continue to dominate. For example, Bosman et al.  \cite{bosmanInnovationsScholarlyCommunication2016} reported from a large-scale online survey of approximately 20k researchers worldwide that reference management tools like EndNote, Mendeley, and Zotero were the most frequently mentioned tools for managing and using literature. These large-scale findings are corroborated by more in-depth qualitative investigations of researcher practices, which generally find the predominance of these document-centric tools, as well as mainstream general-purpose software like Microsoft Word and Excel for note-taking \cite{qianOpeningBlackBox2020,willisDocumentsDistributedScientific2014,hoangOpportunitiesComputerSupport2018}. One notable study by Sawyer and colleagues \cite{sawyerSocialScientistsCyberinfrastructure2012} frames the landscape of tools, workflows, and practices of synthesis as "pre-infrastructural". They note that these are marked by assemblage, fragmentation, and appropriation, and are not mature infrastructure. 

Some of this descriptive work provides empirical validation of the hypothesized benefits of \modelp{} for augmenting synthesis: for example, Ribaupierre et al \cite{ribaupierreExtractingDiscourseElements2017} found that scientists who tested a prototype tool that allowed them to search over a literature by rhetorical elements (e.g., findings, methods, definitions) self-reported higher signal-to-noise ratio in results when searching for specific findings (e.g., "show all findings of studies that have addressed the issue of gender equality in terms of salary"), compared to using a standard keyword search interface. There is also documentation of desire paths of scientists adopting niche tools with affordances that resemble \modelp{}. For example, there is a subculture of academic researchers who repurpose qualitative data analysis tools like NVivo and Atlas.ti to do literature reviews \cite{wolfswinkelUsingGroundedTheory2013,silverCruxLiteratureReviewing2020,anujacabraalWhyUseNVivo2012,morabitoManagingContextScholarly2021}; it is notable that the key affordances of these tools --- layers of contextualizability for excerpts (data) and themes (synthesis/claims) --- are similar to those of \modelp{}. There is also some adoption of niche specialized tools for literature sensemaking, such as LiquidText and Citavi, both of which support the composition of networks of claims that are directly linked to contextualizing excerpts from documents \cite{morabitoManagingContextScholarly2021}.

Because our infrastructure development strategy focuses on developing seeds for infrastructure by integrating into and augmenting local synthesis practices, the prototypes, design patterns, and empirical evidence in this paper also advance research on tools and workflows for synthesis work.


\section{Methodology: Research through Design}
\label{sec:Methods}
We explored our research questions using a Research through Design (RtD) methodology \cite{zimmermanResearchDesignHCI2014}. We chose RtD because the nature of our questions are about "the world that should be brought into being" \cite[p. 178]{zimmermanResearchDesignHCI2014}: our questions cannot be descriptive, because there are no existing examples of a discourse-centric synthesis infrastructure. The ``wicked'' and messy nature of an infrastructure-level problem --- with all of its complex sociotechnical entanglements and entrenched, competing interests --- is also well-matched for the application of a designerly --- vs. scientific --- way of knowing \cite{zimmermanResearchDesignHCI2014,gaverScienceDesignImplications2014}.

\begin{figure}
    \centering
    \includegraphics[width=0.8\linewidth]{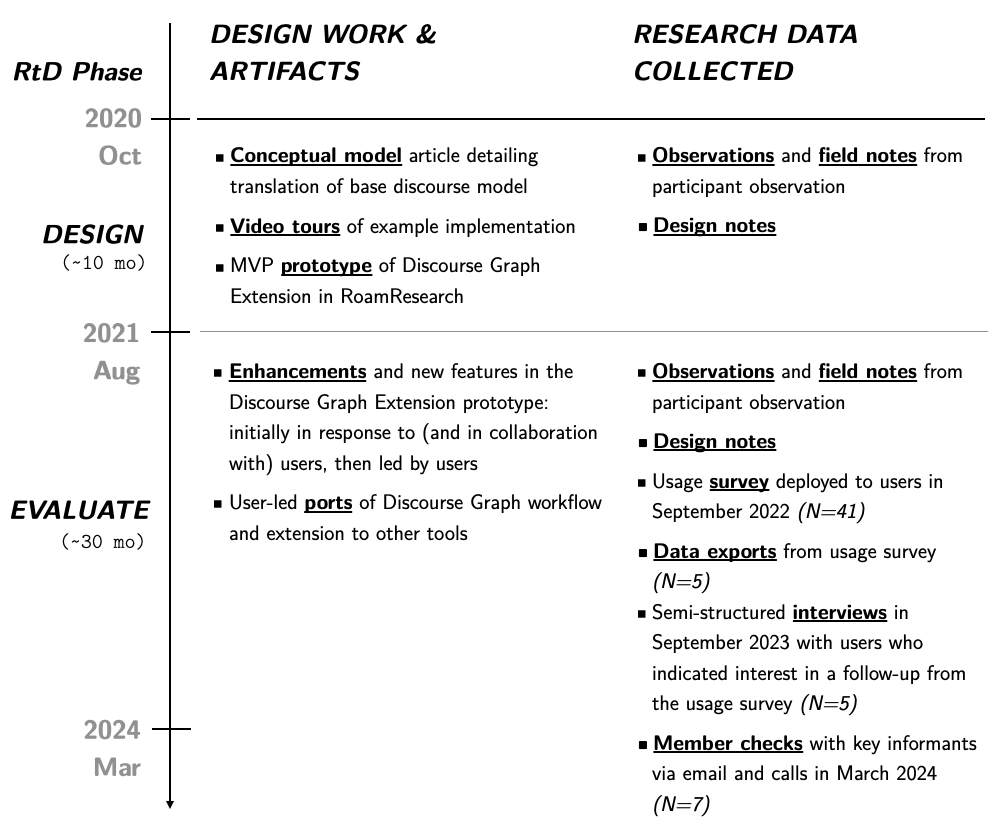}
    \caption{Summary of design work and artifacts and research data collected across the Design and Evaluate steps of our Research through Design work.}
    \label{fig:methods-timeline}
\end{figure}

We followed Zimmerman and Forlizzi's \cite{zimmermanResearchDesignHCI2014} suggested 5-step cycle for Research through Design work: 1) \textit{select} a design problem or opportunity, 2) conduct \textit{design} work to develop an initial framing and prototypes, 3) \textit{evaluate} the design prototype(s) through deployment with users, 4) \textit{reflect} on the design and evaluation work and \textit{disseminate} resulting insights, and 5) repeat. This paper reports analysis of insights from design work and artifacts and research data collected across the \textbf{Design} and \textbf{Evaluate} steps of our Research through Design work in this setting (see Fig. \ref{fig:methods-timeline} for a visual summary). These phases of work spanned approximately 3 years, and involving direct interactions with 48 participants. 

Our RtD approach was also conducted in a field-based and participatory manner. In particular, our work began as participatory in a participant observation sense: the first author is both researcher and participant/user in the community of tool users where we designed and deployed our tool. He drew on his own active participation and experiences to inform the design work and evaluation. Later, the work became participatory in a participatory design \cite{sandersCocreationNewLandscapes2008} and community-based research sense, as the second and fourth authors joined from the community to substantially develop the designs and insights and practice, fundamentally shaping the conceptual contributions of the work. We therefore drew on participatory methodology \cite{vaughnParticipatoryResearchMethods2020} --- including reflexive, detailed field notes, regular research discussions with key community members, and co-authorship of the final manuscript --- to ensure that community participation was rigorous and equitable. 

\subsection{Setting}
\label{sec:methods-design-setting}
The general setting of this study is centered around a distributed community of researchers who have adopted general-purpose ``hypertext notebooks'' for research knowledge management. These notebooks share many features with wikis, with granular subdocuments and bi-directional hyperlinking between these subdocuments serving as a central organizing principle, as opposed to top-down hierarchies of folders and tags. There are also some interesting differences in affordances and values of these communities: for example, several tools, such as RoamResearch\footnote{https://roamresearch.com/}, Logseq\footnote{https://logseq.com/}, and Tana\footnote{https://tana.inc}, cross the hyperlinking wiki data structure with an "outliner" format, such that each bullet itself becomes a uniquely addressable subdocument that can be interlinked into knowledge and organizational structures; others, such as Tinderbox\footnote{http://www.eastgate.com/Tinderbox/}, embed the hyperlinking structure in a visual medium; and others, such as Obsidian\footnote{https://obsidian.md/}, heavily emphasize privacy and data ownership, overlaying wiki technology on plain text markdown files. From a scholarly primitives perspective \cite{palmerScholarlyInformationPractices2009}, usage of these tools concentrates on the primitives of collecting, reading (including especially notetaking) and subsets of the writing process; frequently, traditional tools like word processors are still used for final drafting and typesetting stages of writing, but drawn heavily from the 
knowledge bases constructed in these tools. 

There are no definitive numbers on how many researchers have adapted these general-purpose tools for synthesis work, but the following data points can give a sense of the size and boundaries of the community: as of the time of this writing, there were approximately 12,000 users on the Slack chat community for the RoamResearch tool, and approximately 800 users on a specialized Discord chat community for academic users of RoamResearch. We think the estimate of ~5--10\% of the total user base is a reasonable ballpark across the other tool communities, each of which have discussions and defined subcommunities for academic users, such as an academia channel in the Discord community of ~8,000 users for the Obsidian tool, and ~1500 users for the Logseq tool. Thus, this community of users is still nascent relative to mainstream tools like Google Docs and Microsoft Word, but it is substantial in size and active. 




\subsection{Design Phase}

Design work began in approximately December 2020 (and was mostly completed in summer of 2021), and centered around exploring how to 
instantiate the discourse graph model into a working system that enables scholars to write notes in as close to their normal formats as possible, but author shareable, machine-readable discourse graphs as a byproduct. The design work in this arc culminated in the initial working prototype of the \software{}, which embodied a set of design hypotheses about how to integrate discourse graphing into everyday research workflows and enable downstream development of synthesis infrastructures. For brevity, we summarize the endpoint of this design work and resulting initial prototype in \S\ref{sec:DesignArc}.

\subsection{Evaluation Phase}

Evaluation work began in approximately August 2021, and centered around testing our design hypotheses about the possibility and preconditions of 1) integrating discourse graphs into scholarly workflows, and 2) growing these "grassroots" discourse graphs into a new infrastructure for scholarly synthesis. 
Data collection has continued to the present, through sustained, serious usage of the \software{}, and community engagement around the software prototype. Thus, the Deployment phase overlapped with the Reflection phase. 

\label{sec:deployment-details}
To test our design hypotheses, we released the extension in a number of different channels beginning the Fall of 2021. The first distribution channel was a set of three \textbf{notetaking courses} for academic users of Roam, one of which was led by LK, the fourth author. These courses are all online courses for users of Roam, and are led by lead users in the Roam community. Each lead user was an advanced PhD student at the time of the deployment. This integration came about because, as part of our participatory deployment and design strategy, the first author had solicited early feedback from each of these users on a call in August of 2021 (once an MVP of the prototype had been developed), and each of them expressed immediate strong interest in integrating this into their courses. 
The second distribution channel was the \textbf{wider Roam community}, which congregated on Twitter, blogs, Youtube, and a shared Slack channel. 

At the time of this writing, the extension remains in active use. 
In this paper, however, we report our reflections on the following sources of data collected over the course of 30 months, from August 2021 to March 2024: 1) observations and field notes from participant observation, 2) design notes from design work, 3) a formal usage survey, 4) semi-structured interviews, 5) member checks with key informants, and 6) data exports from participants' lab notebooks.

\subsubsection{Observations and field notes}
The first author participated in wider discussions of workflows and tooling on Twitter, Slack, as well as synchronous Zoom calls. Structured direct observations were also conducted in a Discord channel dedicated to the extension in a Discord server for academic users of Roam. Finally, the first author held bi-weekly ``open office hours'' over Zoom with users of the extension. Field notes were recorded throughout these observations, including links to messages on the social media or chat threads, where appropriate/available, for later analysis.

\subsubsection{Design notes from design work.}
All design and development discussions with the DV, the third author, were documented in a shared Roam notebook.

\subsubsection{Formal usage survey.}
We distributed a usage survey in the Fall of 2022 to get a quantitative picture of usage. Key questions included demographics (e.g., occupation, location, research area, profession, source of learning about the extension), current usage patterns (e.g., current user vs. not, motivations for using/exploring the extension, which features of the extension are core/occasional/unused/unknown). The full survey items are shown in Appendix \ref{ap:usage-survey-items}.

The survey was distributed via a one-time popup in the extension, which users would see upon refresh of their Roam application, and also distributed over Twitter and in the Discord channel for users of the extension. We received $N=41$ total responses to the survey. 

\subsubsection{Semi-structured interviews}
The first author conducted semi-structured interviews, each lasting approximately 60--90 minutes, with $N=5$ users who indicated an interest in a follow-up from the usage survey. The interview focused on whether/how they were still using the extension (and whether they'd be willing to share a csv export and/or screenshots of your discourse graph usage), impacts (positive or negative) the extension has had on their work since the last usage survey, changes in your synthesis practice (with or without the extension), and wishlist items for the extension (for ongoing development work). These interviews were recorded and transcribed.

\subsubsection{Member checks with key informants}
After initial analysis was completed in March 2024, the first author conducted member checks with $N=7$ key informants via email and calls.

\subsubsection{Data exports.}
The usage survey requested an optional data export of the user's discourse graph. We were able to obtain user-provided exports of data from $N=4$ users (5, including the author's lab) from the usage survey approximately 12 months into the deployment (in Fall of 2022). These were neo4j-compatible csv exports of the nodes and edges.

This research study was approved by the University of Maryland's institutional review board.

\subsection{Deployment Participants.}
Since the software extension was released publicly into the community, and we (for privacy reasons) chose not to track specific usage on our servers from participants' machines, we do not have a precise quantitative picture of usage. We can, however, triangulate from at least two sources to give a quantitative sense of the userbase that is informing the insights from this arc. First, because of the way that Roam works with extensions, each time a Roam user reloaded their application, all software extensions would be (re)-downloaded and unpacked. This is not a precise measure of usage/coverage, since extensions that were left installed but were not in active use would still be downloaded at each reload of the application. Thus, the average daily downloads gave us a rough upper bound on the coverage of the extension. Based on the 2nd author's experience with other software extensions for Roam, an approximate estimate of active users given download counts is to divide the number of downloads by 10. As of November 2022, we had 320 daily downloads of the extension, so we estimate that the extension had about ~30 daily active users as of November 2022. 
Second, 23 of the 41 (~56\%) respondents to our usage survey self-reported as current users of the extension, while 9 (~22\%) self-reported as "still exploring" the extension to assess fit, and 9 (~22\%) self-reported as not currently using the extension. Putting these two estimates together, we estimate our total participant base to be on the order of 30 daily active users. 

Based on our usage survey, the majority (14/23, or ~60\%) of current users were academic researchers of some variety, with the remaining (9/23, or ~40\%) of users engaged in consulting, librarianship, medical training, independent research, and venture-building. Users hail from all over the world, including North America, China, Thailand, Germany, Chile, Austria, Spain, and New Zealand. 

Also, in our usage survey, 11 respondents explicitly mentioned one of the three notetaking courses that integrated the extension, and 12 mentioned a mix of blogs, social media, or word-of-mouth as the first places they learned about the extension (the remaining 18 respondents did not describe how they learned about the extension).

\subsection{Positionality and Rapport}

The overall research design and conceptualization was initially led by the first author, an HCI researcher who is also a core user of the tool we extended. He participated as a ``power user'' in workflow and tool exploration, including the first "Roam tour", published in April 2020, where he explored how to implement the "zettelkasten method" in this new hypertext notebook medium. In general, his reputation as a "power user" in the hypertext notebook user community was at least as salient as his reputation as an HCI researcher. This gave him access and credibility he would not otherwise have gotten as outsiders, including partnerships with other community leaders who developed learning resources for the community of users. 
These partnerships were the first initial vector of the deployment of our design prototype. 
The first author's additional identity as a an HCI researcher was made plain to the community throughout the deployment, prominently advertised in primary announcements and resource sharing posts with the community, as well as in 1-on-1 and group calls. 
This is important because we obtained a waiver for documentation of informed consent, due to the distributed and often transient nature of interactions in the community. 

Formative work on design requirements were shaped in part through extensive interactions with the second author --- identified in the manuscript as MA --- as fellow members of the user community: across several calls and interactions on the community forums. MA later led the extension of the discourse graph workflow and system to improve and transform primary research in his cell biology lab (see \S\ref{sec:improving-primary-research}). The design and prototyping work for the software extension (see \S\ref{sec:DesignArc}) was done in close collaboration with the third author, who is a key developer of software extensions for the tool we extended.

As the field study became increasingly participatory, two additional members of the community began to make increasingly integral contributions to the conceptualization and development of the key ideas in this paper, and have therefore been named as co-authors of this work (in keeping with norms of participatory research). The fourth author --- identified in the manuscript as LK --- saw the potential of the discourse graph workflow for enabling the learning of synthesis, and became a data collection partner, deploying the discourse graph software prototype we developed in and from the notetaking course he was running, which became a central nucleating site for the deployment of our prototype for our field study (see \S\ref{sec:deployment-details}). To continue technical advances that enabled this work as well as augmented collaborative synthesis in his lab (see \S\ref{sec:augmenting-collaboration}), the second author later brought on the fifth author to assist with further technical development. 

Data collection and analysis was led by the first author, who led the drafting of the manuscript with elaboration and checks from community-based authors. 

\subsection{Analytic Approach}


Analysis was primarily conducted by the first author, using a general constructivist and interpretivist \cite{schwandtConstructivistInterpretivistApproaches1994} approach. Analysis was interleaved with data collection throughout, including in reflective memos, and, once critical mass of data accumulated, iterative coding and analysis. Analysis was conducted in a theory-informed manner, drawing on sensitizing concepts such as boundary objects \cite{starInstitutionalEcologyTranslations1989}, sensemaking  \cite{russellCostStructureSensemaking1993}, and infrastructure \cite{leeHumanInfrastructureCyberinfrastructure2006}.

\section{Design of our Discourse Graph Extension for Local Synthesis Work}
\label{sec:DesignArc}

We began our design work building on a previous line of work where we explored how to translate discourse-centric models like SEPIO \cite{brushSEPIOSemanticModel2016} into a note-taking convention for synthesis. We wrote up this discourse-graph-as-practice translation in \cite{chanKnowledgeSynthesisConceptual2020}, shared it with the tools for thought community via a set of video demos and social media posts, and tested the process in the first author's lab with collaborative literature reviews for a few ongoing research projects, as well as a collaborative literature review on COVID with other Roam users. These experiences gave us some confidence that the discourse graph model could be integrated as a \textit{practice} into everyday scholarly work; the remaining question from an infrastructure perspective was \textbf{whether it was (socio)technically possible to bridge this discourse-graph-as-practice beyond a single notebook or lab}, to seed new scholarly synthesis infrastructures. 

More concretely, we wanted to explore \textbf{how we might design a system that made it possible to write and use discourse-graph-as-practice in as natural a manner as possible (e.g., in outlines, with a mix of less formal and more formal argumentation) but seamlessly translate to queryable, shareable discourse graph formalisms in the backend}. Reflecting on our experiences building out these discourse-graph-as-practice working examples, and prior technical and conceptual work, our initial design requirements were:
\begin{enumerate}
    \item \textbf{DR1: Enable users to seamlessly create a formal discourse graph alongside their informal notes}. 
    \item \textbf{DR2: Enable users to leverage their formal discourse graph to improve their thinking and writing}. 
\end{enumerate}

\begin{figure}
    \centering
    \includegraphics[width=\linewidth]{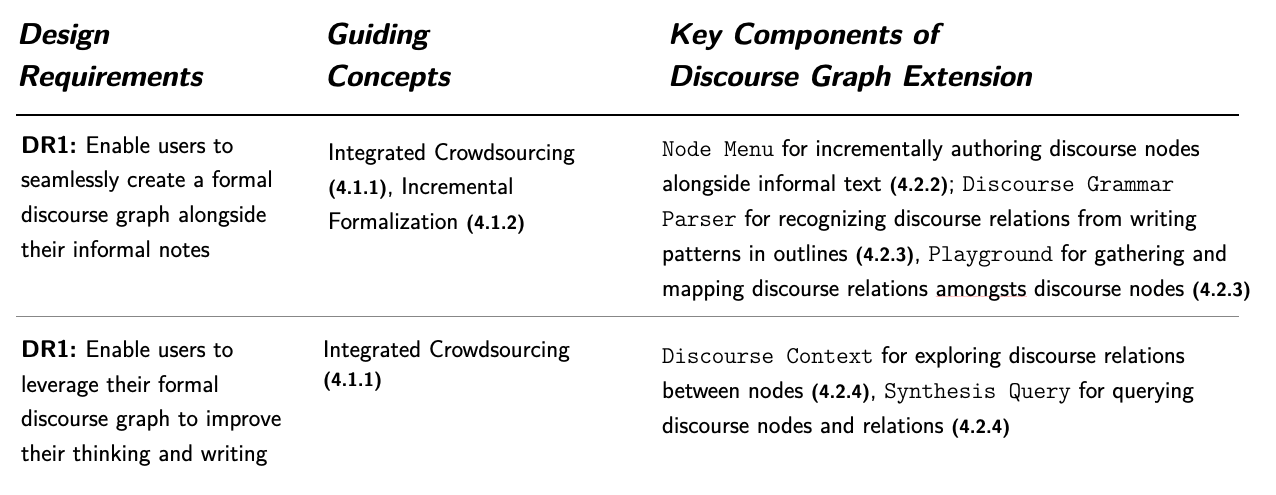}
    \caption{Summary of design requirements and design hypotheses developed in Design Arc (Section \ref{sec:DesignArc})}
    \label{fig:sys-design-summary}
\end{figure}

For brevity, we focus our description here on the \textit{outputs} of our design work, in the form of design rationales and initial hypotheses, to frame what we learned from our field deployment and evaluation. Figure \ref{fig:sys-design-summary} shows a summary of these outputs: guided by key design inspirations from previous work (\S\ref{sec:design-inspirations}), we formulated design hypotheses for how to address our key design requirements, embodied in the key components of a discourse-centric software extension for a hypertext notebook (\S\ref{sec:system-description}). 

\subsection{Guiding Design Inspirations from Previous Work}
\label{sec:design-inspirations}

Our design exploration was particularly informed by the key concepts of integrated crowdsourcing \cite{siangliulueIdeaHoundImprovingLargescale2016} and incremental formalization \cite{shipmanFormalityConsideredHarmful1999} from HCI research, as well as key insights on synthesis tools (summarized in part in \S\ref{sec:rw-synthesis-tools}). 

\subsubsection{Integrated Crowdsourcing}

The first design pattern of \textbf{integrated crowdsourcing} has been proposed and validated as a means of sustaining work that is beneficial for a community --- such as annotating bookmarks on tutorial videos \cite{weirLearnersourcingSubgoalLabels2015} or specifying semantic relationships between ideas on a large-scale innovation platform \cite{siangliulueIdeaHoundImprovingLargescale2016} --- that are experienced as tedious work with no immediate individual benefits. The design pattern is to seamlessly integrate the ``crowdsourcing'' work (that will benefit a community) into tasks that community members are already intrinsically motivated to do. One example of this design pattern is \cite{siangliulueIdeaHoundImprovingLargescale2016}' IdeaHound system that integrates semantic judgments of idea relationships in a large-scale innovation platform into the intrinsically motivating activity of visuospatial brainstorming with inspirations from others' ideas (on computationally augmented digital whiteboards). Another example is Kim et al's work on ``learnersourcing'' subgoal labels on tutorial videos by integrating annotation work into the intrinsically motivating activity of creating video bookmarks and notes for oneself \cite{weirLearnersourcingSubgoalLabels2015,kimCrowdsourcingStepbystepInformation2014}. A final example is the Cobi system that sources judgments of research paper relationships for conference planning by integrating these semantic judgments into the intrinsically motivating activity of selecting papers to see at a conference \cite{andreCommunityClusteringLeveraging2013,chiltonFrenzyCollaborativeData2014}. 

This design pattern motivated and guided our design work for both \textbf{DR1} and \textbf{DR2}: our guiding approach was to develop a core user experience that enabled people to write as close to prose as they normally would (and for intrinsic benefits of advancing their own synthesis work), and author a shareable discourse graph as a natural byproduct.


\subsubsection{Incremental Formalization}

The first concept of incremental formalization comes from foundational research on hypertext systems that have explored integrating formal structure --- such as schemas --- into interactive systems for knowledge management. This line of work has identified how requirements for specifying formal structure --- especially if they are the only means of entering data, or imposed early in open-ended tasks --- can interfere with the very work they are meant to support \cite{conklin1987hypertext,shipmanFormalityConsideredHarmful1999}. For instance, Conklin and Begeman's \cite{conklin1987hypertext} classic work on Group Issue-Based Information Systems (gIBIS), a hypermedia environment for collaborative policy discussions integrated Rittel's \cite{kunzIssuesElementsInformation1970} Issue-Based Information Systems (IBIS) framework for structuring collaborative policy discussions, and found that users found the ontology of issues, arguments, and positions, helpful for structuring their discussions, but also wanted ways to integrate this into their work in a more semi-structured way, such as specifying "proto-nodes", or expanding the ontology to cover more aspects of their thinking, such as ideas and requirements.

Shipman and colleagues \cite{shipmanFormalityConsideredHarmful1999} proposed the concept of \textbf{incremental formalization} as a means of mitigating these risks of formalisms: users enter information in a mostly informal fashion, and then formalize only later in the task when appropriate formalisms become clear and also (more) immediately useful. Notable examples of this pattern include VIKI, a spatial hypertext system that includes heuristic algorithms to find recurring visual/spatial patterns in layout of objects; users can use these to specify schemas if they wish \cite{marshallVIKISpatialHypertext1994}; and the Hyper-Object Substrate (HOS) system, users enter mostly informal text initially, and the system recognizes patterns in the textual information to suggest possible formal attributes or relations for the underlying knowledge base, which the user can then accept/modify/reject as they wish \cite{shipmanSupportingKnowledgebaseEvolution1994}; and the Jourknow system that includes a variety of features that can recognize formal structure (e.g., location, time, meeting information) from (relatively) unstructured notes in pidgin or more lightweight entry format, such as Notation3 \cite{vankleekGuiPhooeyCase2007}. 

We found this design pattern of incremental formalization to be particularly inspirational for our design work on \textbf{DR1}. 

\subsection{The Discourse Graph Software Extension for Hypertext Notebooks}
\label{sec:system-description}


\subsubsection{Hypertext notebook as technical substrate}
Our first design choice was to integrate discourse authoring into the hypertext notebook RoamResearch, for a few reasons. 

First, the underlying data model is a graph, which is nicely aligned with the discourse graph model. In Roam, users write outlines, where each bullet (block), and/or page it is contained in, is a node in a graph. Hypertext links (e.g., references, embeds, backlinks) between blocks and pages constitute edges in the graph. This data structure was closely inspired by Nelson's seminal zippered-list data model for hypertext notebooks \cite{nelsonComplexInformationProcessing1965}. 

Second, the hypertext notebook allows for a mixture of both formal and informal information. Indeed, the base data model in Roam is highly informal and minimally structured: from the user's point of view, it is just an outline with text. But users can (and do!) add structure through pages, tags, and namespaces, among other things. These are highly beneficial technical properties for implementing the incremental formalization design pattern \cite{shipmanFormalityConsideredHarmful1999}. 

Third, as we described in \S\ref{sec:methods-design-setting}, the platform has an existing robust and substantial userbase where scholars are already doing work of reading, note-taking and writing. This is an ideal field site for us to test the hypothesis that we can integrate discourse graph authoring into their workflows. 

Fourth and finally, the platform is highly extensible, with an API that allows for interactions with and modifications of the underlying data model, as well as the UI. This would allow us to not only implement technical means of authoring discourse nodes and edges, but also prototype UI extensions that could provide intrinsic benefits to the users from the discourse graph being computationally reified. Integrating into existing tools was also motivated by the design pattern of integrated crowdsourcing \cite{siangliulueIdeaHoundImprovingLargescale2016}; we hypothesized that we could develop a software extension to Roam that could integrate the work of discourse authoring into the intrinsically motivating synthesis work that we already knew was happening in these tools.

\subsubsection{Authoring formal discourse nodes alongside informal notes}

\begin{figure}
    \centering
    \includegraphics[width=\linewidth]{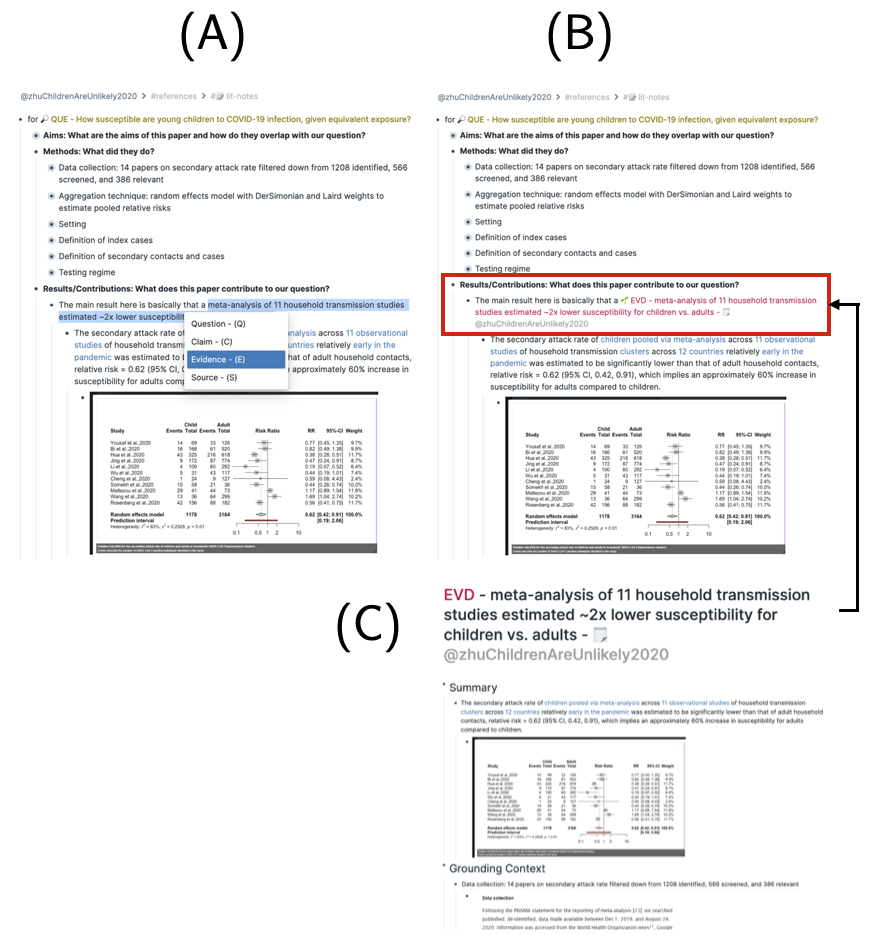}
    \caption{Our \texttt{Node Menu} component is designed to enable incremental formalization of discourse nodes. Similar to the user experience of annotation,  users select the text they want to “mark as” a discourse node, press a hotkey to initiate the node authoring action, then select the appropriate node type, either with the mouse or with the associated shortcut (1). The Discourse Graph extension then automatically creates a page with that title typed as the appropriate discourse node, and replaces the selected text with a reference to that node (2). If needed for further synthesis, users can also elaborate the nodes with more details over time, such as by migrating in screenshots of key tables and figures, into the body of a node page (3).}
    \label{fig:sys-nodecreation}
\end{figure}

Within Roam as a technical substrate, we prototyped a \textbf{\texttt{Node Menu}} extension to the UI as a means of authoring discourse nodes in an incrementally formalized manner. The intended feeling was to be similar to highlighting or annotating (rather than forcing structured data entry): we envisioned use of the \texttt{Node Menu} to factor parts of notes --- which would be a mix of informal and semi-structured notes --- into formal `nodes` of a discourse graph (claims, evidence, etc.). 

As an example, Figure \ref{fig:sys-nodecreation} (leftmost panel) shows an example of in-progress notes on a research paper, with a general content structure that is similar to a Google Doc of reading notes: a mix of informal and formal observations and structure, including general notes about related ideas, key details about methods, and the core results of the paper. Zooming in on the key results here, a user could mark out the key result of estimated 2x lower susceptibility as a key piece of evidence from the paper that might inform her synthesis for a focal question. The user experience of this would be very similar to annotating: she would simply select the text she wants to “mark as” evidence, press a hotkey to initiate the node authoring action, then select the appropriate node type, either with her mouse or with the associated shortcut (in this case, `E`; Fig. \ref{fig:sys-nodecreation}A). 

In doing so, the extension would automatically create a note that is structured with appropriate metadata (and possibly a template) and typed as an "Evidence" (EVD) discourse node, which would appear marked as a formal Evidence node --- the prefix “EVD” is automatically prepended to the page title to denote that this is an evidence node, and the @zhuChildrenAreUnlikely2020 citekey is appended to allow for easier return to the source material from the evidence page --- inside the previously informal notes (Fig. \ref{fig:sys-nodecreation}B). The user could stop in her formalization of the nodes at this point, though as her synthesis progresses, and she begins to need more contextual details while comparing and making sense of multiple evidence nodes, she could elaborate the nodes with more details over time, such as by migrating in screenshots of key tables and figures, or methodological details like participants and measures, into the body of the Evidence node page (Fig. \ref{fig:sys-nodecreation}C). 

\begin{figure}
    \centering
    \includegraphics[width=0.9\linewidth]{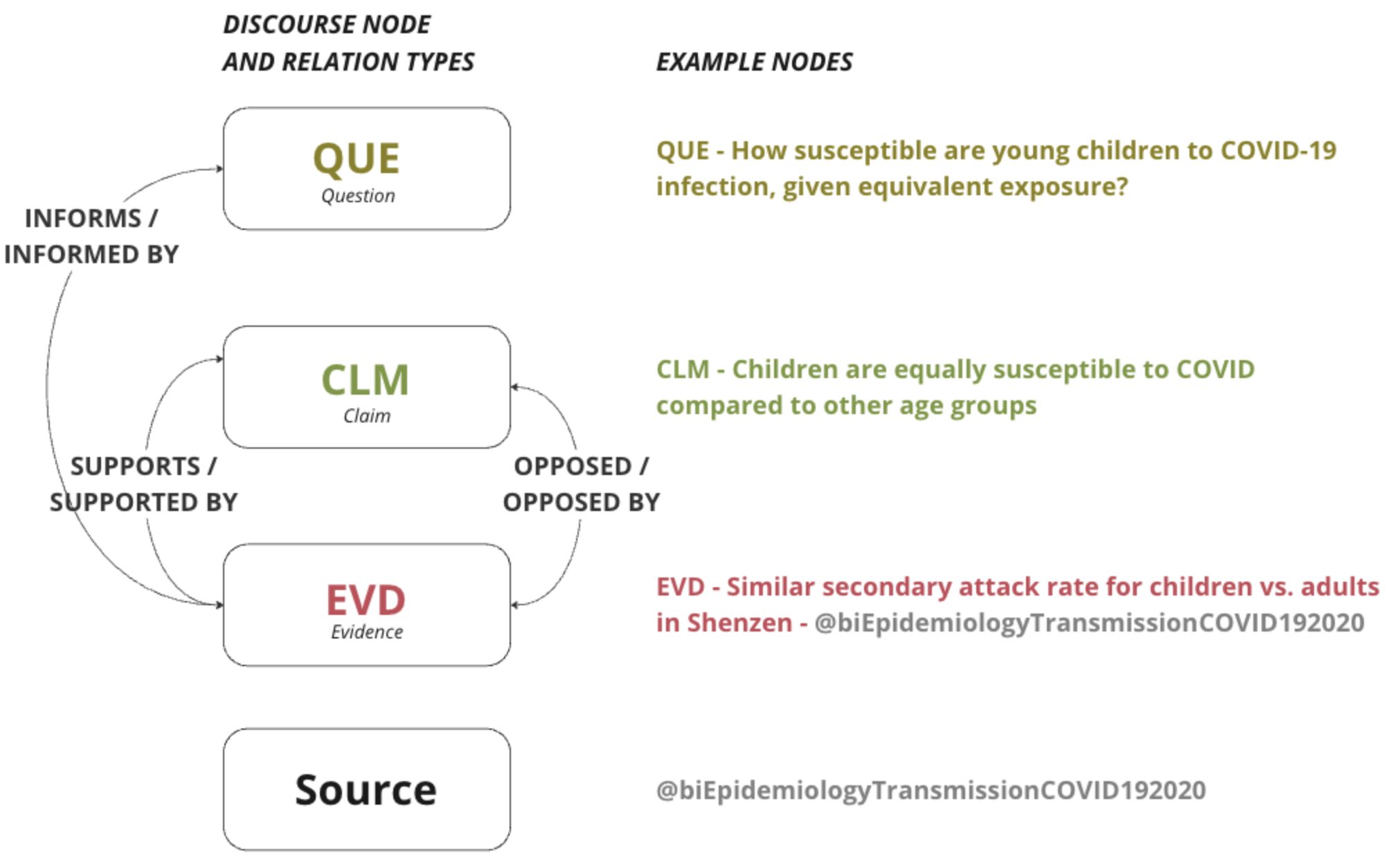}
    \caption{Base \model{} grammar that was shipped with the \software{}.}
    \label{fig:sys-base-grammar}
\end{figure}

At the start of our field deployment, our discourse graph extension shipped with the following base set of node types: 1) Questions, 2) Claims, 3) Evidence, and 4) Sources (see Figure \ref{fig:sys-base-grammar}). Claims and evidence mapped directly to elements of the discourse-oriented models we wanted to integrate with (e.g., hypothesis-evidence from SEPIO \cite{brushSEPIOSemanticModel2016}, claim-data from Micropublications \cite{clarkMicropublicationsSemanticModel2014}); questions were important from the formative field work for helping to organize and motivate synthesis work; and sources were a bridge to the paper-centric infrastructure of publishing (e.g., all evidence was directly tied to a source, which we hypothesized would enable writing and citing from a discourse graph).

The extension also shipped with a means for users to add new discourse nodes to their grammar (see Figure \ref{fig:sys-base-grammar-editor} in Appendix \ref{ap:designs}). This was mostly a debugging feature for the development team, but was also made user-facing, motivated in part by the learnings prior work on incremental formalization that suggested the need for users to expand a formal grammar to structure their thinking while adapting to local conventions and evolving thinking \cite{conklin1987hypertext}, as well as an intuition from infrastructure studies that these seeds of infrastrucure should, where possible, design for flexibility to avoid undue path dependence \cite{edwards2007understanding}). As we will see later, this design choice enabled user behaviors in the field deployment (see \S\ref{sec:extending-grammar}) that led to significant expansions in our design vision. 


\subsubsection{Naturally creating formal discourse relations between nodes by writing and outlining}

\begin{figure}
    \centering
    \includegraphics[width=\linewidth]{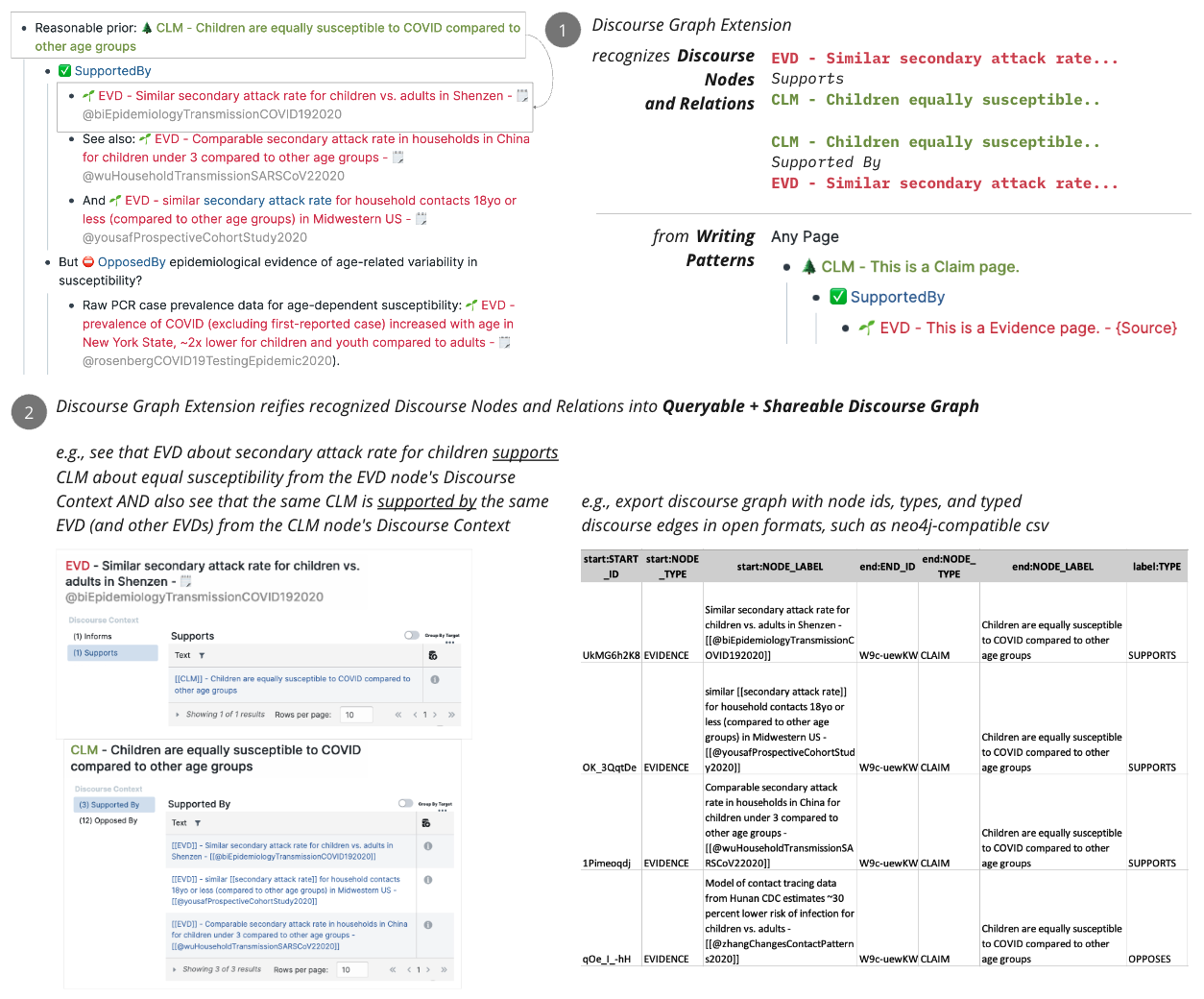}
    \caption{Our \texttt{Discourse Grammar Parser} parses formal discourse relation `edges` (e.g., support, oppose, inform) from user-defined conventions of prose writing and outlining (1) and reifies the recognized discourse nodes and relations into a queryable and shareable discourse graph (2). This makes it possible for users to write and outline very similarly to how you would normally write/outline for their own thinking, and author edges between nodes in a formal discourse graph as a natural byproduct of doing this intrinsically meaningful activity.}
    \label{fig:userstory_inlineannotation-to-graph}
\end{figure}

To enable seamless authoring of discourse \textit{edges}, we prototyped a \textbf{\texttt{Discourse Grammar Parser}} to enable the system to parse formal discourse relation `edges` (e.g., support, oppose, inform) from user-defined conventions of prose writing and outlining. The intent was to make it possible for users to write and outline very similarly to how you would normally write/outline for their own thinking, and author edges between nodes in a formal discourse graph as a natural byproduct of doing this intrinsically meaningful activity (in keeping with the principles of integrated crowdsourcing). 

The discourse grammar consists of a set of defined \textit{relation patterns} that map user patterns of writing, which are represented explicitly in a Datomic graph of blocks/pages, and relationships such as referencing and indentation (creating parent/child relationships between blocks), to discourse relations. 
Figure \ref{fig:userstory_inlineannotation-to-graph} sketches out the core user flow that enables this translation. The top left of Figure \ref{fig:userstory_inlineannotation-to-graph} shows a section of a scholarly outline for synthesis, where a user is writing and outlining her ongoing understanding for a focal question, with a typical mixture of formal and informal notes, and links to resources and references. Here, the user can reference specific results (evidence notes) while making sense of the case for and against a focal claim. Here, each pink EVD link is a link to an evidence node. 

In this outline, the first block of text is actually referencing a \textbf{claim} that children are equally susceptible to COVID compared to other age groups. 
This is done through a linking feature in Roam that is not so different from hyperlinking in Web-based communications, such as email. A child bullet (or block, in Roam parlance) references an \textbf{evidence} node from Bi et al (highlighted in pink); in this way, the inline citation to Bi et al is actually to a specific result from that paper in an evidence node, rather than a reference to the whole paper. 
We also see references to pink evidence nodes in the text. To record and reason through the evidence that supports the claim, the user could note in writing in the second block that the claim is "Supported By" some evidence, by referencing a tag with the title "Supported By". 
If so, the discourse graph grammar could then recognize the formal discourse relations amongst these nodes (see Figure \ref{fig:userstory_inlineannotation-to-graph}, Step 1), and reify them into a queryable and shareable formal graph, from which the user can reap immediate benefits, e.g., by exploring and querying their discourse nodes and relations (described in \S\ref{sec:enabling-queries}; see Figure \ref{fig:userstory_inlineannotation-to-graph}, Step 2). Indeed, this translation of implicit, textual specifications of discourse relations in writing would be in effect for all the discourse nodes referenced in the synthesis outline. Figure \ref{fig:userstory_inlineannotation-to-graph} (bottom right) shows a portion of the exported subgraph of discourse nodes and relations that could be parsed from the writing in the outline, in a Neo4j-compatible csv export from the Roam notebook. 

\begin{figure}
    \centering
    \includegraphics[width=\linewidth]{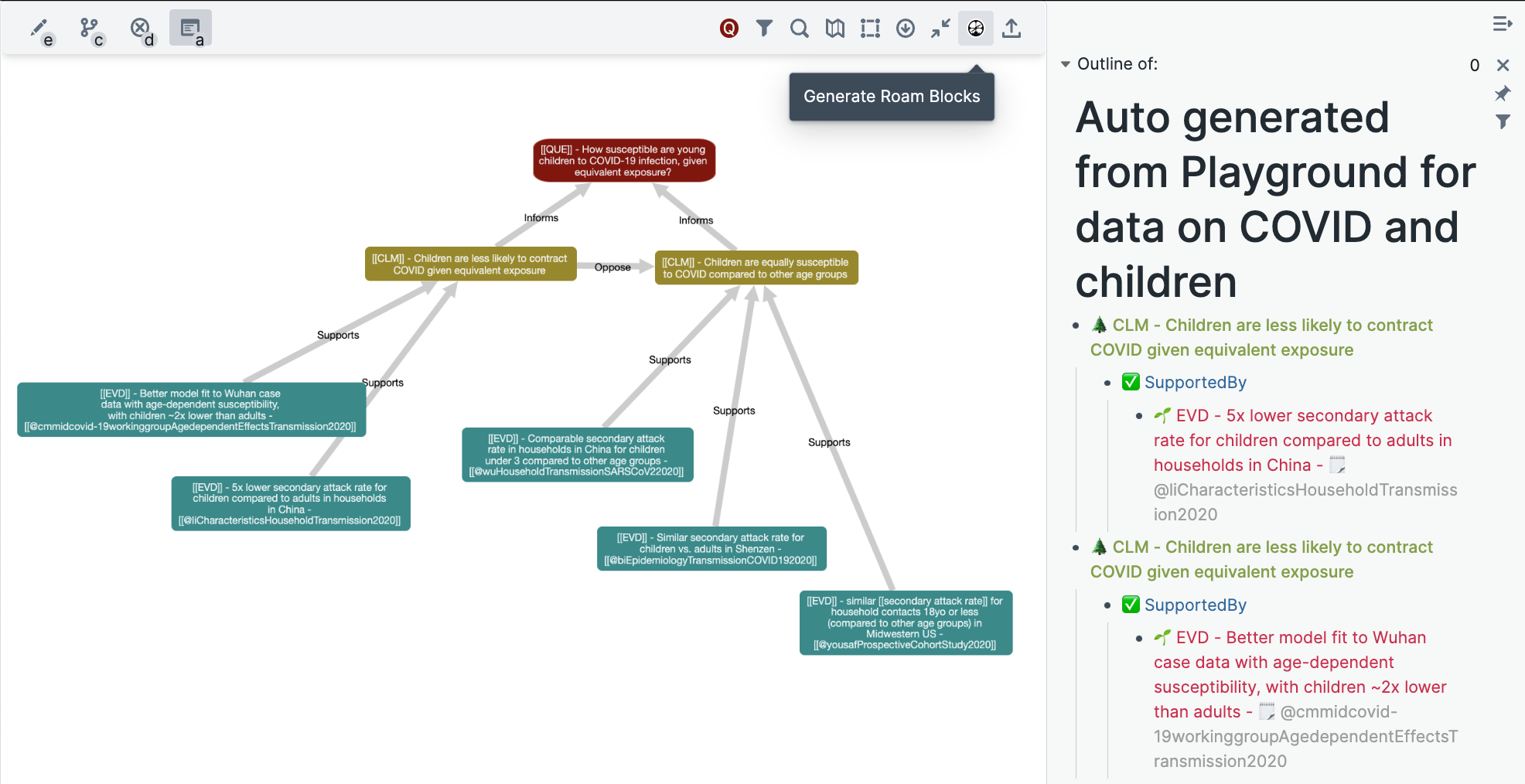}
    \caption{Our \texttt{Playground} interface provides an infinite canvas component where users could insert or create discourse nodes, visually cluster them, and draw edges between nodes; they could then make the nodes and edges "real" by triggering the parser to generate the associated Roam blocks in the formats that specify the discourse nodes and edges according to the relevant relation patterns defined in the grammar.}
    \label{fig:sys-playground}
\end{figure}

Separately, inspired by previous work on visuospatial sensemaking and visual hypertext systems (e.g., IBIS \cite{conklin1987hypertext}), we also prototyped a \texttt{Playground} infinite canvas component where users could insert or create discourse nodes, visually cluster them, and draw edges between nodes; they could then make the nodes and edges "real" by triggering the parser to generate the associated Roam blocks in the formats that specify the discourse nodes and edges according to the relevant relation patterns defined in the grammar.

At the start of our field deployment, our discourse graph extension shipped with the following base set of discourse relation types, focusing on relationships between evidence and claims/questions:
\begin{itemize}
    \item Evidence \textbf{INFORMS} Question
    \item Evidence \textbf{SUPPORTS} Question
    \item Evidence \textbf{OPPOSES} Claim
\end{itemize}

As with the node types, the extension also included a means for users to edit the in-built relation patterns, as well as add their own to match their own styles of writing and outlining (see Figure \ref{fig:sys-base-grammar-editor} in Appendix \ref{ap:designs}). 

\subsubsection{Enabling users to derive immediate benefits from their queryable and shareable formal discourse graph}
\label{sec:enabling-queries}

The final components of our design were designed to enable users to reap immediate benefits from their local discourse graph.

\begin{figure}
    \centering
    \includegraphics[width=\linewidth]{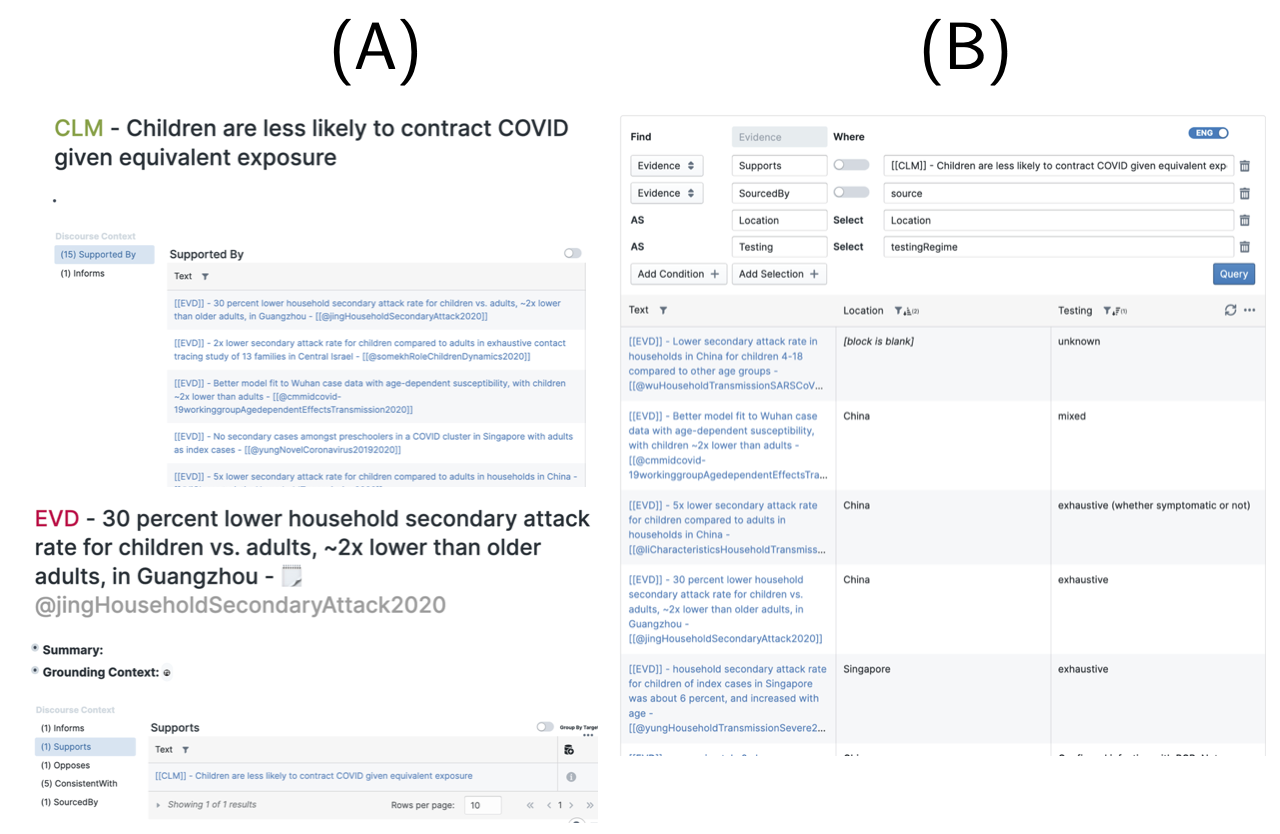}
    \caption{Our \texttt{Discourse Context} component was designed to provide associative trail navigation between notes based on discourse relations like inform/support/oppose (A). We also prototyped a \texttt{Synthesis Query} component to enable users to perform specialized querying based on discourse graph elements and relations (e.g., find all evidence that informs my question, but opposes some claim; find all evidence that informs a question and display key methodological attributes for each piece of evidence that I have specified (B). These queries could be used as collections, or as a means to gather discourse nodes to insert into a Playground or outline for synthesis.}
    \label{fig:sys-discourse-context-queries}
\end{figure}

We hypothesized that having a formal discourse graph in a hypertext notebook could help with synthesis in many ways, such as querying and sorting between claims based on evidentiary support/opposition. To explore this, we prototyped a \textbf{\texttt{Discourse Context}} component that provided additional associative trail navigation between notes based on discourse relations like inform/support/oppose (see Figure \ref{fig:sys-discourse-context-queries}A). This discourse context was designed as a slightly more formalized/focused version of the native Roam backlinks feature that enabled users to traverse bi-directional links between notes. Importantly, because the discourse relations are computationally 'real", discourse relations can be traversed from source to target, or vice versa, without requiring the user to tediously specify both directions of a discourse relation (e.g., EVD Supports CLM, CLM SupportedBy CLM): as Figure \ref{fig:sys-discourse-context-queries}A shows, once the parser has reified a discourse relation between nodes, such as a "Supports" relation between an EVD and a CLM, the user can see the EVDs a CLM is SupportedBy (from the CLM) page, and also navigate from any given EVD back to the CLM that it Supports. 

We also prototyped a \textbf{\texttt{Synthesis Query}} component to enable users to perform specialized querying based on discourse graph elements and relations (e.g., find all evidence that informs my question, but opposes some claim; find all evidence that informs a question and display key methodological attributes for each piece of evidence that I have specified; see Figure \ref{fig:sys-discourse-context-queries}B for an example). These queries could be used as collections, or as a means to gather discourse nodes to insert into a Playground or outline for synthesis.





\section{Evaluation I: Intrinsic Benefits of Discourse Graphs for Local Scientific Practices}

\label{sec:DeployReflections}

Our deployment yielded a rich picture of a possible future of a discourse-centric infrastructure being grown from transformed local scientific practices: a thriving ecosystem of personal and lab hypertext notebooks that researchers use to author local discourse graphs because they improve synthesis (\S\textbf{\ref{sec:improving-synthesis}}), enhance primary research and research training (\S\textbf{\ref{sec:improving-primary-research}}), and augment collaborative research (\S \textbf{\ref{sec:augmenting-collaboration}}). 

\subsection{Augmenting Local Synthesis Work with Discourse Graphs}
\label{sec:improving-synthesis}

\begin{figure}
    \centering
    \includegraphics[width=\linewidth]{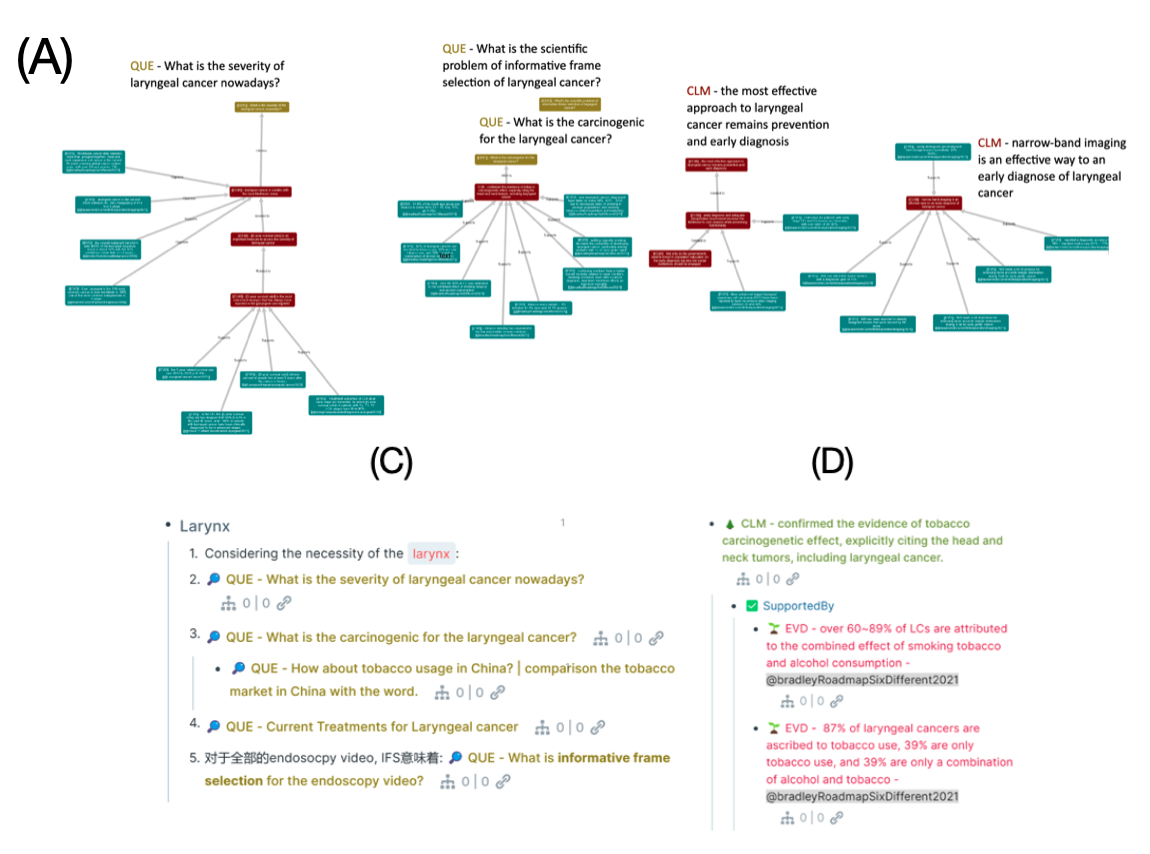}
    \caption{Snapshots of LZ's usage of the \sysshort{}. (A) shows how LZ visually maps questions (gold boxes), claims (maroon boxes) and evidence (teal boxes) in the Discourse Graph \texttt{Playground} to make sense of the "logic line" of a research project's contribution to the literature. The snapshot is annotated with the text of key questions and claims in the argument. (C) and (D) show LZ's question-centric usage of the \sysshort{}: he tracks questions for topics of interest, and elaborates on them with relevant claims and evidence. Here, (C) shows a series of questions around the importance, causes, and diagnostic approaches and issues for laryngeal cancer, and (D) shows elaboration of the question of the carcinogenesis of laryngeal cancer with relevant claims and evidence.}
    \label{fig:lz-snapshot}
\end{figure}

\subsubsection{Opening Vignette}
LZ is a PhD student in medical image processing, based in Shanghai, China. He came to know about the extension and approach via the initial whitepaper we published on the synthesis workflow \cite{chanKnowledgeSynthesisConceptual2020}. He learned how to use the extension specifically from one of the Roam notetaking courses in which we deployed the extension, as well as tutorials and examples from Youtube videos and other resources from the community of users.

He had been taking notes in Roam for awhile, but without much explicit structure. In one of his main writing projects, he was struggling to conceptualize the contribution of his work against the existing literature, and the extension enabled him to write out and trace the "logic line" that articulated the contribution of his work, enabling him to write a compelling introduction to his paper. In an interview, he shared\footnote{Zoom call 2023-09-29}:

\begin{quote}
   \textit{In my former studies...I read from about dozens of papers Then...I find that this logic line...I find that, and through another one and through another one, I find this is the... Oh, my God! If I not[sic] write it down, all these lines, my my brain will not understand this this story, like there's a line...this is amazing!}
\end{quote}

His notes on the questions, claims, and evidence were written in outlines and rough scratch notes throughout his notebook, but he was able to use the extension to gather the major discourse elements together to visually map out the "logic line" for his paper visually on the \texttt{Playground} (see Fig. \ref{fig:lz-snapshot}A)\footnote{Graph snapshots shared by LZ 2023-10-20}. We can see here how the project's core argument and motivation about the scientific problem of informative frame selection of laryngeal cancer is mapped out in a series of core questions and claims, from the severity of laryngeal cancer, to its carcinogenesis, to motivating early diagnosis as an effective approach to addressing this disease, and the specific technique of narrow band imaging that is studied in the project.

He shared in an interview in September 2023 --- following up approximately 1 year after he responded to our usage survey --- that he was still actively using the extension for his work, and has used the extension to assist with mapping out the argument for an additional publication. More generally, the "bread and butter" of his work with the extension (aside from intensive writing for a focused publication) is heavily question-centric: he writes regularly in his "daily notes", noting down new questions that motivate him, or elaborating on existing questions with new sources, claims or evidence (see Fig. \ref{fig:lz-snapshot} C and D)\footnote{Field notes and Zoom call 2023-09-29}.

\subsubsection{Using Discourse Graphs to Plan and Produce Research Artifacts.}

LZ's usage of discourse graphs to plan and produce research artifacts was echoed by a number of other participants. For example, as of Fall of 2023, AP had used the extension to research a chapter to be published in an academic collection of essays, and also to prepare a compilation of positions, extracted from statements made by governments at the United Nations in the last several years, to serve as a basic guiding paper for negotiations on a potential international treaty regulating international cooperation in disaster relief assistance and risk reduction. He shared that using the extension to see \textit{"some of the patterns of positions was particularly useful"}\footnote{Discord message 2023-09-25}. Another striking example was MB, who used the extension to deepen and speed up the drafting of his dissertation, including being able to write a draft of the first chapter of his dissertation (the introduction) in less than 2 days. He described the mechanics of it like this on a user call in terms of enabling him to map out and think through the structure of his thinking sufficiently that putting words on the page for a draft became much easier\footnote{Open office hours call 2022-04-07}: 
\begin{quote}
    \textit{"So when you go to writing I'm like, I'm not going to pick up that article again. I'm going to write from here, so if it's not here I'm in trouble right. So my first paragraph might be, there's a vast amount of theories that you know could supply the literature: Boyd discusses bargaining power, Bertrand discusses, you know, parental attribute theory, Gomez discusses whatever Gomez discusses right um. And boom paragraph done just because I have, you know, put this together."}
\end{quote} 
Similarly, GX shared in a usage survey that the extension enabled him to \textit{"easily sort out the framework and context of the article I want to write"}, especially because he was branching out from his prior background in public relations to criminology for his PhD work\footnote{Usage survey GX}.

\subsubsection{Using discourse graphs to address specific process issues in synthesis work.}
From our user calls and discussions and surveys, we got more of a picture of how the extension enabled users to \textbf{improve their synthesis work by addressing specific weaknesses or desire paths in their synthesis process}. One common dimension of this was mitigating issues from relying solely on memory for details and context. MB, elaborating more on his usage of the extension, described how he was able to \textit{"pull things up in seconds"}, in contrast to his peers in his comprehensive exam course, who often used programs like Word, and struggled to pull up details over the course of minutes\footnote{Field note 2022-03-11}. Additionally, in a call describing how the usage of the extension was impacting his work, MA said that he appreciated how the clearer documentation of takeaways from previous papers in terms of claims and evidence enabled him to rely less on only his memory from reading papers, and instead be far more rigorous and justified (and less wasteful, not running simulations on questions that already have good answers) when thinking and planning out parameters for computational models\footnote{Zoom call 2021-11-05}: 
\begin{quote}
    \textit{"[M]ore or less previously I was only sort of remembering as much as I could from the literature, from like an initial read through the literature: what's, what's, what's been done and what hasn't. And and I was running simulations that I thought would be interesting, but it wasn't rigorously justified that.. let's say I'm varying parameter X, maybe parameter X is already known from the literature, but I, I just didn't read the literature carefully enough."}.
\end{quote}
Similarly, in our usage survey, PO, a clinical researcher, said (in response to the question "Why did you start using the extension? What value is it providing to you now?"\footnote{Usage survey PO}: 
\begin{quote}
    \textit{"I read a lot of medical papers. A lot of times I don't remember who wrote this, how I reference that ... where did I read that ... I thought it would be an excellent tool to organize knowledge."}.
\end{quote} 
And QQ --- a physics PhD student --- noted, for the same usage survey question about value\footnote{Usage survey QQ}: 
\begin{quote}
    \textit{"I need to connect the claim with both research question and literature. I think it is helpful for me to check which literature it comes from."}
\end{quote}

Another prominent theme was the usage of the extension to \textbf{increase the signal to noise ratio in their personal knowledge bases}. For instance, AP said, in response to the usage survey question about value\footnote{Usage survey AP}:
\begin{quote}
    \textit{"it has helped me apply a metadata "language" to the links in my graph such that I can see patterns and relationships between different nodes (containing information), that would simply not be discernible from the "noise" of thousands of links."}.
\end{quote} 
Similarly, in response to the same question, CV, an independent sociologist, said \footnote{Usage survey CV}: 
\begin{quote}
    \textit{"This extension responded to my need to overcome the fragmentation of pieces of information that I had in Roam (pages named after a single concept, different words to refer to similar concepts, etc.). To solve this problem, Discourse Graph has been a wonderful tool, because it has allowed me to organize the information on a more aggregate level than the one allowed by the use of hashtags/pages."}.
\end{quote} 
And BO, a qualitative social researcher, described how the \texttt{Discourse Context} allowed for higher-signal ways to explore relations to other nodes and ideas compared to the in-build "backlinks"\footnote{Usage survey BO}:     
\begin{quote}
    \textit{"The "Discourse Context" section allows me to focus on important relations between my notes. In the Linked References section there is much more 'noise'."}.
\end{quote}
This comment about the \texttt{Discourse Context} providing a higher signal to noise ratio than the native backlinks feature is consistent with observations of usage, as well as data from the usage survey. In response to a question about the extent to which various \sysshort{} features were currently in use (with response options ranging from "Didn't know about this", "Never used", "Use on occasion", and "Core part of my workflow"), 70\% of respondents who self-identified as current active users of the extension reported that the \texttt{Discourse Context} feature was a core part of their workflow, and 17\% reporting that it was used at least on occasion. This was by far the most used extension feature (with the closest second being the \texttt{Node Menu}: 57\% reporting that it was core to their workflow, and 26\% reporting that it was used on occasion).

\begin{figure}
    \centering
    \includegraphics[width=\linewidth]{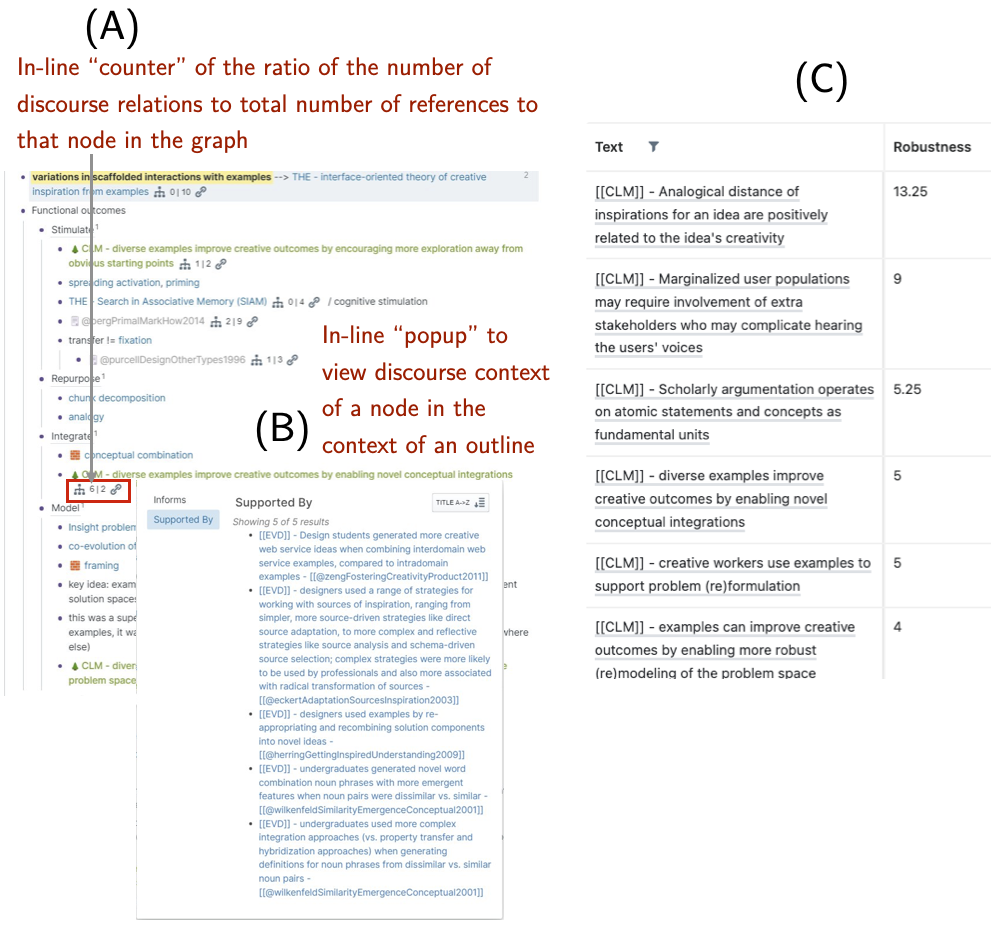}
    \caption{Snapshots of usage of the experimental \texttt{Discourse Overlay} and \texttt{Discourse Attributes} features for enhancing users' ability to leverage discourse nodes and relations in their synthesis work. 
    (A) Using the inline ratio of discourse relations to number of references for a discourse node to keep track of the "strength" of different pieces of an argument while outlining with discourse nodes. 
    Specific discourse relations and nodes are easily accessible via the overlay popup of the Discourse Context for each node (B). An experimental "discourse attributes" feature allows users to define named mathematical functions over the discourse relations of a node (e.g., weighted sum over number of supporting claims, evidence, and sources) that could be displayed in the overlay and used in queries, such as showing the most "robust" claims in a graph (C).}
    \label{fig:overlay-snapshots}
\end{figure}

One key design iteration that happened during the field deployment sheds further light on the potential of integrated discourse graphing to address specific weaknesses or desire paths in users' synthesis processes: we prototyped a \textbf{\texttt{Discourse Overlay}} feature that enabled users to see an in-line preview of the discourse context of a discourse node --- in the form of a count of the number of discourse relations for a discourse node, as well as the number of references to that node, appended to the right of each reference to that node in an outline --- as well as explore the discourse context for  the node --- by opening a "popup" overlay that displayed the \textbf{\texttt{Discourse Context}} for that node --- without leaving the context of a larger outline or argument they were being used in. The "ratio" of discourse relations to references was inspired by the citation "badge" from Scite.AI that shows separate counts of "higher signal" Supporting or Opposing citations to a paper, relative to the number of "mentioning" citations, to enable more nuanced reasoning over the impact of research papers than just raw citation counts. 

One core goal --- in response to user feedback --- was simply to increase the accessibility of the Discourse Context of each node, enabling more efficient retrieval and traversal/exploration of these "higher signal" relations and ideas. One comment from a user --- who was collaborating with the first author on a project --- illustrated this dimension of value, describing the Discourse Context "popup" as extremely helpful for quickly exploring the discourse context around a set of claims and questions in an outline (see Figure \ref{fig:overlay-snapshots}A)\footnote{Discord message 2022-01-04}. Another core goal --- also in response to user feedback about wanting to be better able to leverage their discourse nodes in their synthesis --- was to enable more "talkback" \cite{schonReflectivePractitionerHow1983,yamamotoRepresentationalTalkbackApproach1998} from the discourse nodes to guide and support development of ideas: for example, discourse nodes that are heavily used or referenced in an argument but with very few discourse relations might represent ideas that need further development; nodes that have many discourse relations might indicate well-developed ideas that are more ready for use in synthesis. The first author used this in his own writing and outlining for this exact purpose, to keep track of the "strength" of different pieces of his argument while outlining with discourse nodes (see Figure \ref{fig:overlay-snapshots}B). This value proposition was echoed by AP, who used the overlay extensively during synthesis, and shared this comment with us\footnote{Discord message 2022-01-14}:
\begin{quote}
    \textit{"[I]nitially the ratio did not register with me. However, as I started using the overlay more and more I started finding myself opening up claims etc that had interesting reference-to-link ratios. It is the "pull" or "weight" you refer to. It is really powerful and is at the core of the d/graph. You have set up a complex system, and it deserves a more attenuated or granular representation to really convey the power of the underlying relationship structure that is being developed"}
\end{quote}

We also prototyped an enhancement of this overlay to enable specification of custom \textbf{\texttt{Discourse Attributes}} to display in the overlay and use in queries of the discourse nodes (see Figure \ref{fig:overlay-snapshots}C). These discourse attributes could be a mathematical function of a node's discourse relations: for instance, one could define the "robustness" of a claim as a weighted sum of the number of supporting claims, evidence, and sources, perhaps also including a "penalty" for opposing claims/evidence.

Unfortunately, the overlay was very performance-intensive, and we were unable to find workable solutions for optimizing it, given the limitations of the querying mechanisms we had technical access to for the extension, which led to very noticeable and disruptive slowdowns in the user interface. These slowdowns led to the feature becoming mostly left unused due to performance issues. Even here, though, user reports of these performance issues underscored the value of the feature. AP's comment from the usage survey (in response to the question "What are your biggest pain points or wishlist items at the moment with the extension?") is particularly illustrative\footnote{Usage survey AP}:
\begin{quote}
    \textit{"I have stopped using the d/graph overlay because it slows my graph down significantly (an issue that has been there from the beginning). It would be great to have a solution for that because I have found that accessing the traditional context menu add extra clicks, i.e friction, which slows things down. After setting it up, there are two components of working: adding d/graph nodes and then afterwards processing the information from the d/graph (i.e. either looking at the relationships in the context menu or doing queries). The former is much easier/more efficient than the latter. While search/querying is powerful, it is sometimes overkill. If the overlay worked on large graphs/questions pages it would be a really good intermediate solution."}
\end{quote}

\subsection{Improving and Transforming Primary Research Work with Discourse Graphs}
\label{sec:improving-primary-research}

\subsubsection{Opening Vignette}
MA is a cell biologist who came to learn about the extension and discourse graphing approach while preparing to start up a new lab as a tenure-track faculty member at an R1 university in the United States. He uses computational modeling and fluorescence microscopy of stem cells to study cell biology and biophysics questions, like the mechanisms of cellular membrane bending by self-organizing cytoskeletal protein assemblies. The goal is to make experimentally verifiable, predictive computational models of self-organizing processes inside cells.

One motive for usage was to better bridge his computational theoretical work with experimental literature. In his experience, there was often very little formal connections between the two, such that modeling was not systematically backed by experimentally derived knowledge, limiting productive intercommunication between models and experiments. A concrete example of this: in models of the role of type-I myosin in endocytosis, MA needed to choose and study plausible value for such parameters as the binding radius, binding rate, and length of myosin-I. He used the extension to systematically map and collate experimental observations of this angle to inform his modeling work. 

\begin{figure}
    \centering
    \includegraphics[width=\linewidth]{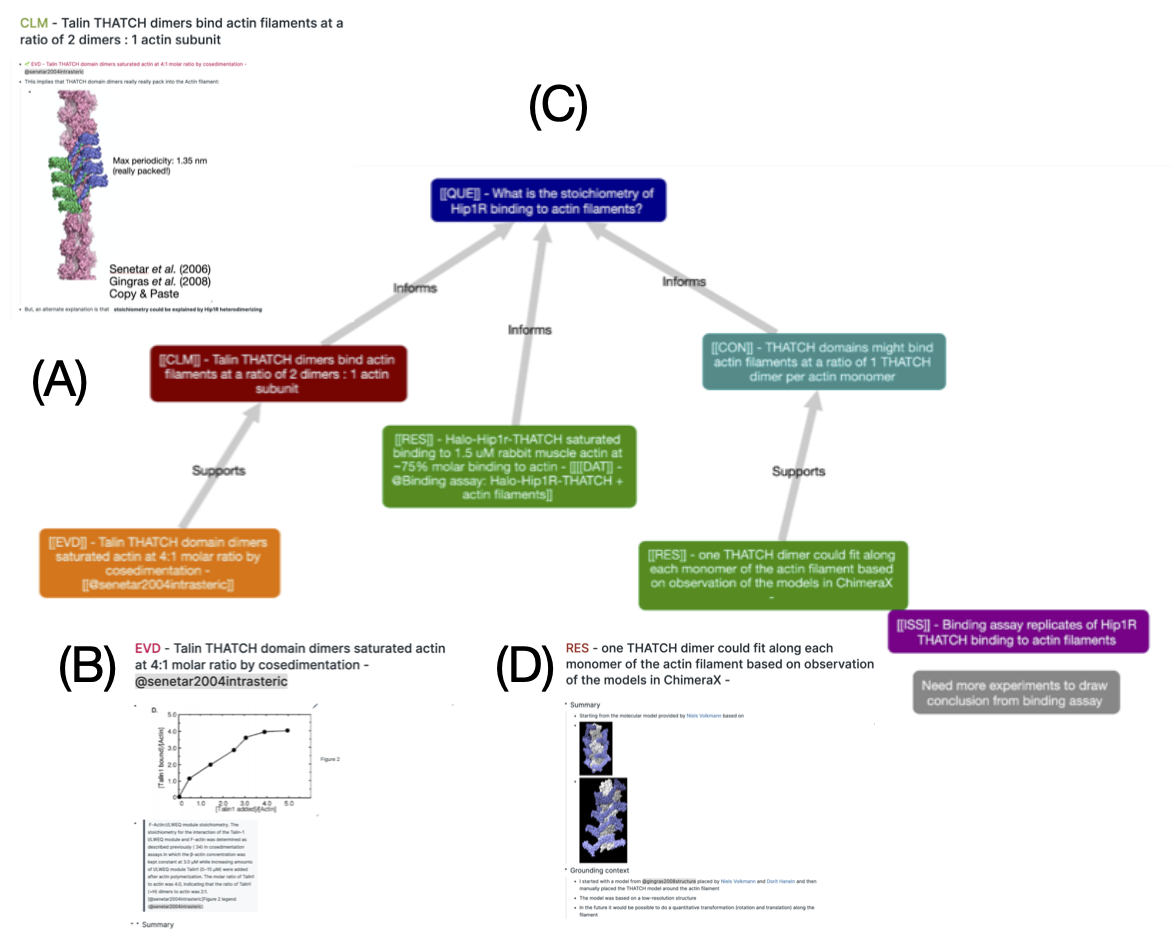}
    \caption{Snapshots of MA's usage of the \sysshort{} for structuring synthesis and primary research for a single project. (A) and (B) show a Claim and Evidence node from the literature that informs a driving question about the stoichiometry of Hip1R binding to actin filaments. This prior discourse is shown in context of a \texttt{Playground} snapshot (C) alongside Results and Conclusions for the same driving question. Note how the playground also includes an "Issue" node that describes an experiment that needs to be run to continue the inquiry, and a freeform note about what other experiments are needed. (D) shows the contents of a Result node about how one THATCH dimer could fit along each monomer of the actin filament based on observation of the models in ChimeraX.}
    \label{fig:ma-snapshot-personal}
\end{figure}

Very quickly, however, he began to adapt the extension to integrate into his day-to-day experimental work in simulations and at the  bench. He realized that he could adapt the ontological distinction between claims and evidence to his day-to-day work of forming and testing \textit{hypotheses} against experimental \textit{results}. This enabled him to retain rich discourse context for his inquiry, tracing his original experimental results directly to the context of past claims and evidence that bear on the same overarching question. He was able to extend the discourse grammar with new nodes (\texttt{HYP} (hypothesis) and \texttt{RES} (result) nodes to formalize this extension) using the grammar editor. Figure \ref{fig:ma-snapshot-personal} shows a snapshot of how he used the \sysshort{} to structure inquiry from literature and primary experimental work for a question about the stoichiometry of Hip1R binding to actin filaments. In the playground (Fig. \ref{fig:ma-snapshot-personal}C) he is tracking an overall question about the stoichiometry of Hip1R binding to actin filaments. He is questioning a specific claim about the ratio being 2 dimers to 1 actin subunit, based on a specific result from a prior study (indicated by the citekey @senetar2004intraasteric) that "*Talin THATCH domain dimers saturated actin at 4:1 molar ratio by cosedimentation*". The EVD node itself (\ref{fig:ma-snapshot-personal}A) contains a screenshot of the key figure that depicts the result, as well as a snippet of text from the paper that describes the result. MA has also elaborated on the specific methods context of the result ("muscle beta actin was used"), as well as some freeform notes about possible interpretations. Note also the new \texttt{HYP}, \texttt{RES}, \texttt{CON}, and \texttt{ISS} nodes that are added to the grammar to enable this integration of synthesis with primary research. From the structure, it is clear that \texttt{HYP} and \texttt{CON} are the primary research analogs to \texttt{CLM} nodes, while \texttt{RES} is the analog of \texttt{EVD} nodes.

\begin{figure}
    \centering
    \includegraphics[width=\linewidth]{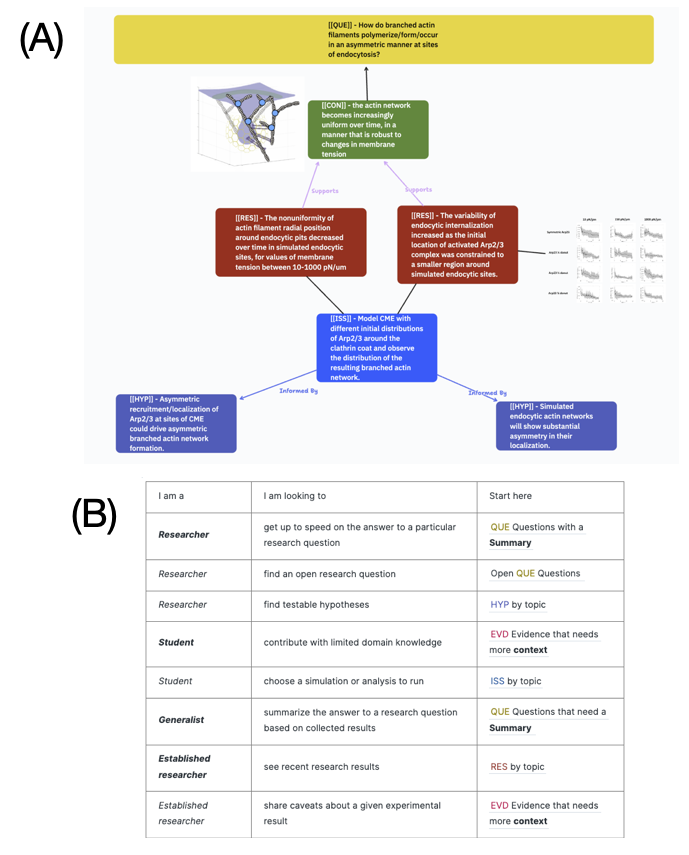}
    \caption{Snapshots of the usage of the \sysshort{} for structuring work in MA's cell biology lab. (A) shows how the extension enables a succinct documentation of the hypotheses, experiments, results, and conclusions of a single quarter rotation by a rotating PhD student, enabling modular attribution and credit tracking. (B) shows the construction of an onboarding "dashboard" that uses discourse nodes as entry points for contributors of varying levels of expertise, established researchers (who can find open research questions or testable hypotheses or specific experiments they can contribute to), to generalists (who can summarize what is learnt for a given \texttt{QUE} (question)), to students (who can elaborate on evidence nodes or run a scope simulation or analysis).}
    \label{fig:ma-snapshot-lab-usage}
\end{figure}

MA has since deeply integrated discourse graphs into the practice of his new lab, which he started up in the Summer of 2022. At the time of this writing, his lab includes 1 research scientist and lab manager, 2 PhD students, and 3 undergraduate students. MA is using the extension and the shared Roam graph to implement his vision for a more deliberate, nonhierarchical approach to lab science, that emphasizes empowerment and ownership of work by all participants, and a pace and openness of work that opposes the prevailing publish-or-perish hypercompetitive culture of academic science. His overall theory of change is that new collaboration tools can facilitate cultural and organizational change, and that using discourse graphs to break research into modular parts that can be shared and cited could facilitate getting up to speed on a new field of research, identifying holes in research to motivate new experiments, create new theories of understanding that synthesize interdisciplinary data, and regular exchange of discretized results between researchers. One concrete use of the extension for this goal was to more concretely structure and document contributions of students who rotate in for a single quarter (see Figure \ref{fig:ma-snapshot-lab-usage}A). MA also hired a research assistant "cybrarian" to specifically focus on the work of structuring the lab graph and developing and enforcing (software) conventions, including the onboarding dashboard shown in Figure \ref{fig:ma-snapshot-lab-usage}B that describes entry points into contributing to the lab's science for researchers of varying expertise and experience levels.

\begin{figure}
    \centering
    \includegraphics[width=\linewidth]{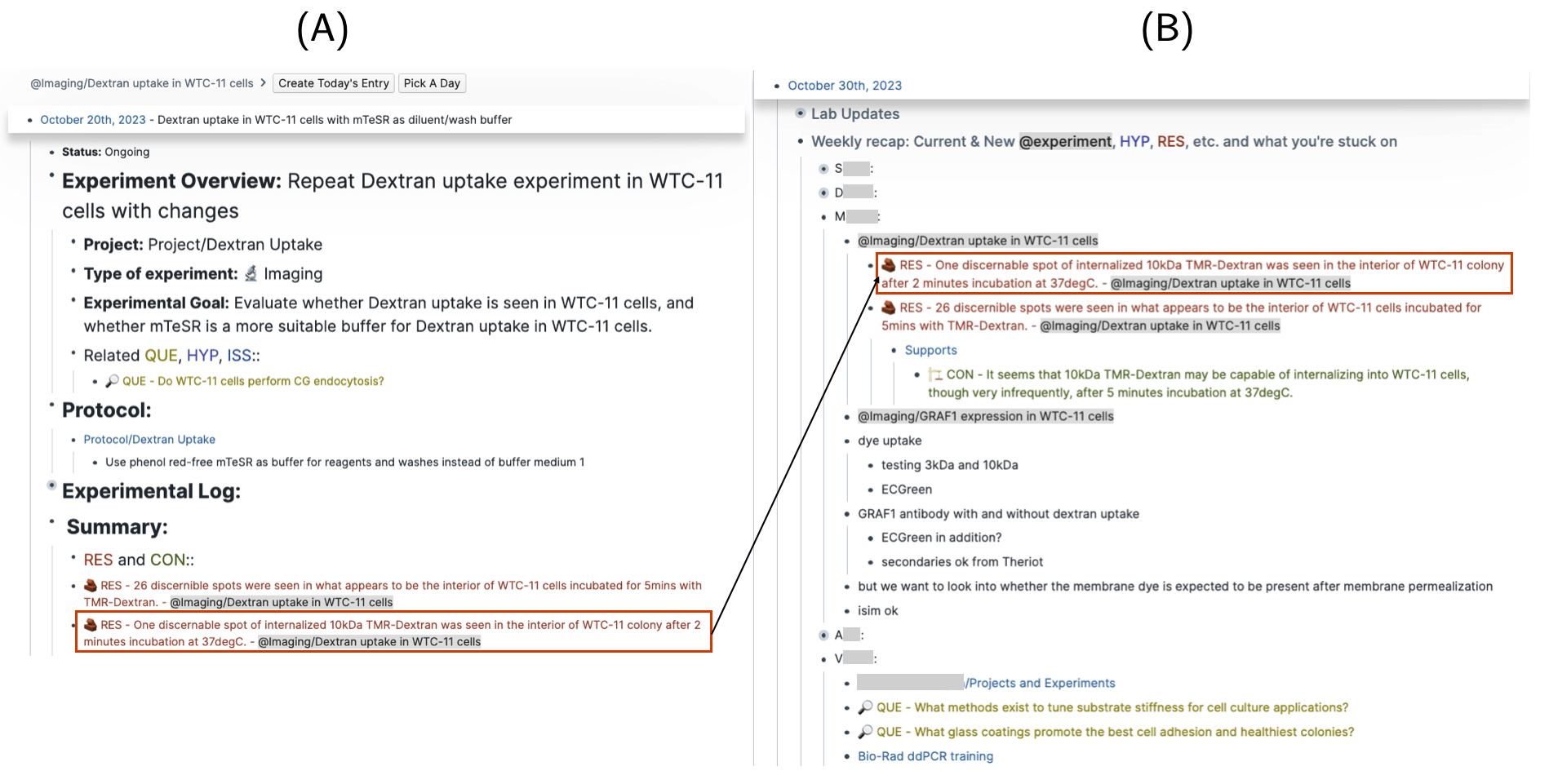}
    \caption{Snapshots of MA's lab's usage of the extension to structure primary research: (A) shows a recently completed section of an experiment documented in MA's lab graph by a lab manager, documenting the questions that motivate the experiment, and the tangible outputs of the experiment summarized as result nodes. (B) shows how the result nodes are then shared and discussed in lab meeting notes.}
    \label{fig:results-graph-matsu}
\end{figure}

Figure \ref{fig:results-graph-matsu} shows how the lab now uses individual experiment pages --- analogous to individual papers, though yet to be published --- , where questions are motivating the experiment, and the tangible outputs of the experiment are summarized as result nodes (Fig. \ref{fig:results-graph-matsu}A)\footnote{Page 2\_Yv78JFx in MA's lab graph}, that can then be shared and discussed in lab meetings (Fig. \ref{fig:results-graph-matsu}B)\footnote{Page iGKQlvlO- in MA's lab graph}.

\subsubsection{Using Discourse Graphs to Structure Ongoing Primary Research}

MA's vignette illustrates a surprising observation from the deployment: researchers could integrate the \sysshort{} to not only structure synthesis of prior literature, but also structure ongoing primary research. This was not surprising from a logical information model perspective: the information models that are translated in the Discourse Graph workflow, such as micropublications \cite{clarkMicropublicationsSemanticModel2014} and nanopublications \cite{kuhnBroadeningScopeNanopublications2013}, were after all proposed as an alternative model for reporting on primary research (instead of traditional narrative publications); as such, the elements of the base grammar map readily to ongoing primary research (e.g., a claim could be a "hypothesis" in development, and evidence could be a "result" from an experiment that was just completed). The surprise came from our initial sense that the workflows and practices for writing up primary research were too entrenched to change.


\begin{figure}
    \centering
    \includegraphics[width=\linewidth]{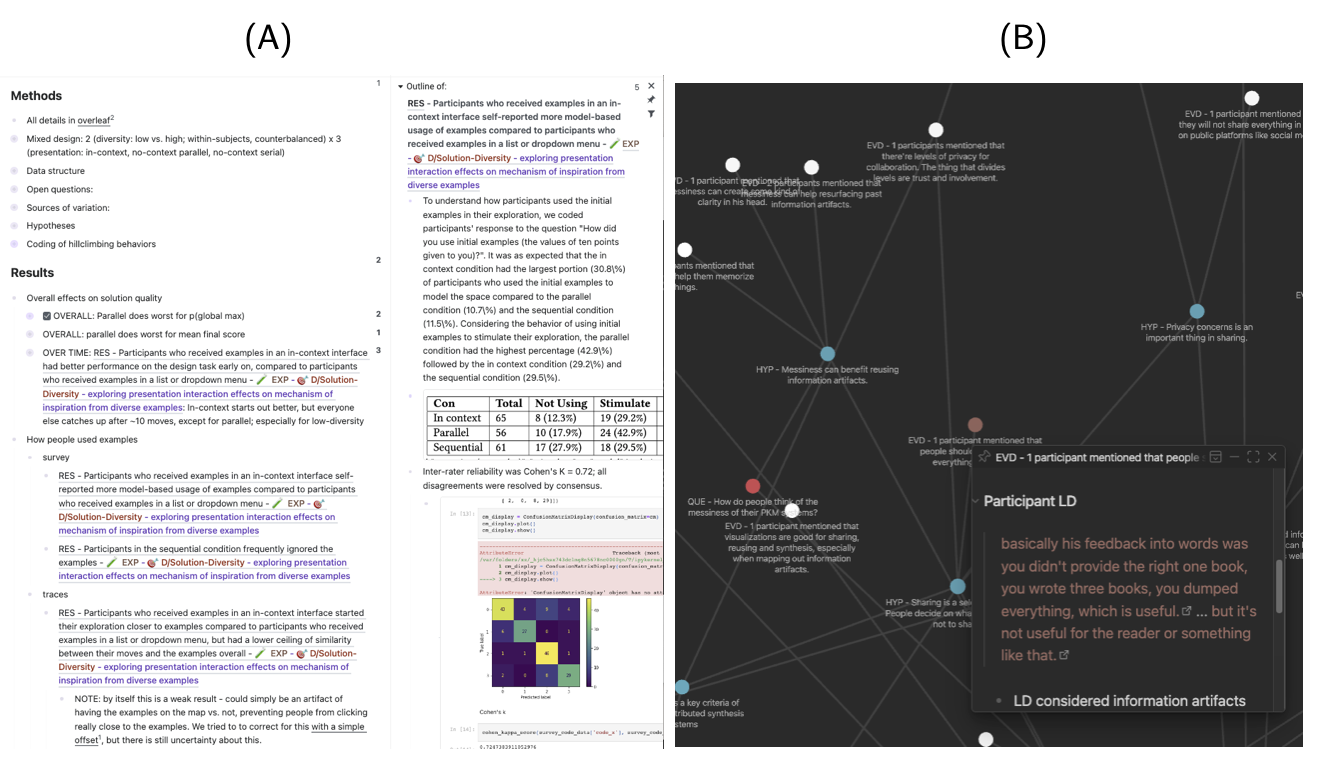}
    \caption{Snapshots of the first author's lab's usage of the extension to structure primary research: first for tracking the soldiv experiment results (A), and then for structuring ongoing qualitative analysis in terms of Hypotheses and Evidence (with links to qualitative data points).}
    \label{fig:results-graph-oasis}
\end{figure}

The first author has since adopted this ontology in his own work, starting with tracking documentation of results from an experiment and then thinking through how to sequence those results into a coherent larger contribution (Fig. \ref{fig:results-graph-oasis}A). More recently, the first author and one of his students have been using the discourse graph workflow to structure analysis of a qualitative study (Fig. \ref{fig:results-graph-oasis}B)\footnote{SIF analysis graph}; it should also be noted that the analyses for this paper itself were also structured in a discourse graph workflow, albeit in a different hypertext tool, Obsidian, and independently initiated by the student (more on this in \S\ref{sec:tool-transfer}). A related desire to use the extension to support primary qualitative research was also independently expressed by GM, an independent researcher, on the Roam Discord\footnote{Discord message 2021-11-20}:
\begin{quote}
    \textit{Are there any qualitative researchers here using the discourse graph for coding? I am conducting an interpretative phenomenological analysis on psychotherapists who use psychedelics in their private life: Using the discourse graph and query search to exhaust my data. I could do that quite easily, using the questions from the semi structured interview as QUE's, data from the QUE's as EVD to support claims for the themes.}
\end{quote}

\subsubsection{Using Discourse Graphs to Augment Research Training}

Related to this structuring of ongoing research, we observed the potential of the extension to \textbf{help beginning researchers learn the process of doing research and synthesis}. Indeed, the potential of discourse graphing to do this was the impetus for launching the field study. In the initial demo of the MVP system in August 2021 to a small set of power users in the Roam community to get a sense of the potential interest and value of the extension, LM, a senior PhD student who was teaching a dissertation writing in Roam in an online course 
said\footnote{Field note 2021-08-11}:
\begin{quote}
    \textit{I see this as a teaching tool. so many people come into a degree, with no idea how to even begin researching yes, so it could be sold as a, this is not a tool to use as you research, but this is a tool to teach you how to research.}
\end{quote}
That call led to an initial demo of the extension to a set of students the next day by LK, another senior PhD student who also taught an online course on academic workflows in Roam. 
The positive initial feedback from the students and the demo spurred a decision to deploy it for the next cohort of his course.

Separately, we also saw evidence of this "learning effect" from other sources of data from the deployment. For example, CV, an independent sociologist, described how the extension enabled her to think differently about how to organize her thoughts\footnote{Usage survey CV}:
\begin{quote}
    \textit{Discourse Graph has not only helped me as a tool, but it has made me conceive the process of organizing thought in a different way. It has helped me collapse different related pieces of information around one question/goal.}
\end{quote}
Similarly, SD, a cognitive science PhD student, said\footnote{Usage survey SD}:
\begin{quote}
    \textit{I liked the way it made me *think* about the literature, in terms of Questions/Answers}
\end{quote}

\subsection{Augmenting Collaborative Research with Discourse Graphs}
\label{sec:augmenting-collaboration}

\begin{figure}
    \centering
    \includegraphics[width=\linewidth]{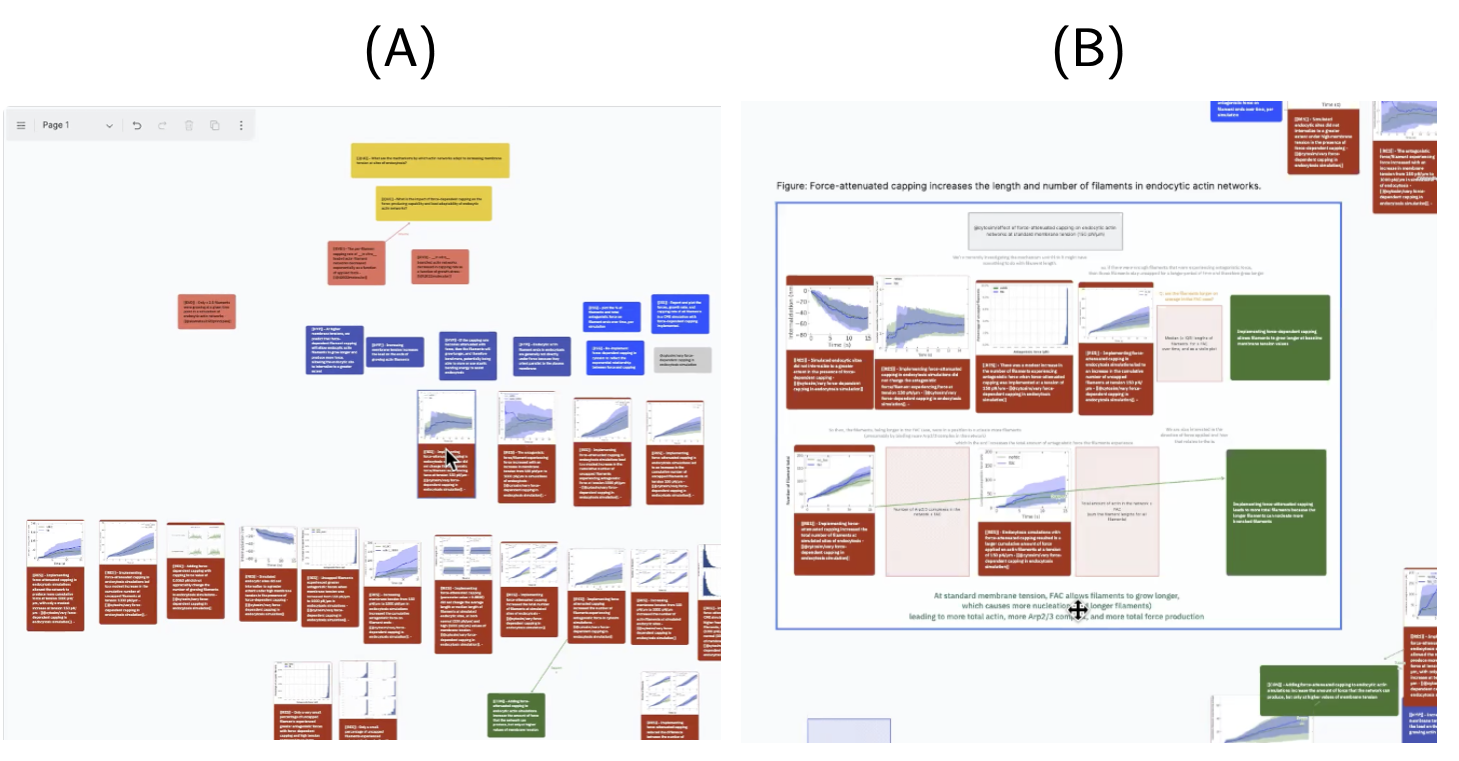}
    \caption{Snapshots of MA using the discourse graph to collaboratively synthesize experimental results with a PhD student to formulate conclusions for a core research question. MA and the student visually cluster result nodes, shown in maroon here (A), into the beginnings of a "compound figure" with results that substantiate shared conclusions, suitable for inclusion in a publication, shown in dark green here (B).}
    \label{fig:enter-label}
\end{figure}

\subsubsection{Opening Vignette} 
In the Fall of 2023, approximately 1 year into using the discourse graph to structure ongoing primary research, MA shared how he had used the discourse graph to work with a PhD student to synthesize and summarize a series of results from experiments. The process began with a series of queries over the results discourse nodes associated with their ongoing questions, experiments, and hypotheses for the project. After ``dumping'' these nodes onto a \texttt{Playground} canvas, MA and the student then used the canvas and nodes to formulate their conclusions for a core research question, structured in terms of a cluster of result nodes that would become the basis of a later compound figure in a publication (see Figure \ref{fig:ma-structuring-collaborative-work} C and D). Reflecting on the process, MA noted \footnote{Discord message 2023-11-25}:
\begin{quote}
    \textit{I did a 3-hour synthesis session with...(a PhD student) yesterday. I had him convert his results into result nodes (which he had been resisting) and we used a query to drop those nodes + related discourse nodes onto a canvas, to compose into what ended up being two figures. He commented a couple of times how the process called out where he thought he was keeping everything in his head but there were actually parts he had confused or forgot about or duplicated. Especially for the purpose of introducing and communicating the findings to someone else.} 
\end{quote}

\begin{figure}
    \centering
    \includegraphics[width=\linewidth]{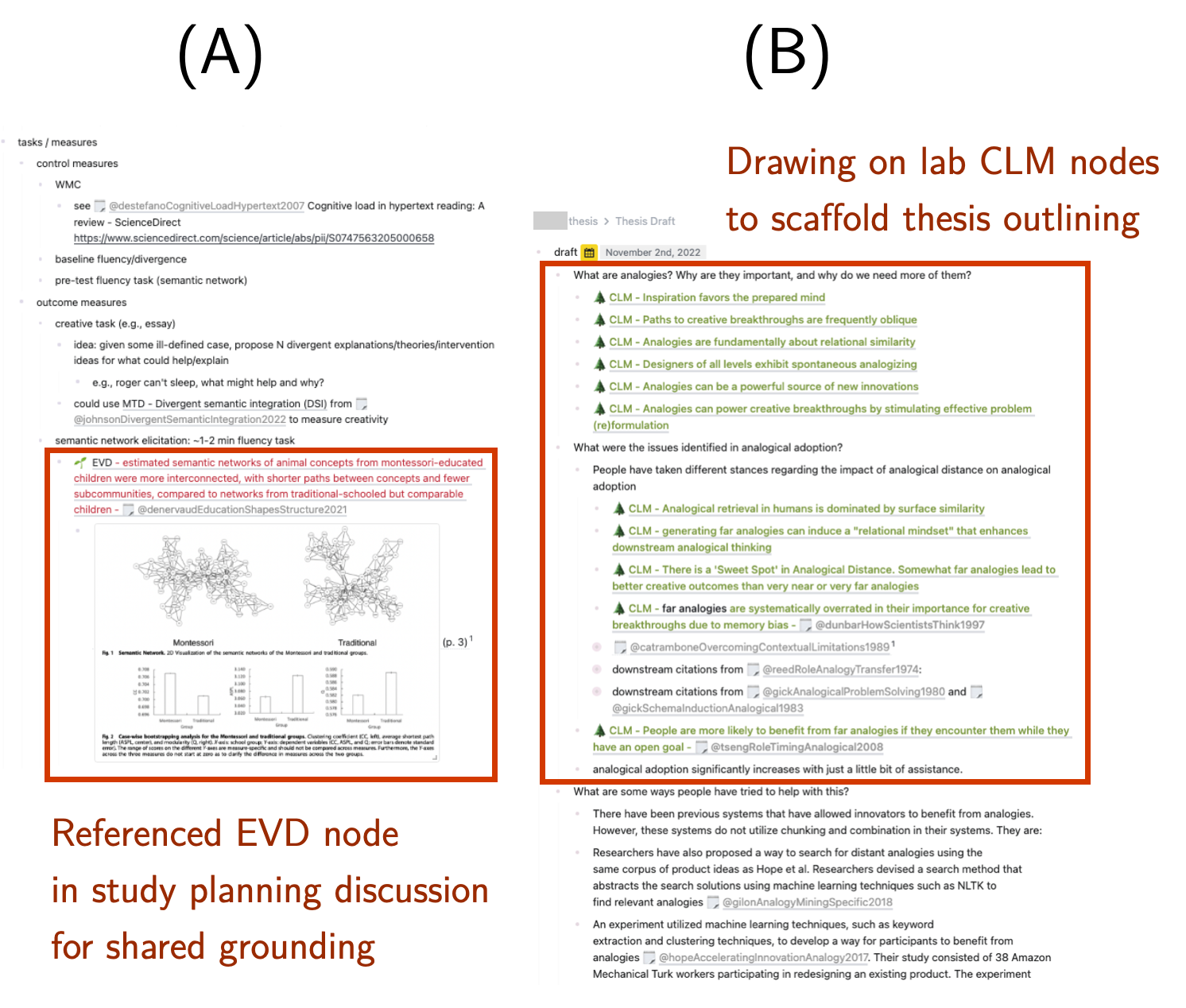}
    \caption{Snapshots of discourse graphs augmenting collaborative work in the first author's lab: (A) referencing a specific result as an example of a specific type of analysis, to reached shared grounding for planning studies with student advisees, and (B) a master's student scaffolding his outlining of his thesis introduction by building on the lab's library of claims for the formative steps of his argument, allowing him to focus on exploring and expanding the literature review closest to the unique contribution of his thesis.}
    \label{fig:collab-snapshots-oasis}
\end{figure}

\subsubsection{Discourse Graphs as Shared Context}

Collaborative usage of the discourse graph --- as illustrated by the preceding vignette --- was primarily directly observed in he first and fourth author's labs. 
In the first author's lab, the first author developed a practice of routinely \textbf{"bringing in" discourse nodes in collaborative 1-on-1 and group meetings to provide shared context, spur ideation, and leave pointers for followup work}. Figure \ref{fig:collab-snapshots-oasis}A shows one example of the first author referencing a network analysis result encapsulated in an EVD node in a group meeting to assist with the development of a study plan; the EVD node was used as a quick access point to retrieve an example of the semantic network analysis --- including concrete screenshots and methodological details that were attached to the EVD node --- that was being proposed for the study \footnote{Page sKd3L\_WST in first author's lab Roam graph}. In this specific example, the EVD note was created in March 2022, and referenced in the meeting more than a year later, in June 2023. 
In a note reflecting on this practice, the first author described its impact on collaborative work like this \footnote{Field note 2022-05-23}:
\begin{quote}
    \textit{really tough to kind of... push things through a straw (speaking), and really ineffective and inefficient to try to push through writing. can get alignment through shared sketching. but what about all that context? ...d/graphing is a way to structure your environment to make resources easier to access...bringing it to the table instead of a mere mention}
\end{quote}
Students and collaborators were also able to \textbf{draw on the growing discourse graph in the lab to scaffold their thinking and writing}. Figure \ref{fig:collab-snapshots-oasis}B shows an example of how AS, a master's student in the lab, was able to draw on the lab's discourse graph of claims about the role of analogies in creative work, and barriers to benefiting from cross-domain analogies, to scaffold his outlining of his thinking for his thesis: the CLM nodes provided efficient access points to key sources and empirical results to set the context for his problematization, and allowed him to focus on the more immediate literature context of his thesis contributions \footnote{Page a2sXaJP4u from first author's lab Roam graph}. RH and BL, with whom the first author is collaborating on a research project, also built on existing CLM nodes to propose new hypotheses \footnote{SS hypertext notebook} and also connect analysis of data to existing CLMs \footnote{Email 2023-11-18}.

\begin{figure}
    \centering
    \includegraphics[width=\linewidth]{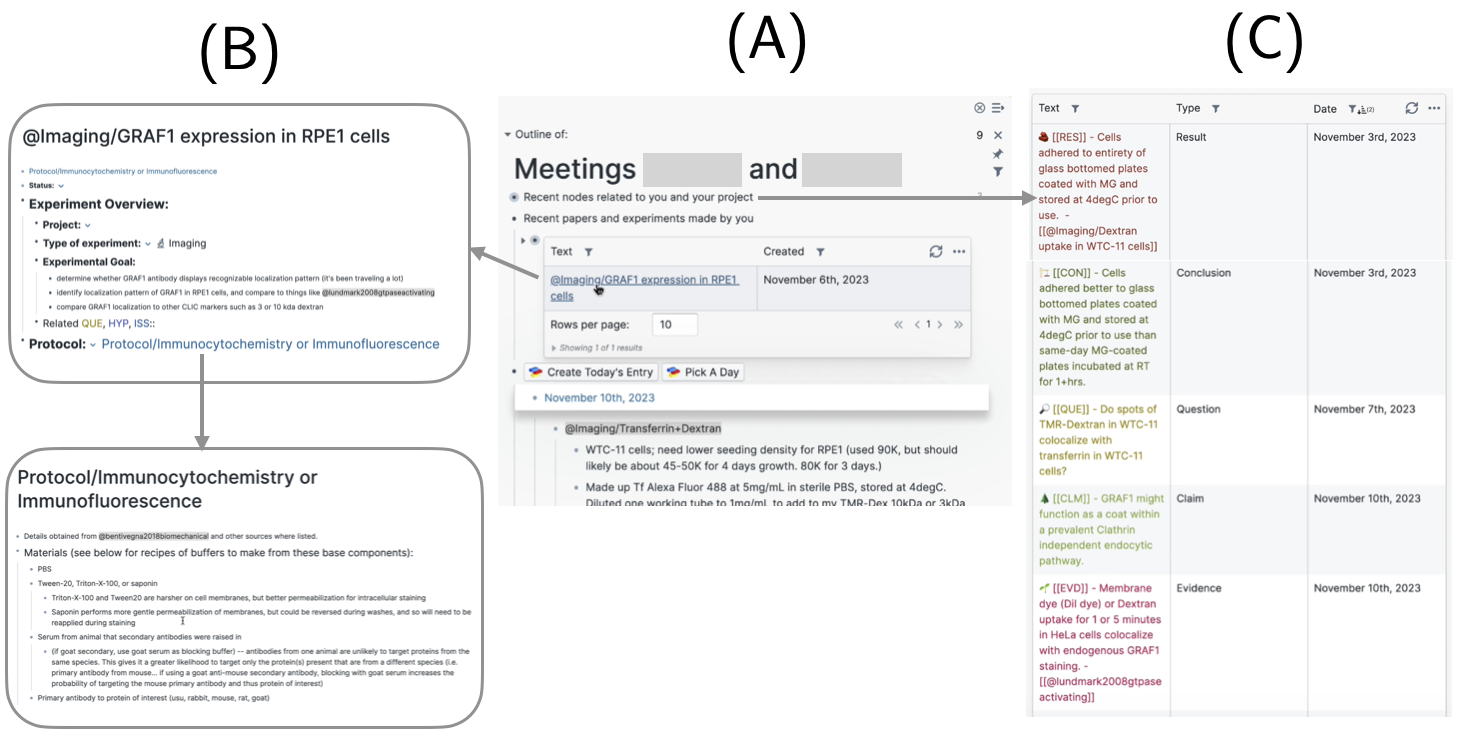}
    \caption{Snapshot of discourse graph usage enabling shared context in collaborative research meetings: MA uses meeting templates with structured queries over discourse nodes to "load context" before meetings (A). The associated questions, claims, evidence, experiments in the structured queries (C), as well as linked experiment nodes provide efficient access points for relevant details, such as experiment plans and protocols (B), enhancing contextualization.}
    \label{fig:ma-contextualizing-meetings}
\end{figure}

Similar examples of discourse graphs augmenting collaboration were observed in MA's lab. For example, MA has developed meeting templates that rely on \textbf{structured queries over discourse nodes to help him "load context" before meetings} (Figure \ref{fig:ma-contextualizing-meetings}A): the questions, claims, evidence, experiments, and results that are associated with a given meeting (Figure \ref{fig:ma-contextualizing-meetings}C) provide context and efficient access points for needed details, such as experiments and protocols associated with a specific result (Figure \ref{fig:ma-contextualizing-meetings}B)\footnote{Screen recording 2023-11-17; Page UA6e6XHVq from MA's lab Roam graph}:
\begin{quote}
    \textit{we have this running query about what types of discourse nodes they or we created, which are the scans for discourse nodes created by them, or that I created in this page, or that were created during our group meetings page under heading corresponding to their name. And it's sort of a running scan for the last two weeks, something like that. So that gives me that sort of continual instant query gives me a lot of useful information that I can look into...And so for example, if I say like, actually, I don't know enough about this experiment to comment on it, I can go click and like look at what has been done so far, like what protocol did they use, etc, etc. So all of a sudden I have enough context loaded, enough context loaded to be useful in the meeting.}
\end{quote}

\subsubsection{Discourse Graphs as Coordination Structure}

\begin{figure}
    \centering
    \includegraphics[width=\linewidth]{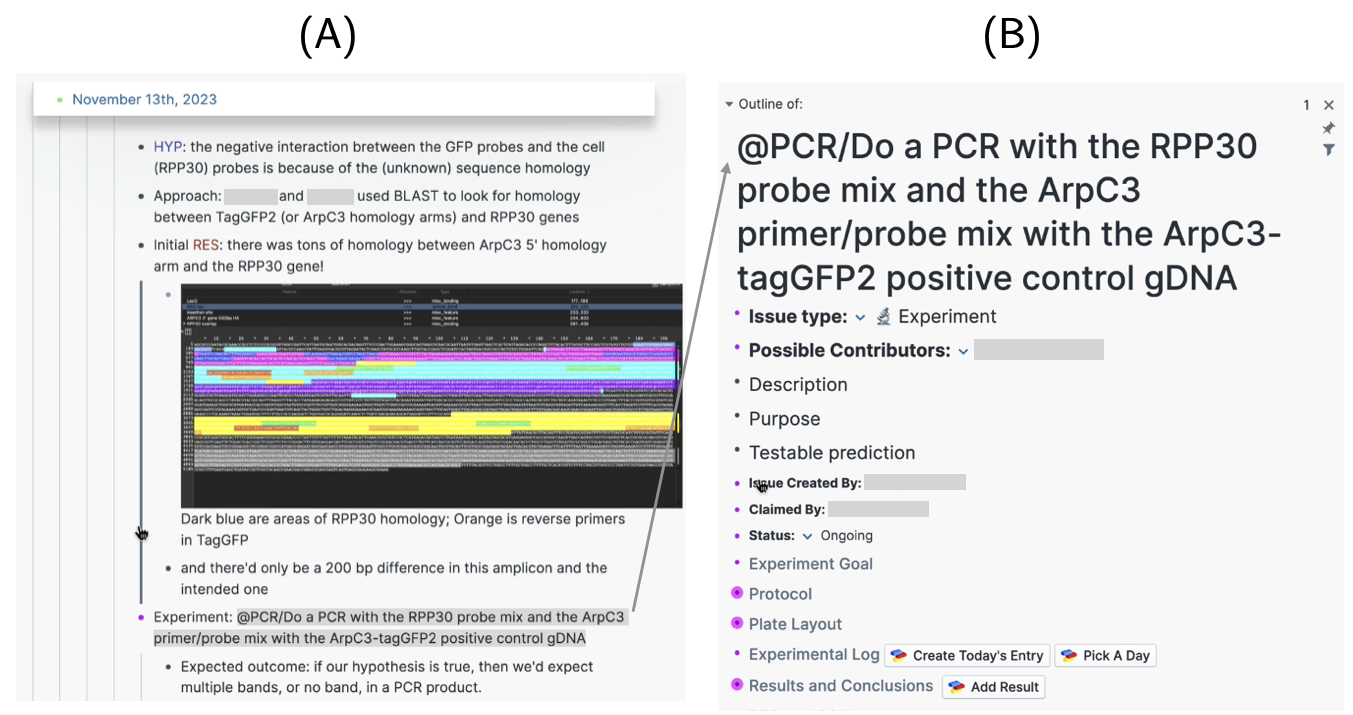}
    \caption{Snapshots of discourse graphs providing common language and context to structure collaborative work: weekly updates are summarized in terms of hypotheses and results (A), with issue nodes structuring follow-up tasks to test new hypotheses (B).}
    \label{fig:ma-structuring-collaborative-work}
\end{figure}

Elaborating more on observations of lab practices for structuring collaboration around discourse graphs, 
MA's group meetings often \textbf{use discourse graphs to provide common language and context to structure collaborative work (moving forward)}. One concrete example was describing and summarizing a weekly update in terms of hypotheses and results, and then creating an issue node to structure needed follow-up work, which was subsequently claimed by an undergraduate researcher, while retaining links to the context of questions and hypotheses (see Figure \ref{fig:ma-structuring-collaborative-work}A and B):
\begin{quote}
    \textit{my favorite example that happened recently is that I, you know, here's like a future issue for later. [student's name] started describing her project, and I converted it. I created an issue. Highlights, converts to an issue, and then you get like all this like information like who might contribute, and then somebody can claim the issue. And if they do that, then it gets converted into an experiment that they're running. And so the like the time lag between me creating and this issue and [the student] claiming it was like 30 seconds, which I think was great.}
\end{quote}

This structuring and focusing effect of discourse graphs for collaborative research was mentioned by a few other users, including enhancing collaboration of PhD students with their advisors \footnote{Usage survey QQ; Field note 2022-04-27}, as well as in a biomedical R\&D lab \footnote{Field note 2022-02-08}. Together, these data points provide an initial proof of concept that discourse nodes can be useful to someone other than the author, albeit within collaborations with a high degree of shared context, and often in synchronous settings.


\section{Evaluation II: Sociotechnical Preconditions for a Discourse-Centric Infrastructure}
\label{sec:DeployReflections-2}
In this section of our reflections, we describe how the local transformations of scientific practice described in \S\ref{sec:DeployReflections} were enabled by technical approaches for incremental formalization of lab and reading notes into discourse nodes and edges (\S\textbf{\ref{sec:incremental-formalization}}), as well as local extensions of the discourse grammar to cover local nuances of practice, including structuring primary research, and supporting synthesis work beyond empirical science (\S\textbf{\ref{sec:extending-grammar}}). We then describe evidence that these discourse-enabled lab notebooks can also be implemented in a range of technical substrates (\S\textbf{\ref{sec:tool-transfer}}), pointing the way to a future where a diverse set of discourse-enabled lab notebooks can be networked in a peer-to-peer fashion using a discourse graph model as a common data exchange protocol.

\subsection{The Possibilities and Limits of Incremental Formalization for Authoring Discourse Nodes and Relations}
\label{sec:incremental-formalization}
In our design work on \textbf{DR1}, we hypothesized that discourse graphing can be integrated into everyday scholarly work if it feels as easy as, and is integrated into, existing work like annotating and outlining. Our field data yielded partial confirmation of this hypothesis, mostly for node authoring, with some major design feedback around the continued need for smoother recognition of relations.

\begin{figure}
    \centering
    \includegraphics[width=\linewidth]{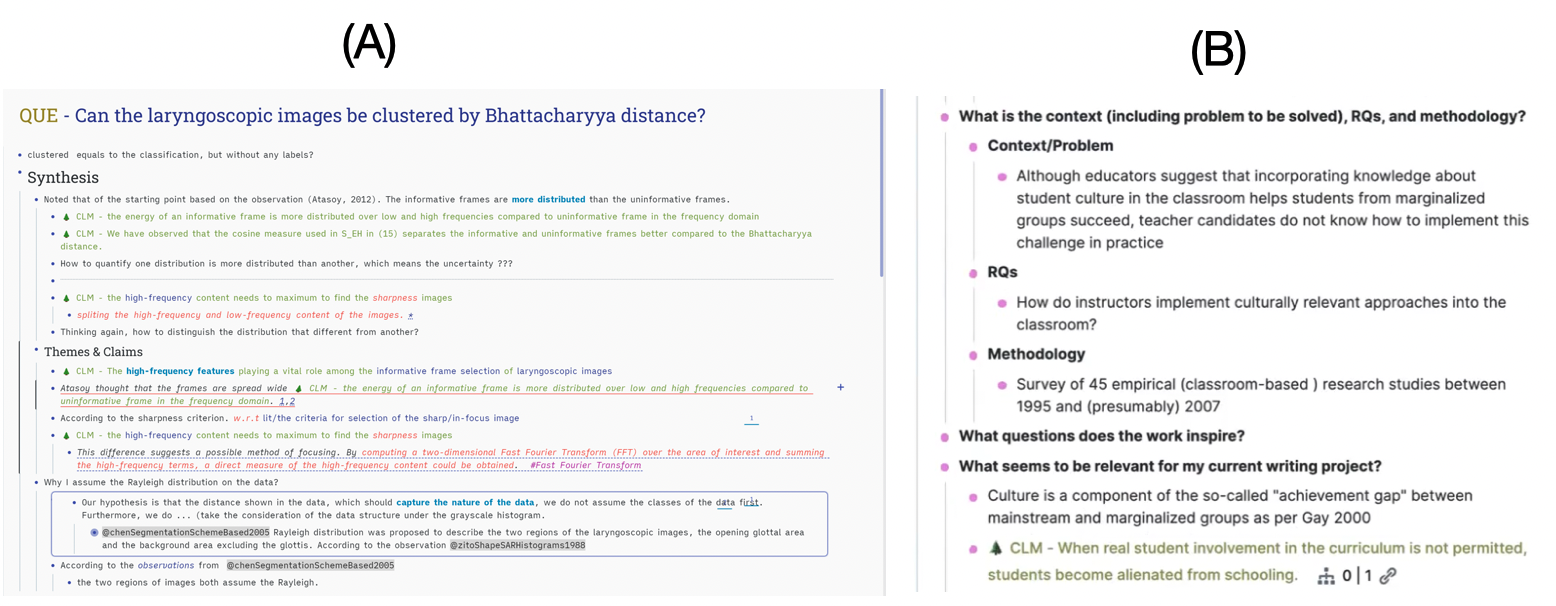}
    \caption{Snapshot examples of users integrating discourse nodes into informal and semi-formal writing in their Roam notes, collating informal notes alongside claims and evidence within a question page (A), and integrating a key claim while taking notes on potential takeaways for a writing project from a single paper (B).}
    \label{fig:node-edge-incremental-formalization}
\end{figure}

A range of data points suggest that \textbf{users were able to use the \texttt{node menu} to integrate authoring of discourse nodes into their synthesis work}. First, of the four users who shared a csv export of their discourse graph with us in the usage survey, we saw that these users created an average of 1.4k nodes, within the first year of usage (see Figure \ref{fig:node-frequencies} in Appendix \ref{ap:supplemental-data}). Additionally, from the self-report data from the usage survey on how important each feature was to their workflow, 60\% of users who (at the time) identified as active users reported that the \texttt{node menu} --- which enabled the create-node-as-annotation authoring experience --- was core to their workflow. We also saw frequent examples of integration of discourse nodes into semi-formal or informal text, indicating the intended use case of incremental formalization of ideas into discourse nodes where appropriate (as opposed to forcing all authoring to be only discourse nodes; see Figure \ref{fig:node-edge-incremental-formalization}\footnote{Snapshot shared in usage survey LZ; Open office hours call with MP 2022-04-20}). 

\begin{figure}
    \centering
    \includegraphics[width=\linewidth]{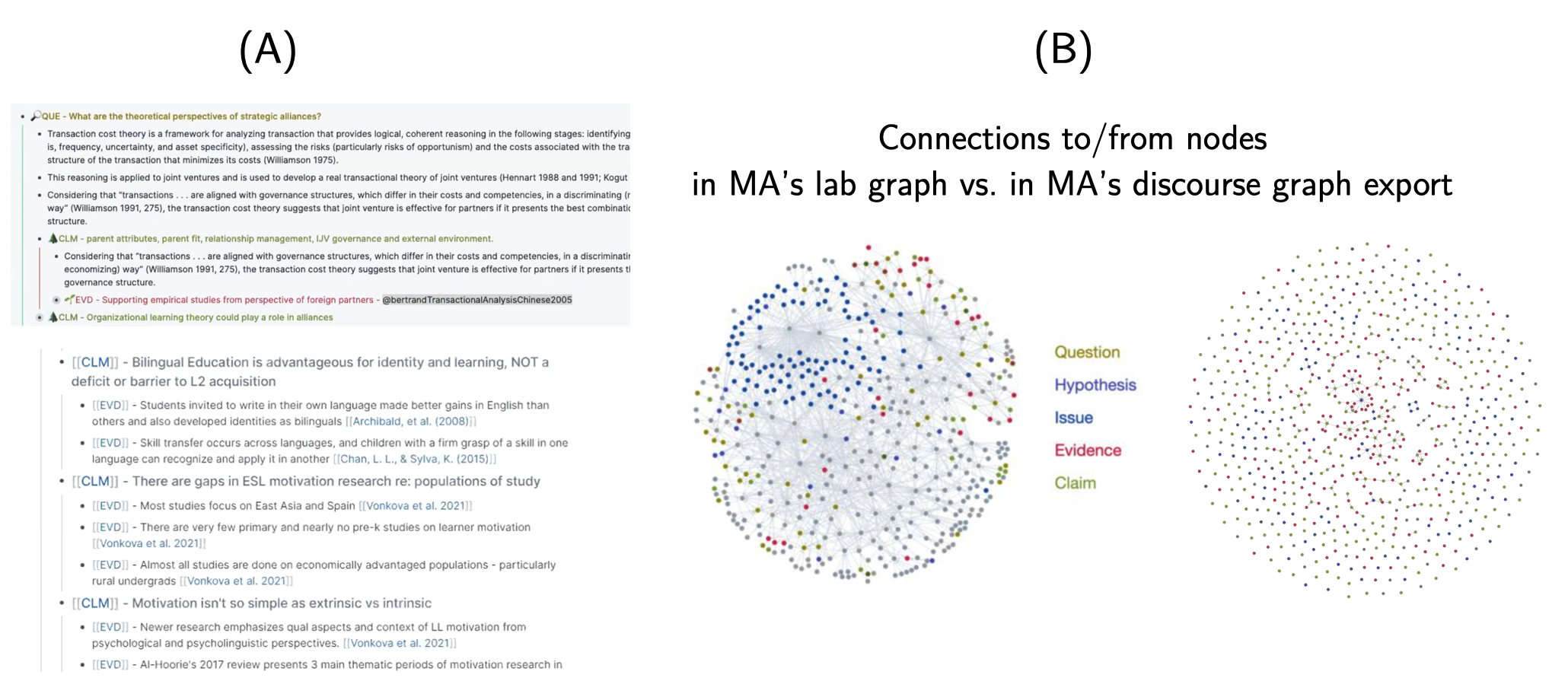}
    \caption{Snapshot examples of users relating discourse nodes in implicit or informal ways, such as indentation, rather than explicit discourse relations such as "SupportedBy" (A), and an example of the contrast between the richness and density of interconnections between nodes in MA's lab graph (as defined by these less formalized connections; here shown in a screenshot of graph overview visualization in their Roam graph) and the relative sparsity of discourse relations in the discourse graph export from the same lab graph (B).}
    \label{fig:limited-discourse-relations}
\end{figure}

The authoring experience for discourse relations was a little more complicated. In both user behaviors and feedback, \textbf{we observed desire paths for smoother and less effortful means of specifying discourse relations than was currently possible in the extension}. First, in discussions with users and observation of their usage patterns, we noticed that, while discourse nodes were frequently created and integrated into various forms of notes, such as notes on research papers or outlines for an argument, explicit relations such as \texttt{SupportedBy} were less frequently present. Instead, users relied on the built in low-friction means of relating nodes, such as simply indenting one node under another, and relying on context or (often unwritten) convention to interpret the relations. Figure \ref{fig:limited-discourse-relations} shows some examples of this behavior (relating nodes simply by indenting them; A)\footnote{Open office hours call 2022-04-27; Zoom call 2022-09-29}, as well as an example of the consequence of this for the density of the discourse relations that are actually encoded vs. implicit in the less formal connections in the graph based on indentation and other forms of connections (B)\footnote{Data exports from MA lab 2023-11-04; Field note 2023-11-04}. 

This lack of explicit discourse relations was less of a limiting factor than we anticipated: users were still able to use the native "bi-directional links" in the Roam graph to navigate and retrieve valuable information from their discourse nodes, as documented in \S\ref{sec:improving-synthesis}. CS, a library and information science PhD student, illustrated how this was possible. She was using the extension to collate key takeaways from across literatures on second-language learning, brokerage, and immigrant acculturation, but was not able to share a usable export of the discourse graph for the usage survey because there were essentially no discourse relations. She explained, though, in a followup call, how she was still able to get value out of her discourse nodes\footnote{Zoom call 2022-09-29}:
\begin{quote}
    \textit{"I know that I haven't done it properly in this case, but it doesn't matter, because I can still effectively click on this and see them all, and I can see what ones they're attached to. So it doesn't matter to me that I can't necessarily see, because I can still see TESOL claim evidence. I have my pop. Pop. The things I need to know, plus where it came from. So I don't care that I did it wrong."} 
\end{quote} 





\subsection{The Importance of Extending and Personalizing the Discourse Grammar}
\label{sec:extending-grammar}

As noted in \S\ref{sec:DesignArc}, we anticipated some extension of the discourse grammar, chiefly to cover ways of expressing discourse relations between the base set of questions, claims, and evidence that we may have missed in our initial design. We did not, however, anticipate the \textbf{extensive expansion of the base discourse grammar by our users, along with intensive use of the grammar editor}. To orient with some basic statistics from the usage survey, ~40\% of users self-reported the grammar editor as core to their workflow, with an additional ~40\% reporting at least occasional usage. Less strictly quantifiable was the frequent discussions of extending/personalizing the discourse grammar on our main Discord channel, and in open office hours and other forums; as a quantitative snapshot, a search for the term "grammar" in our main Discord channel showed 57 individual messages referencing the term "grammar"; many of these messages were in the context of conversations or led to threads discussing how to define or extend the grammar. Qualitatively, one of the core users, AP, created his own video describing his grammar expansion for legal research\footnote{AP video walkthrough of discourse graph set up}, which was mentioned by several other users as a helpful resource\footnote{Open office hours call 2022-04-04}.

\begin{figure}
    \centering
    \includegraphics[width=\linewidth]{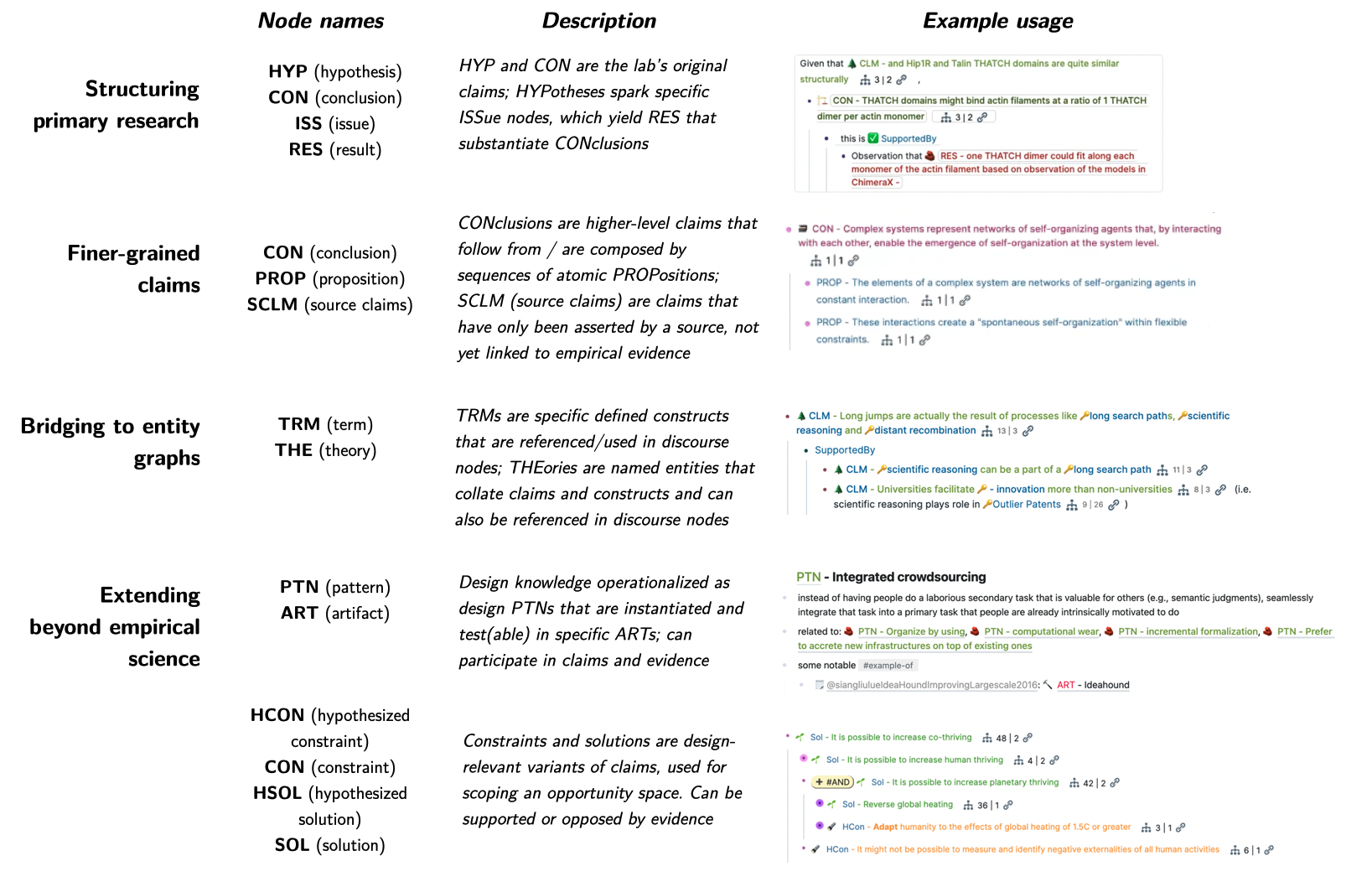}
    \caption{Example snapshots of key user-driven expansions to the base discourse grammar in terms of new discourse nodes, spanning functions such as structuring primary research, finer-grained claims and argumentation, bridging to entity graphs, and expanding beyond empirical science.}
    \label{fig:grammar-expansion-snapshots}
\end{figure}

The primary way that users extended the discourse grammar was by \textbf{defining new types of nodes}. Figure \ref{fig:grammar-expansion-snapshots} summarizes some key new types of nodes we observed in usage \footnote{Usage survey MA; Open office hours call with MP 2022-04-20}. Some nodes --- such as HYPotheses, or RESults --- were created to support primary research (described in more detail in MA's usage vignette in \S\ref{sec:usage-vignettes}, as well as in \S\ref{sec:improving-primary-research}. Others were created to enable finer-grained claims and argumentation, such as distinguishing between higher-level CONclusions and more atomic PROPositions. TRM (term) nodes were created to bridge the discourse nodes to more defined entities and constructs. 

\begin{figure}
    \centering
    \includegraphics[width=\linewidth]{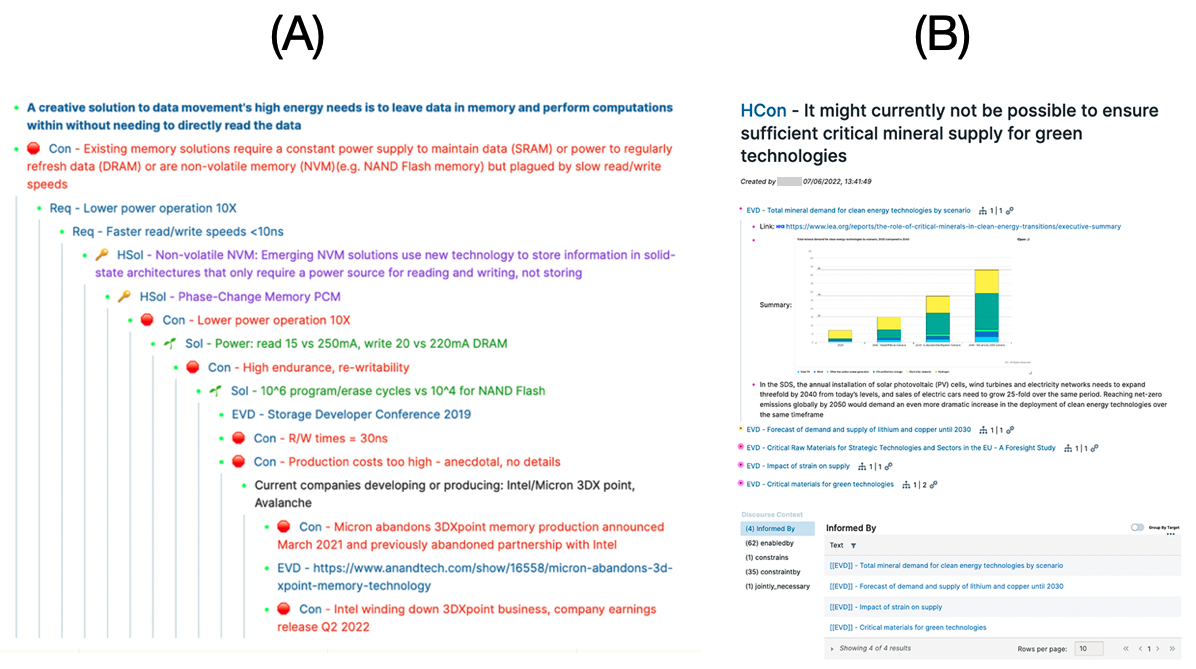}
    \caption{Snapshots of DSV's usage of the \sysshort{} to scope \textit{constraints}, \textit{requirements}, and \textit{solutions} in an opportunity area for potential investment, alongside scientific \textit{evidence}. (A) shows a scoping tree of requirements, constraints, and solutions for the high level grand challenge of energy efficient data systems, and (B) shows the contents of a page describing a hypothesized constraint around the uncertainty of being able to ensure sufficient critical mineral supply for green technologies, which is grounded in references to specific evidence nodes. The page also surfaces (via the \texttt{Discourse Context} component) specific discourse relations of "enabledby", "constrains", and "jointly necessary", as well "informed by" relations to important pieces of scientific evidence.}
    \label{fig:dsv-snapshot}
\end{figure}

Some nodes were also created to extend beyond empirical science. 
For example, DSV is a venture capital firm that focuses on investing in companies that translate frontier science into products with high impact on grand challenges such as regenerative agriculture, computation-assisted collective intelligence, climate change, and curative therapeutics. They extended the discourse grammar to enable explicit tracking of \textit{opportunities}, \textit{constraints}, and \textit{solutions}, and their grounding in scientific evidence nodes. This enabled them to reason much more carefully in the process of \textit{scoping} opportunity areas, including assessment of the necessity and sufficiency of particular venture areas (based on interrelations between constraints and solutions) for grand challenges alongside their level of readiness based on scientific evidence (based on relationships between constraints and solutions and evidence; see Figure \ref{fig:dsv-snapshot}). In a document describing their motivation for developing their usage of the extension, they wrote\footnote{Document 2022-09-22}:
\begin{quote}
    \textit{[W]e need systems to help us make structured claims and arguments about the possibilities of applied science. We want to be able to answer questions like, "What are the known constraints, both technical and commercial, on achieving societal outcomes?", "What specific pieces of knowledge currently exist related to those constraints?", "How might we uncover counterintuitive or non-obvious intervention points?", "What kinds of expertise are likely to interact effectively with these pieces of knowledge, and who has this expertise?", and "How can we stimulate combinatorial innovation - getting people across domains to speak the same language and making knowledge easy to integrate?"}
\end{quote}

This grammar extension was developed with the first author's assistance on a number of user calls, and inspired improvements to the \sysshort{} grammar editor (see \S\ref{sec:extending-grammar} for more details). While they found value in the way that the extension enabled them to structure their thinking, they have since moved on to use their enhanced discourse grammar in a less formal way to structure their thinking and conversations, and their new experiments with a language-model-powered scoping assistant, which also operates on opportunities, constraints, solutions, and evidence (along with connections and chaining between these node types) as primitives. EM, the team member who had led the implementation of the discourse grammar shared in a semi-structured interview in Fall 2023 that a major reason that they left was the frictions of the Roam software itself, including a lack of affordances for structuring collaborative work, such as version control, comments, and granular permissions; yet, the enduring impact of the extension on their work has been, in her words, \textit{``more of a cultural change than a procedural change''}\footnote{Zoom call 2023-10-03}.

\begin{figure}
    \centering
    \includegraphics[width=\linewidth]{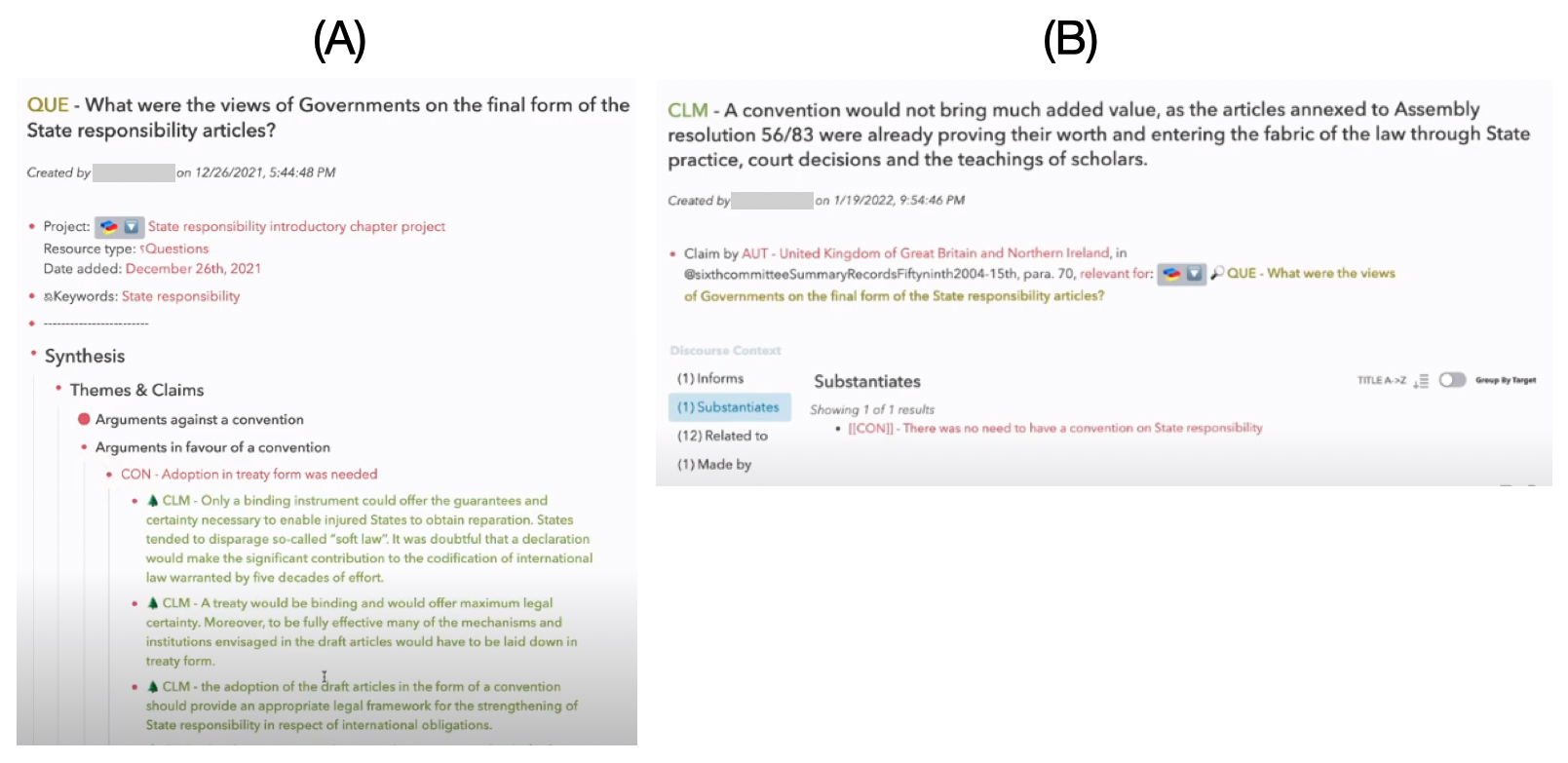}
    \caption{Snapshots of AP's usage of the \sysshort{} to map out positions and issues for an international law research project. (A) shows a question page that was used to track and synthesize various \textit{claims} and \textit{conclusions} about the views of Governments on the final form of State responsibility articles, and (B) shows a specific page for the claim "A convention would not bring much added value, as the articles annexed to Assembly resolution 56/83 were already proving their worth and entering the fabric of the law through State practice, court decisions and the teachings of scholars". The page tracks the source of the claim (the \textit{Author} being the UK, documented in the @sixthcommittee SummaryRecordsFiftyninth2004-15th source, as well as the claim's discourse relations to other discourse nodes that are highlighted in the \texttt{Discourse Context} component, such as its \textit{substantiates} relation to the \textit{conclusion} that "There was no need to have a convention on State responsibility"}
    \label{fig:ap-snapshot}
\end{figure}

AP also extended the discourse grammar to cover \textit{conclusion} nodes, similar to MA, and also specific relations such as \textit{substantiates}, to enable integration of synthesis into his primary research work in international law (see Figure \ref{fig:ap-snapshot})\footnote{AP video walkthrough of discourse graph set up}.

To enable this expansion of the grammar, a substantial portion of our development effort during the deploy period was focused on enhancements to the grammar editor. In addition to small quality-of-life improvements like adding color distinctions between node types, and preserving the layout of nodes in the relation pattern editor, one major design enhancement is worth highlighting here to illustrate the nature of users' expansion of the grammar: \textbf{we added the ability to "clone" or "copy-paste" relation patterns} (see Figure \ref{fig:grammar-editor-changes} in Appendix \ref{ap:designs}). The main motivation for this feature request from users was to ease the cost of defining new discourse node types that were variants or subtypes of existing node types.

For example, MVD, in a direct message on Twitter, described a desire to create subtypes of claims (alluding to the nuances he proposed, as described above) that inherited relation patterns of the base "claim" node\footnote{Twitter direct message 2021-11-22}:
\begin{quote}
    \textit{What I'd really like is to be able to define subtype of relations (i.e. different kinds of claims). BUT that each subtype automatically inherits the properties of its parent type. i.e. you don't have to add every new claim-relation when you make a new subtype of claim.}
\end{quote}
MA expressed a similar request in a Discord message, to assist him in setting up the expansion of the grammar to cover primary research discourse nodes, such as RES (result)\footnote{Discord message 2022-01-08}:
\begin{quote}
    \textit{I was messing around with getting the grammar to work with RES, didnt explore it thoroughly, wondering if you have thoughts about that (e.g. being able to inherit the grammar from EVD).}
\end{quote}


There was also some \textbf{expansion of the discourse graph grammar to cover new types of discourse relations and relation patterns}. For example, AP added new relations such as "\textit{Substantiates}" between Claims and Conclusions, and "\textit{Makes}" between Authors and Claims (to be able to trace claims back to specific authors)\footnote{Field note 2021-12-16}. And DSV added numerous relations to express their such as "\textit{Constrains}" (between Constraints and Solutions) "\textit{Enables}" (between Solutions and Constraints, and between Solutions), and "\textit{Jointly Necessary}" (between Solutions and Constraints/Solutions)\footnote{}. The first author's lab also implemented a "\textit{ConsistentWith}" relation to see "sibling" pieces of evidence that support the same claim (expressing a sort of "conceptual replication" relationship, to give a sense of how "robust" a piece of evidence is in the network of evidence)\footnote{Zoom call 2022-07-22}. 

These new relations, though, were less common than the new nodes created. It is hard to determine the extent to which this was a lack of desire for more sophisticated and expressive relations, vs. technical difficulties. Given the expressed and observed desire for smoother and less effortful specification of relations described in the previous section (\S\ref{sec:incremental-formalization}), as well as some user data on the continued desire for easier ways to express these patterns to the extension's parser to be able to pick them up (including several instances of the first-author assisting users in defining relation patterns on user calls), we infer that user-friendly recognition of discourse relations remained an open problem in the extension. AP's note about the burden of specifying relation patterns in the editor shed some light on this (here his mention of the ``playground'' refers to the visual editor we provided for specifying relation patterns)\footnote{Usage survey AP}:
\begin{quote}
    \textit{It would be great to have a quick way of editing existing relationships, other than returning to the playground. I on occasion "tweak" my existing relationships. For example, Sometimes I want to change a from "child" to "descendant" so as to pick relatinships between nodes nested further down the stack. Going back to the extension page and tweaking can take time and can disrupt the flow. Once a relationship is set up, it would be great to have some sort of abbreviated view of the components of the relationship, each of which could be tweaked. In fact, this latter idea could be the basis for eventually abandoning the playground metaphor. While I like the playground, especially since the visual metaphor is useful, it still suffers from the problem that it loses the layout each time you go out and then comeback (at least it only keeps the layout if you go out and return quickly. It is a real pain point to have to reorganize the relationships each time in the visual editor. Either this has to be fixed, or the playground might need to be replaced with another method. }
\end{quote}

\subsection{Transferring the Discourse Graph System across Diverse Technical Settings}
\label{sec:tool-transfer}

\begin{figure}
    \centering
    \includegraphics[width=\linewidth]{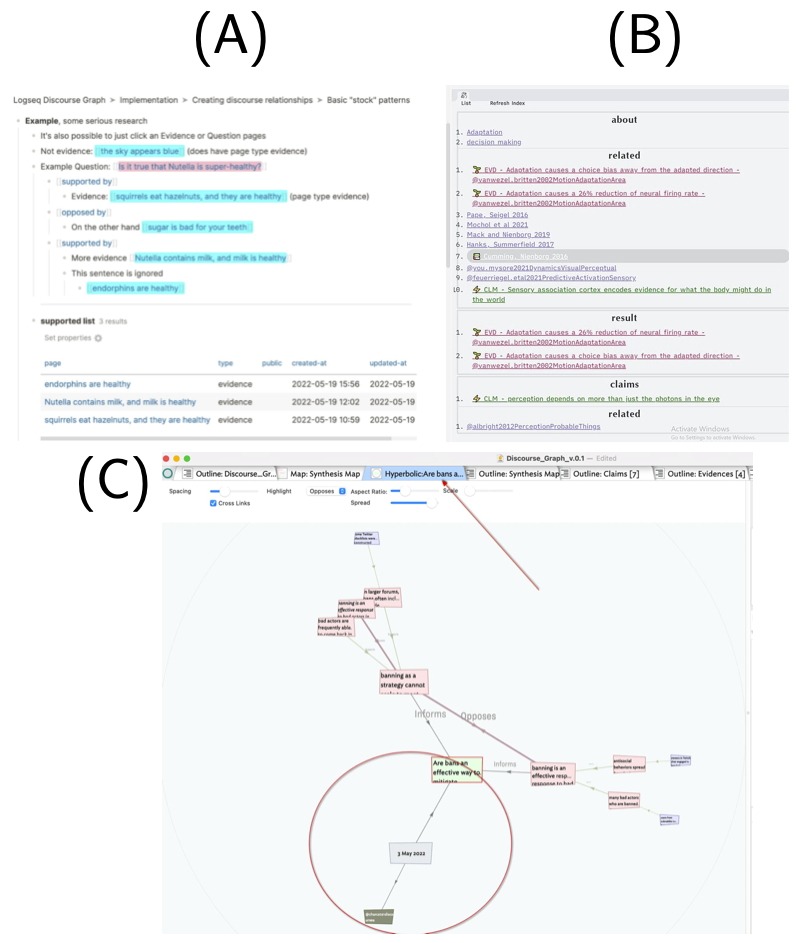}
    \caption{Snapshots of observed instances of user-driven implementation of discourse graphs in a similar outliner hypertext notebook (Logseq) (A), 
    a markdown-based hypertext notebook (Obsidian) (B), and a visual hypertext notebook (Tinderbox) (C).}
    \label{fig:transfer-systems}
\end{figure}

One barrier to observation of collaborative usage of discourse graphs was the fact that many scholars with whom our users collaborated already had their own established workflows. This is not surprising, given what we know about the diversity and idiosyncracy of tooling for scholarly work \cite{sawyerSocialScientistsCyberinfrastructure2012,bosmanInnovationsScholarlyCommunication2016}. To elaborate on this further, we observed several instances of users requesting a "port" of the \sysshort{} to different tools they were more familiar with or preferred to work in, for various reasons. For example, an early-career faculty member, explained in their usage survey how they preferred to do their synthesis work in Obsidian, a markdown-based hypertext notebook, and was attempting to (manually) adapt the Discourse Graph workflow in their work there\footnote{Usage survey MD}:
\begin{quote}
    \textit{I haven't used discourse graph extension as much as I might just because I've been using Roam a lot less recently. This is because I do my synthesis work in Obsidian because it is easier to transition to writing there. But if there were an extension for Obsidian, I would 100\% use it because my solution is kind of a hack.}
\end{quote}

We were heartened, therefore, to observe \textbf{organic spread of the Discourse Graph practice and software beyond Roam}: through our involvement in the "tools for thought" community on forums and community chats, social media, and calls with users, we observed implementations of Discourse Graphs in Obsidian \footnote{Field note 2022-01-13; Field note 2022-04-13} (see Figure \ref{fig:transfer-systems}B), Tinderbox (a longstanding visual hypertext notebook; see Figure \ref{fig:transfer-systems}C), Tana (an outliner hypertext notebook similar to Roam but with more powerful in-built primitives for applying schemas to notes\footnote{Field note 2022-08-20}. Notably, users from the Logseq community organized an \$8,000 bounty (funded in part by a grant from one of their institutions) for a full port of the \sysshort{} to Logseq; over the Spring through Fall of 2022, those users, and the developers they recruited, involved the first author in several working meetings to develop specifications for a minimum viable prototype port\footnote{Field note 2022-04-26}; Figure \ref{fig:transfer-systems}A shows the progress of the prototype, which was released to users for testing in December 2023\footnote{Discord message 2023-12-09}. MA also sparked efforts to prototype a discourse graph authoring markup in a web publishing framework
\footnote{Field note 2022-02-11}.


\section{Discussion}
\label{sec:Discussion}
In this paper, we explored the HCI problem of how to grow new knowledge infrastructures optimized for sharing, reusing, and synthesizing knowledge.  
Over a 2.5-year span of participatory Research through Design with a distributed community of researchers using hypertext notebooks, we sketched out a design vision for how to integrate the seeds of a discourse-centric infrastructure into local scientific practices, and the collaborative and institutional structures of research. Our field deployment highlighted the following key aspects of this vision:
\begin{itemize}
    \item \textbf{Intrinsic benefits} of discourse graphing for improved synthesis (\textbf{\S\ref{sec:improving-synthesis}}), transformed primary research and research training (\textbf{\S\ref{sec:improving-primary-research}}), and augmented collaboration (\textbf{\S\ref{sec:augmenting-collaboration}}) 
    \item \textbf{Sociotechnical preconditions} for successful integration, including smooth(er) incremental formalization \cite{shipmanFormalityConsideredHarmful1999} of notes into discourse nodes and edges (\textbf{\S\ref{sec:incremental-formalization}}), and local personalization of discourse graph grammars (\textbf{\S\ref{sec:extending-grammar}})
    \item \textbf{Technical means of networking} these local discourse graphs into a larger infrastructure by integrating discourse graphs into a diversity of notetaking tools, while retaining a shared data model (\textbf{\S\ref{sec:tool-transfer}})
\end{itemize} 

We can map our results to Edwards et al's \cite{edwards2007understanding} historical model of infrastructure development to understand how these results contribute steps towards growing an infrastructure for synthesis. From the perspective of this model, the discourse-enhanced lab notebooks are \textbf{local systems} that augment synthesis. The authoring of discourse graphs in an interoperable format, and evidence of their use in collaborative research work, are seeds of distributed \textbf{networking} of these local synthesis systems. And the organic spread of the discourse-enhanced lab notebook pattern from our original technical setting of Roam Research to other technical settings, and from our original empirical science setting to other substantive domains, constitute seeds of \textbf{technology transfer} across different contexts. We are especially excited about this point, since infrastructure studies have observed that technology transfer is often a key catalyst for growing local systems into infrastructure \cite{edwards2007understanding}. For these reasons, we claim that our results go beyond illuminating a design vision: 
they also contribute concrete \textbf{steps towards an infrastructure for synthesis}.

We close by discussing the open HCI research problems our design vision illuminates for continuing the growth of a discourse-centric research synthesis infrastructure, as well as broader implications of our work for HCI. 

\subsection{How HCI can Contribute to Continued Growth of an Infrastructure for Synthesis}
As we have argued, the effective local systems for synthesis and transfer across technical contexts constitute a trajectory towards infrastructure. We see the next phase of work being fruitfully focused on \textit{networking} local discourse graphs: what work remains to continue this trajectory, and what can HCI contribute? 


First, as described in \S\ref{sec:incremental-formalization}, we observed limitations in terms of being able to smoothly author discourse edges. This is in part because the mechanisms by which our system was able to recognize patterns of discourse was limited to fairly coarse primitives of the hypertext medium (of indentation or linking, for example): to enable more specific semantics, such as differentiating between support and oppose relations, users had to manually specify and execute the relevant patterns. We wonder whether users might be able to more smoothly recognize latent discourse edges in their existing work with systems that integrate AI approaches for transforming unstructured (text) data into more structured formats. There is already a history of work like this in the field of (semi-)automated evidence synthesis \cite{thomasLivingSystematicReviews2017,jonnalagaddaAutomatingDataExtraction2015,marshallSemiAutomatedEvidenceSynthesis2020}: we are curious whether the apparent leaps in capacity from transformer models and large pretrained models \cite{bommasaniOpportunitiesRisksFoundation2021} might make AI assistance more immediately useful. This could be an interesting problem to connect with research on mixed-initiative intelligent user interfaces \cite{horvitzPrinciplesMixedinitiativeUser1999,heerAgencyAutomationDesigning2019}. This problem could also be framed as one of empowering users --- often with little to no existing programming expertise --- to "program" their notes, thereby connecting with research on end-user programming \cite{myersInvitedResearchOverview2006}.

Second, we also observed some ``non-atomic'' authoring of discourse nodes. For example, some nodes used undefined abbreviations, or were written with unresolved references. Evidence nodes were also often "empty" placeholders: they lacked direct descriptions or links to methodological details or contextualizing snippets from papers. These nodes were still useful for individuals and within synchronous, tight-knit collaborative work contexts, and provided some ready-made on-ramps for conceptually significant ``legitimate peripheral participation'' \cite{laveSituatedLearningLegitimate1991} for beginning research assistants (as described in \S\ref{sec:improving-primary-research} and Fig. \ref{fig:ma-snapshot-lab-usage}B). However, CSCW research on the knowledge sharing \cite{ackermanSharingKnowledgeExpertise2013} suggests that additional labor is likely needed to make them useful beyond these tight-knit settings. Here, again, we think there may be interesting research directions around mixed-initiative interfaces that integrate advances in argument mining \cite{schneiderReviewArgumentationSocial2013}, claim generation \cite{wrightGeneratingScientificClaims2022}, or scientific document information extraction \cite{kilicogluBiomedicalTextMining} to assist with rewriting and contextualization of discourse graphs for sharing.

Third, as described in \S\ref{sec:extending-grammar}, we saw the core data model of questions, claims, and evidence, being frequently locally extended, while still preserving core properties across the local variations. For example, the core "claim" node type bridged between a past-focused variant of synthesis work (similar to systematic reviews) and the emerging practice of micropublishing ongoing primary research. And the core "evidence" node type also bridged between empirical science and design innovation. This dynamic of minimal shared global semantics paired with local variation the core notion of \textit{boundary objects} \cite{starInstitutionalEcologyTranslations1989} that facilitate coordination across social worlds and can be developed into standards for infrastructure formation \cite{leighstarThisNotBoundary2010}. 
We are curious to see more direct tests of whether discourse graphs could act as boundary objects in actual practice. For example, what would happen if we enabled peer-to-peer exchange of local discourse graphs across, say, evidence synthesis groups like Cochrane, biomedical research labs, and biotech innovation companies and investing firms? Would these groups be able to coordinate and share information usefully via the common nodes of claims and evidence, while extending and connecting them to local extensions, such as hypotheses and issues (for primary research), or constraints and solutions (for innovation)? This research could illuminate the next set of (socio)technical problems that HCI researchers might be able to tackle to enable federation or networking across local discourse graph systems. 
From one perspective, this could be framed as a purely technical problem, and approach it similarly to research on ontology alignment and integration. But given the deeply sociotechnical nature of negotiation and alignment that constitutes standards development \cite{edwards2007understanding}, HCI research might make important contributions by drawing on existing design patterns for supporting collective deliberation \cite{kripleanSupportingReflectivePublic2012} and infrastructuring \cite{pipekInfrastructuringIntegratedPerspective2009,ledantecInfrastructuringFormationPublics2013}. 


\subsection{Broader implications for HCI}
Our work also has broader implications of for HCI beyond metascience. For instance, the successful application of incremental formalization as a design pattern in the context of research synthesis further validates the power of that design pattern for building interactive systems that support creative knowledge work. The power of the discourse graph formalism can also be understood as convergent validation of the power of domain-specific languages (DSL) \cite{fischerTurningBreakdownsOpportunities1994} for augmenting individual and collaborative intelligence, such as the IBIS system \cite{kunzIssuesElementsInformation1970,conklin1987hypertext}, Fischer's work on the Envisionment and Discovery Collaboratory (EDC) \cite{ariasTranscendingIndividualHuman2000,ariasEnvisionmentDiscoveryCollaboratory2015}, Liddo et al's work on applying similar discourse-centric models to structure and augment contested collective deliberation \cite{liddoContestedCollectiveIntelligence2012a,deliddoUnderstandingFailuresPotentials2021}, and models of cross-domain analogical retrieval based on an underlying schema of matching problems and mechanisms \cite{kitturScalingAnalogicalInnovation2019} to name just a few. We think that more exploration of varieties of domain-specific languages might broaden the settings of creative work that HCI is able to augment. These domain-specific languages might also form the basis of more effective ``shared representations'' \cite{heerAgencyAutomationDesigning2019} that enable more transparent, controllable and effective development of mixed-initiative systems, or more user-controlled retrieval-augmented generation (RAG) workflows \cite{gaoRetrievalAugmentedGenerationLarge2024} for creativity support.

To close, we reiterate that we 
found key concepts from sociotechnical and critical studies of infrastructure and knowledge sharing --- such as the frame of growing infrastructure, or boundary objects --- to be immensely helpful for guiding our design work. These concepts went beyond descriptions of what should \textit{not} be built or what cannot work (e.g., naive repository models for knowledge sharing, as described in \cite{ackermanSharingKnowledgeExpertise2013}): they helped us envision what \textit{could} be built: in this case, a new infrastructure for synthesis grown from a new installed based of enhanced synthesis and research notetaking systems.  
We see our work therefore as contributing an additional example of how HCI research might fruitfully bridge the ``great divide'' between system building and social analysis \cite{bowkerSocialScienceTechnical2014}. We believe the integration of critique and systems building is increasingly necessary as HCI seeks to respond to pressing sociotechnical and infrastructure-scale problems, such as misinformation, and privacy and security amid the proliferation of generative AI. 

\begin{acks}
This work was made possible in part by grants from Protocol Labs Research to the first, second, and third authors. We are deeply indebted also to the members of the RoamResearch and larger ``tools for thought'' community, and to Wayne Lutters, Katrina Fenlon, Susan Winter, Jodi Schneider, Evan Miyazono, Karola Kirsanow, Silvia diBessa, SJ Klein, Robert Haisfield, Joseph Chang, Pao Siangliulue, and Aniket Kittur for helpful discussions of this work.
\end{acks}

\bibliographystyle{ACM-Reference-Format}
\bibliography{references}


\begin{thebibliography}{118}


\ifx \showCODEN    \undefined \def \showCODEN     #1{\unskip}     \fi
\ifx \showDOI      \undefined \def \showDOI       #1{#1}\fi
\ifx \showISBNx    \undefined \def \showISBNx     #1{\unskip}     \fi
\ifx \showISBNxiii \undefined \def \showISBNxiii  #1{\unskip}     \fi
\ifx \showISSN     \undefined \def \showISSN      #1{\unskip}     \fi
\ifx \showLCCN     \undefined \def \showLCCN      #1{\unskip}     \fi
\ifx \shownote     \undefined \def \shownote      #1{#1}          \fi
\ifx \showarticletitle \undefined \def \showarticletitle #1{#1}   \fi
\ifx \showURL      \undefined \def \showURL       {\relax}        \fi
\providecommand\bibfield[2]{#2}
\providecommand\bibinfo[2]{#2}
\providecommand\natexlab[1]{#1}
\providecommand\showeprint[2][]{arXiv:#2}

\bibitem[\protect\citeauthoryear{Aanestad, Grisot, Hanseth, and Vassilakopoulou}{Aanestad et~al\mbox{.}}{2017}]%
        {aanestadInformationInfrastructuresChallenge2017}
\bibfield{author}{\bibinfo{person}{Margunn Aanestad}, \bibinfo{person}{Miria Grisot}, \bibinfo{person}{Ole Hanseth}, {and} \bibinfo{person}{Polyxeni Vassilakopoulou}.} \bibinfo{year}{2017}\natexlab{}.
\newblock \showarticletitle{Information infrastructures and the challenge of the installed base}.
\newblock In \bibinfo{booktitle}{\emph{Information {Infrastructures} within {European} {Health} {Care}}}, \bibfield{editor}{\bibinfo{person}{Margunn Aanestad}, \bibinfo{person}{Miria Grisot}, \bibinfo{person}{Ole Hanseth}, {and} \bibinfo{person}{Polyxeni Vassilakopoulou}} (Eds.). \bibinfo{publisher}{Springer International Publishing}, \bibinfo{address}{Cham}, \bibinfo{pages}{25--33}.
\newblock
\showISBNx{978-3-319-51018-7 978-3-319-51020-0}
\urldef\tempurl%
\url{https://doi.org/10.1007/978-3-319-51020-0}
\showDOI{\tempurl}


\bibitem[\protect\citeauthoryear{Ackerman, Dachtera, Pipek, and Wulf}{Ackerman et~al\mbox{.}}{2013}]%
        {ackermanSharingKnowledgeExpertise2013}
\bibfield{author}{\bibinfo{person}{Mark~S. Ackerman}, \bibinfo{person}{Juri Dachtera}, \bibinfo{person}{Volkmar Pipek}, {and} \bibinfo{person}{Volker Wulf}.} \bibinfo{year}{2013}\natexlab{}.
\newblock \showarticletitle{Sharing {{Knowledge}} and {{Expertise}}: {{The CSCW View}} of {{Knowledge Management}}}.
\newblock \bibinfo{journal}{\emph{Computer Supported Cooperative Work (CSCW)}} \bibinfo{volume}{22}, \bibinfo{number}{4-6} (\bibinfo{date}{Aug.} \bibinfo{year}{2013}), \bibinfo{pages}{531--573}.
\newblock
\showISSN{0925-9724, 1573-7551}
\urldef\tempurl%
\url{https://doi.org/10.1007/s10606-013-9192-8}
\showDOI{\tempurl}


\bibitem[\protect\citeauthoryear{Alton-Lee}{Alton-Lee}{1998}]%
        {alton-leeTroubleshooterChecklistProspective1998}
\bibfield{author}{\bibinfo{person}{Adrienne Alton-Lee}.} \bibinfo{year}{1998}\natexlab{}.
\newblock \showarticletitle{A {Troubleshooter}'s {Checklist} for {Prospective} {Authors} {Derived} from {Reviewers}' {Critical} {Feedback}}.
\newblock \bibinfo{journal}{\emph{Teaching and Teacher Education}} \bibinfo{volume}{14}, \bibinfo{number}{8} (\bibinfo{year}{1998}), \bibinfo{pages}{887--90}.
\newblock
\showISSN{0742-051X}


\bibitem[\protect\citeauthoryear{Alvesson and Sandberg}{Alvesson and Sandberg}{2011}]%
        {alvessonGeneratingResearchQuestions2011}
\bibfield{author}{\bibinfo{person}{Mats Alvesson} {and} \bibinfo{person}{Jörgen Sandberg}.} \bibinfo{year}{2011}\natexlab{}.
\newblock \showarticletitle{Generating research questions through problematization}.
\newblock \bibinfo{journal}{\emph{Academy of management review}} \bibinfo{volume}{36}, \bibinfo{number}{2} (\bibinfo{year}{2011}), \bibinfo{pages}{247--271}.
\newblock
\showISSN{0363-7425}
\newblock
\shownote{Publisher: Academy of Management Briarcliff Manor, NY.}


\bibitem[\protect\citeauthoryear{André, Zhang, Kim, Chilton, Dow, and Miller}{André et~al\mbox{.}}{2013}]%
        {andreCommunityClusteringLeveraging2013}
\bibfield{author}{\bibinfo{person}{Paul André}, \bibinfo{person}{Haoqi Zhang}, \bibinfo{person}{Juho Kim}, \bibinfo{person}{Lydia Chilton}, \bibinfo{person}{Steven~P. Dow}, {and} \bibinfo{person}{Robert~C. Miller}.} \bibinfo{year}{2013}\natexlab{}.
\newblock \showarticletitle{Community clustering: {Leveraging} an academic crowd to form coherent conference sessions}. In \bibinfo{booktitle}{\emph{First {AAAI} {Conference} on {Human} {Computation} and {Crowdsourcing}}}.
\newblock


\bibitem[\protect\citeauthoryear{{anujacabraal}}{{anujacabraal}}{2012}]%
        {anujacabraalWhyUseNVivo2012}
\bibfield{author}{\bibinfo{person}{{anujacabraal}}.} \bibinfo{year}{2012}\natexlab{}.
\newblock \bibinfo{title}{Why use {NVivo} for your literature review?}
\newblock
\newblock
\urldef\tempurl%
\url{https://anujacabraal.wordpress.com/2012/08/01/why-use-nvivo-for-your-literature-review/}
\showURL{%
\tempurl}


\bibitem[\protect\citeauthoryear{Arias, Eden, Fischer, Gorman, and Scharff}{Arias et~al\mbox{.}}{2000}]%
        {ariasTranscendingIndividualHuman2000}
\bibfield{author}{\bibinfo{person}{Ernesto Arias}, \bibinfo{person}{Hal Eden}, \bibinfo{person}{Gerhard Fischer}, \bibinfo{person}{Andrew Gorman}, {and} \bibinfo{person}{Eric Scharff}.} \bibinfo{year}{2000}\natexlab{}.
\newblock \showarticletitle{Transcending the individual human mind—creating shared understanding through collaborative design}.
\newblock \bibinfo{journal}{\emph{ACM Transactions on Computer-Human Interaction}} \bibinfo{volume}{7}, \bibinfo{number}{1} (\bibinfo{date}{March} \bibinfo{year}{2000}), \bibinfo{pages}{84--113}.
\newblock
\showISSN{1073-0516, 1557-7325}
\urldef\tempurl%
\url{https://doi.org/10.1145/344949.345015}
\showDOI{\tempurl}


\bibitem[\protect\citeauthoryear{Arias, Eden, and Fischer}{Arias et~al\mbox{.}}{2015}]%
        {ariasEnvisionmentDiscoveryCollaboratory2015}
\bibfield{author}{\bibinfo{person}{Ernesto~G. Arias}, \bibinfo{person}{Hal Eden}, {and} \bibinfo{person}{Gerhard Fischer}.} \bibinfo{year}{2015}\natexlab{}.
\newblock \showarticletitle{The {Envisionment} and {Discovery} {Collaboratory} ({EDC}): {Explorations} in {Human}-{Centered} {Informatics}}.
\newblock \bibinfo{journal}{\emph{Synthesis Lectures on Human-Centered Informatics}} \bibinfo{volume}{8}, \bibinfo{number}{5} (\bibinfo{date}{Oct.} \bibinfo{year}{2015}), \bibinfo{pages}{i--216}.
\newblock
\showISSN{1946-7680}
\urldef\tempurl%
\url{https://doi.org/10.2200/S00670ED1V01Y201509HCI032}
\showDOI{\tempurl}
\newblock
\shownote{Publisher: Morgan \& Claypool Publishers.}


\bibitem[\protect\citeauthoryear{Bhurke, Cook, Tallant, Young, Williams, and Raftery}{Bhurke et~al\mbox{.}}{2015}]%
        {bhurkeUsingSystematicReviews2015}
\bibfield{author}{\bibinfo{person}{Sheetal Bhurke}, \bibinfo{person}{Andrew Cook}, \bibinfo{person}{Anna Tallant}, \bibinfo{person}{Amanda Young}, \bibinfo{person}{Elaine Williams}, {and} \bibinfo{person}{James Raftery}.} \bibinfo{year}{2015}\natexlab{}.
\newblock \showarticletitle{Using systematic reviews to inform {NIHR} {HTA} trial planning and design: a retrospective cohort}.
\newblock \bibinfo{journal}{\emph{BMC Medical Research Methodology}} \bibinfo{volume}{15}, \bibinfo{number}{1} (\bibinfo{date}{Dec.} \bibinfo{year}{2015}), \bibinfo{pages}{108}.
\newblock
\showISSN{1471-2288}
\urldef\tempurl%
\url{https://doi.org/10.1186/s12874-015-0102-2}
\showDOI{\tempurl}
\newblock
\shownote{00000.}


\bibitem[\protect\citeauthoryear{Blake and Pratt}{Blake and Pratt}{2006}]%
        {blakeCollaborativeInformationSynthesis2006}
\bibfield{author}{\bibinfo{person}{Catherine Blake} {and} \bibinfo{person}{Wanda Pratt}.} \bibinfo{year}{2006}\natexlab{}.
\newblock \showarticletitle{Collaborative information synthesis {I}: {A} model of information behaviors of scientists in medicine and public health}.
\newblock \bibinfo{journal}{\emph{Journal of the American Society for Information Science and Technology}} \bibinfo{volume}{57}, \bibinfo{number}{13} (\bibinfo{year}{2006}), \bibinfo{pages}{1740--1749}.
\newblock
\showISSN{1532-2890}
\urldef\tempurl%
\url{https://doi.org/10.1002/asi.20487}
\showDOI{\tempurl}
\newblock
\shownote{00097 \_eprint: https://onlinelibrary.wiley.com/doi/pdf/10.1002/asi.20487.}


\bibitem[\protect\citeauthoryear{Bommasani, Hudson, Adeli, Altman, Arora, von Arx, Bernstein, Bohg, Bosselut, Brunskill, Brynjolfsson, Buch, Card, Castellon, Chatterji, Chen, Creel, Davis, Demszky, Donahue, Doumbouya, Durmus, Ermon, Etchemendy, Ethayarajh, Fei-Fei, Finn, Gale, Gillespie, Goel, Goodman, Grossman, Guha, Hashimoto, Henderson, Hewitt, Ho, Hong, Hsu, Huang, Icard, Jain, Jurafsky, Kalluri, Karamcheti, Keeling, Khani, Khattab, Koh, Krass, Krishna, Kuditipudi, Kumar, Ladhak, Lee, Lee, Leskovec, Levent, Li, Li, Ma, Malik, Manning, Mirchandani, Mitchell, Munyikwa, Nair, Narayan, Narayanan, Newman, Nie, Niebles, Nilforoshan, Nyarko, Ogut, Orr, Papadimitriou, Park, Piech, Portelance, Potts, Raghunathan, Reich, Ren, Rong, Roohani, Ruiz, Ryan, Ré, Sadigh, Sagawa, Santhanam, Shih, Srinivasan, Tamkin, Taori, Thomas, Tramèr, Wang, Wang, Wu, Wu, Wu, Xie, Yasunaga, You, Zaharia, Zhang, Zhang, Zhang, Zhang, Zheng, Zhou, and Liang}{Bommasani et~al\mbox{.}}{2021}]%
        {bommasaniOpportunitiesRisksFoundation2021}
\bibfield{author}{\bibinfo{person}{Rishi Bommasani}, \bibinfo{person}{Drew~A. Hudson}, \bibinfo{person}{Ehsan Adeli}, \bibinfo{person}{Russ Altman}, \bibinfo{person}{Simran Arora}, \bibinfo{person}{Sydney von Arx}, \bibinfo{person}{Michael~S. Bernstein}, \bibinfo{person}{Jeannette Bohg}, \bibinfo{person}{Antoine Bosselut}, \bibinfo{person}{Emma Brunskill}, \bibinfo{person}{Erik Brynjolfsson}, \bibinfo{person}{Shyamal Buch}, \bibinfo{person}{Dallas Card}, \bibinfo{person}{Rodrigo Castellon}, \bibinfo{person}{Niladri Chatterji}, \bibinfo{person}{Annie Chen}, \bibinfo{person}{Kathleen Creel}, \bibinfo{person}{Jared~Quincy Davis}, \bibinfo{person}{Dora Demszky}, \bibinfo{person}{Chris Donahue}, \bibinfo{person}{Moussa Doumbouya}, \bibinfo{person}{Esin Durmus}, \bibinfo{person}{Stefano Ermon}, \bibinfo{person}{John Etchemendy}, \bibinfo{person}{Kawin Ethayarajh}, \bibinfo{person}{Li Fei-Fei}, \bibinfo{person}{Chelsea Finn}, \bibinfo{person}{Trevor Gale}, \bibinfo{person}{Lauren Gillespie}, \bibinfo{person}{Karan
  Goel}, \bibinfo{person}{Noah Goodman}, \bibinfo{person}{Shelby Grossman}, \bibinfo{person}{Neel Guha}, \bibinfo{person}{Tatsunori Hashimoto}, \bibinfo{person}{Peter Henderson}, \bibinfo{person}{John Hewitt}, \bibinfo{person}{Daniel~E. Ho}, \bibinfo{person}{Jenny Hong}, \bibinfo{person}{Kyle Hsu}, \bibinfo{person}{Jing Huang}, \bibinfo{person}{Thomas Icard}, \bibinfo{person}{Saahil Jain}, \bibinfo{person}{Dan Jurafsky}, \bibinfo{person}{Pratyusha Kalluri}, \bibinfo{person}{Siddharth Karamcheti}, \bibinfo{person}{Geoff Keeling}, \bibinfo{person}{Fereshte Khani}, \bibinfo{person}{Omar Khattab}, \bibinfo{person}{Pang~Wei Koh}, \bibinfo{person}{Mark Krass}, \bibinfo{person}{Ranjay Krishna}, \bibinfo{person}{Rohith Kuditipudi}, \bibinfo{person}{Ananya Kumar}, \bibinfo{person}{Faisal Ladhak}, \bibinfo{person}{Mina Lee}, \bibinfo{person}{Tony Lee}, \bibinfo{person}{Jure Leskovec}, \bibinfo{person}{Isabelle Levent}, \bibinfo{person}{Xiang~Lisa Li}, \bibinfo{person}{Xuechen Li}, \bibinfo{person}{Tengyu Ma},
  \bibinfo{person}{Ali Malik}, \bibinfo{person}{Christopher~D. Manning}, \bibinfo{person}{Suvir Mirchandani}, \bibinfo{person}{Eric Mitchell}, \bibinfo{person}{Zanele Munyikwa}, \bibinfo{person}{Suraj Nair}, \bibinfo{person}{Avanika Narayan}, \bibinfo{person}{Deepak Narayanan}, \bibinfo{person}{Ben Newman}, \bibinfo{person}{Allen Nie}, \bibinfo{person}{Juan~Carlos Niebles}, \bibinfo{person}{Hamed Nilforoshan}, \bibinfo{person}{Julian Nyarko}, \bibinfo{person}{Giray Ogut}, \bibinfo{person}{Laurel Orr}, \bibinfo{person}{Isabel Papadimitriou}, \bibinfo{person}{Joon~Sung Park}, \bibinfo{person}{Chris Piech}, \bibinfo{person}{Eva Portelance}, \bibinfo{person}{Christopher Potts}, \bibinfo{person}{Aditi Raghunathan}, \bibinfo{person}{Rob Reich}, \bibinfo{person}{Hongyu Ren}, \bibinfo{person}{Frieda Rong}, \bibinfo{person}{Yusuf Roohani}, \bibinfo{person}{Camilo Ruiz}, \bibinfo{person}{Jack Ryan}, \bibinfo{person}{Christopher Ré}, \bibinfo{person}{Dorsa Sadigh}, \bibinfo{person}{Shiori Sagawa},
  \bibinfo{person}{Keshav Santhanam}, \bibinfo{person}{Andy Shih}, \bibinfo{person}{Krishnan Srinivasan}, \bibinfo{person}{Alex Tamkin}, \bibinfo{person}{Rohan Taori}, \bibinfo{person}{Armin~W. Thomas}, \bibinfo{person}{Florian Tramèr}, \bibinfo{person}{Rose~E. Wang}, \bibinfo{person}{William Wang}, \bibinfo{person}{Bohan Wu}, \bibinfo{person}{Jiajun Wu}, \bibinfo{person}{Yuhuai Wu}, \bibinfo{person}{Sang~Michael Xie}, \bibinfo{person}{Michihiro Yasunaga}, \bibinfo{person}{Jiaxuan You}, \bibinfo{person}{Matei Zaharia}, \bibinfo{person}{Michael Zhang}, \bibinfo{person}{Tianyi Zhang}, \bibinfo{person}{Xikun Zhang}, \bibinfo{person}{Yuhui Zhang}, \bibinfo{person}{Lucia Zheng}, \bibinfo{person}{Kaitlyn Zhou}, {and} \bibinfo{person}{Percy Liang}.} \bibinfo{year}{2021}\natexlab{}.
\newblock \showarticletitle{On the {Opportunities} and {Risks} of {Foundation} {Models}}.
\newblock \bibinfo{journal}{\emph{arXiv:2108.07258 [cs]}} (\bibinfo{date}{Aug.} \bibinfo{year}{2021}).
\newblock
\urldef\tempurl%
\url{http://arxiv.org/abs/2108.07258}
\showURL{%
\tempurl}
\newblock
\shownote{00021 arXiv: 2108.07258.}


\bibitem[\protect\citeauthoryear{Boote and Beile}{Boote and Beile}{2005}]%
        {booteScholarsResearchersCentrality2005}
\bibfield{author}{\bibinfo{person}{David~N. Boote} {and} \bibinfo{person}{Penny Beile}.} \bibinfo{year}{2005}\natexlab{}.
\newblock \showarticletitle{Scholars {Before} {Researchers}: {On} the {Centrality} of the {Dissertation} {Literature} {Review} in {Research} {Preparation}}.
\newblock \bibinfo{journal}{\emph{Educational Researcher}} \bibinfo{volume}{34}, \bibinfo{number}{6} (\bibinfo{date}{Aug.} \bibinfo{year}{2005}), \bibinfo{pages}{3--15}.
\newblock
\showISSN{0013-189X, 1935-102X}
\urldef\tempurl%
\url{https://doi.org/10.3102/0013189X034006003}
\showDOI{\tempurl}


\bibitem[\protect\citeauthoryear{Borah, Brown, Capers, and Kaiser}{Borah et~al\mbox{.}}{2017}]%
        {borahAnalysisTimeWorkers2017}
\bibfield{author}{\bibinfo{person}{Rohit Borah}, \bibinfo{person}{Andrew~W. Brown}, \bibinfo{person}{Patrice~L. Capers}, {and} \bibinfo{person}{Kathryn~A. Kaiser}.} \bibinfo{year}{2017}\natexlab{}.
\newblock \showarticletitle{Analysis of the time and workers needed to conduct systematic reviews of medical interventions using data from the {PROSPERO} registry}.
\newblock \bibinfo{journal}{\emph{BMJ Open}} \bibinfo{volume}{7}, \bibinfo{number}{2} (\bibinfo{date}{Feb.} \bibinfo{year}{2017}), \bibinfo{pages}{e012545}.
\newblock
\showISSN{2044-6055, 2044-6055}
\urldef\tempurl%
\url{https://doi.org/10.1136/bmjopen-2016-012545}
\showDOI{\tempurl}
\newblock
\shownote{Publisher: British Medical Journal Publishing Group Section: Health informatics.}


\bibitem[\protect\citeauthoryear{Borgman, Edwards, Jackson, Chalmers, Bowker, Ribes, Burton, and Calvert}{Borgman et~al\mbox{.}}{2013}]%
        {borgman2013knowledge}
\bibfield{author}{\bibinfo{person}{Christine~L Borgman}, \bibinfo{person}{Paul~N Edwards}, \bibinfo{person}{Steven~J Jackson}, \bibinfo{person}{Melissa~K Chalmers}, \bibinfo{person}{Geoffrey~C Bowker}, \bibinfo{person}{David Ribes}, \bibinfo{person}{Matt Burton}, {and} \bibinfo{person}{Scout Calvert}.} \bibinfo{year}{2013}\natexlab{}.
\newblock \showarticletitle{Knowledge Infrastructures: {{Intellectual}} Frameworks and Research Challenges}.
\newblock  (\bibinfo{year}{2013}).
\newblock


\bibitem[\protect\citeauthoryear{Bosman and Kramer}{Bosman and Kramer}{2016}]%
        {bosmanInnovationsScholarlyCommunication2016}
\bibfield{author}{\bibinfo{person}{Jeroen Bosman} {and} \bibinfo{person}{Bianca Kramer}.} \bibinfo{year}{2016}\natexlab{}.
\newblock \bibinfo{title}{Innovations in scholarly communication - data of the global 2015-2016 survey}.
\newblock
\newblock
\urldef\tempurl%
\url{https://doi.org/10.5281/zenodo.49583}
\showDOI{\tempurl}
\newblock
\shownote{00014 type: dataset.}


\bibitem[\protect\citeauthoryear{Bowker, Star, Gasser, and Turner}{Bowker et~al\mbox{.}}{2014}]%
        {bowkerSocialScienceTechnical2014}
\bibfield{author}{\bibinfo{person}{Geoffrey Bowker}, \bibinfo{person}{Susan~Leigh Star}, \bibinfo{person}{Les Gasser}, {and} \bibinfo{person}{William Turner}.} \bibinfo{year}{2014}\natexlab{}.
\newblock \bibinfo{booktitle}{\emph{Social {Science}, {Technical} {Systems}, and {Cooperative} {Work}: {Beyond} the {Great} {Divide}}}.
\newblock \bibinfo{publisher}{Psychology Press}.
\newblock
\showISBNx{978-1-317-77876-9}
\newblock
\shownote{Google-Books-ID: XSaPAwAAQBAJ.}


\bibitem[\protect\citeauthoryear{Bowker}{Bowker}{2000}]%
        {bowkerBiodiversityDatadiversity2000}
\bibfield{author}{\bibinfo{person}{Geoffrey~C. Bowker}.} \bibinfo{year}{2000}\natexlab{}.
\newblock \showarticletitle{Biodiversity {Datadiversity}}.
\newblock \bibinfo{journal}{\emph{Social Studies of Science}} \bibinfo{volume}{30}, \bibinfo{number}{5} (\bibinfo{date}{Oct.} \bibinfo{year}{2000}), \bibinfo{pages}{643--683}.
\newblock
\showISSN{0306-3127}
\urldef\tempurl%
\url{https://doi.org/10.1177/030631200030005001}
\showDOI{\tempurl}
\newblock
\shownote{Publisher: SAGE Publications Ltd.}


\bibitem[\protect\citeauthoryear{Brush, Shefchek, and Haendel}{Brush et~al\mbox{.}}{2016}]%
        {brushSEPIOSemanticModel2016}
\bibfield{author}{\bibinfo{person}{Matthew~H. Brush}, \bibinfo{person}{Kent Shefchek}, {and} \bibinfo{person}{Melissa Haendel}.} \bibinfo{year}{2016}\natexlab{}.
\newblock \showarticletitle{{SEPIO}: {A} semantic model for the integration and analysis of scientific evidence}.
\newblock \bibinfo{journal}{\emph{CEUR Workshop Proceedings}}  \bibinfo{volume}{1747} (\bibinfo{year}{2016}).
\newblock
\showISSN{1613-0073}
\urldef\tempurl%
\url{https://ohsu.pure.elsevier.com/en/publications/sepio-a-semantic-model-for-the-integration-and-analysis-of-scient}
\showURL{%
\tempurl}


\bibitem[\protect\citeauthoryear{Bucur, Kuhn, Ceolin, and van Ossenbruggen}{Bucur et~al\mbox{.}}{2022}]%
        {bucurNanopublicationBasedSemanticPublishing2022}
\bibfield{author}{\bibinfo{person}{Cristina-Iulia Bucur}, \bibinfo{person}{Tobias Kuhn}, \bibinfo{person}{Davide Ceolin}, {and} \bibinfo{person}{Jacco van Ossenbruggen}.} \bibinfo{year}{2022}\natexlab{}.
\newblock \showarticletitle{Nanopublication-{Based} {Semantic} {Publishing} and {Reviewing}: {A} {Field} {Study} with {Formalization} {Papers}}.
\newblock \bibinfo{journal}{\emph{arXiv:2203.01608 [cs]}} (\bibinfo{date}{March} \bibinfo{year}{2022}).
\newblock
\urldef\tempurl%
\url{http://arxiv.org/abs/2203.01608}
\showURL{%
\tempurl}
\newblock
\shownote{arXiv: 2203.01608.}


\bibitem[\protect\citeauthoryear{Chan}{Chan}{2020}]%
        {chanKnowledgeSynthesisConceptual2020}
\bibfield{author}{\bibinfo{person}{Joel Chan}.} \bibinfo{year}{2020}\natexlab{}.
\newblock \showarticletitle{Knowledge synthesis: {A} conceptual model and practical guide}.
\newblock \bibinfo{journal}{\emph{Open and Sustainable Innovation Systems (OASIS) Lab}} (\bibinfo{date}{Dec.} \bibinfo{year}{2020}).
\newblock
\urldef\tempurl%
\url{https://oasislab.pubpub.org/pub/54t0y9mk/release/2}
\showURL{%
\tempurl}
\newblock
\shownote{00000 Publisher: PubPub.}


\bibitem[\protect\citeauthoryear{Chandrasekharan, Jhaver, Bruckman, and Gilbert}{Chandrasekharan et~al\mbox{.}}{2022}]%
        {chandrasekharanQuarantinedExaminingEffects2022}
\bibfield{author}{\bibinfo{person}{Eshwar Chandrasekharan}, \bibinfo{person}{Shagun Jhaver}, \bibinfo{person}{Amy Bruckman}, {and} \bibinfo{person}{Eric Gilbert}.} \bibinfo{year}{2022}\natexlab{}.
\newblock \showarticletitle{Quarantined! {Examining} the {Effects} of a {Community}-{Wide} {Moderation} {Intervention} on {Reddit}}.
\newblock \bibinfo{journal}{\emph{ACM Transactions on Computer-Human Interaction}} \bibinfo{volume}{29}, \bibinfo{number}{4} (\bibinfo{date}{Aug.} \bibinfo{year}{2022}), \bibinfo{pages}{1--26}.
\newblock
\showISSN{1073-0516, 1557-7325}
\urldef\tempurl%
\url{https://doi.org/10.1145/3490499}
\showDOI{\tempurl}


\bibitem[\protect\citeauthoryear{Chilton, Kim, André, Cordeiro, Landay, Weld, Dow, Miller, and Zhang}{Chilton et~al\mbox{.}}{2014}]%
        {chiltonFrenzyCollaborativeData2014}
\bibfield{author}{\bibinfo{person}{Lydia~B. Chilton}, \bibinfo{person}{Juho Kim}, \bibinfo{person}{Paul André}, \bibinfo{person}{Felicia Cordeiro}, \bibinfo{person}{James~A. Landay}, \bibinfo{person}{Daniel~S. Weld}, \bibinfo{person}{Steven~P. Dow}, \bibinfo{person}{Robert~C. Miller}, {and} \bibinfo{person}{Haoqi Zhang}.} \bibinfo{year}{2014}\natexlab{}.
\newblock \showarticletitle{Frenzy: {Collaborative} {Data} {Organization} for {Creating} {Conference} {Sessions}}. In \bibinfo{booktitle}{\emph{Proceedings of the {SIGCHI} {Conference} on {Human} {Factors} in {Computing} {Systems}}} \emph{(\bibinfo{series}{{CHI} '14})}. \bibinfo{publisher}{ACM}, \bibinfo{address}{New York, NY, USA}, \bibinfo{pages}{1255--1264}.
\newblock
\showISBNx{978-1-4503-2473-1}
\urldef\tempurl%
\url{https://doi.org/10.1145/2556288.2557375}
\showDOI{\tempurl}


\bibitem[\protect\citeauthoryear{Ciccarese, Wu, Wong, Ocana, Kinoshita, Ruttenberg, and Clark}{Ciccarese et~al\mbox{.}}{2008}]%
        {ciccareseSWANBiomedicalDiscourse2008}
\bibfield{author}{\bibinfo{person}{Paolo Ciccarese}, \bibinfo{person}{Elizabeth Wu}, \bibinfo{person}{Gwen Wong}, \bibinfo{person}{Marco Ocana}, \bibinfo{person}{June Kinoshita}, \bibinfo{person}{Alan Ruttenberg}, {and} \bibinfo{person}{Tim Clark}.} \bibinfo{year}{2008}\natexlab{}.
\newblock \showarticletitle{The {{SWAN}} Biomedical Discourse Ontology}.
\newblock \bibinfo{journal}{\emph{Journal of Biomedical Informatics}} \bibinfo{volume}{41}, \bibinfo{number}{5} (\bibinfo{date}{Oct.} \bibinfo{year}{2008}), \bibinfo{pages}{739--751}.
\newblock
\showISSN{1532-0464}
\urldef\tempurl%
\url{https://doi.org/10.1016/j.jbi.2008.04.010}
\showDOI{\tempurl}


\bibitem[\protect\citeauthoryear{Clark, Ciccarese, and Goble}{Clark et~al\mbox{.}}{2014}]%
        {clarkMicropublicationsSemanticModel2014}
\bibfield{author}{\bibinfo{person}{Tim Clark}, \bibinfo{person}{Paolo~N. Ciccarese}, {and} \bibinfo{person}{Carole~A. Goble}.} \bibinfo{year}{2014}\natexlab{}.
\newblock \showarticletitle{Micropublications: A Semantic Model for Claims, Evidence, Arguments and Annotations in Biomedical Communications}.
\newblock \bibinfo{journal}{\emph{Journal of Biomedical Semantics}}  \bibinfo{volume}{5} (\bibinfo{date}{July} \bibinfo{year}{2014}), \bibinfo{pages}{28}.
\newblock
\showISSN{2041-1480}
\urldef\tempurl%
\url{https://doi.org/10.1186/2041-1480-5-28}
\showDOI{\tempurl}


\bibitem[\protect\citeauthoryear{Clark and Kinoshita}{Clark and Kinoshita}{2007}]%
        {clarkAlzforumSWANPresent2007}
\bibfield{author}{\bibinfo{person}{Tim Clark} {and} \bibinfo{person}{June Kinoshita}.} \bibinfo{year}{2007}\natexlab{}.
\newblock \showarticletitle{Alzforum and {SWAN}: the present and future of scientific web communities.}
\newblock \bibinfo{journal}{\emph{Briefings in Bioinformatics}} \bibinfo{volume}{8}, \bibinfo{number}{3} (\bibinfo{date}{May} \bibinfo{year}{2007}), \bibinfo{pages}{163--171}.
\newblock
\showISSN{1467-5463, 1477-4054}
\urldef\tempurl%
\url{https://doi.org/10.1093/bib/bbm012}
\showDOI{\tempurl}
\newblock
\shownote{00074.}


\bibitem[\protect\citeauthoryear{Conklin}{Conklin}{1987}]%
        {conklin1987hypertext}
\bibfield{author}{\bibinfo{person}{Jeff Conklin}.} \bibinfo{year}{1987}\natexlab{}.
\newblock \showarticletitle{Hypertext: An Introduction and {{SurvevJ}}}.
\newblock \bibinfo{journal}{\emph{Computer}} \bibinfo{volume}{20}, \bibinfo{number}{9} (\bibinfo{year}{1987}), \bibinfo{pages}{17--41}.
\newblock


\bibitem[\protect\citeauthoryear{Dagiral and Peerbaye}{Dagiral and Peerbaye}{2016}]%
        {dagiralMakingKnowledgeBoundary2016}
\bibfield{author}{\bibinfo{person}{Éric Dagiral} {and} \bibinfo{person}{Ashveen Peerbaye}.} \bibinfo{year}{2016}\natexlab{}.
\newblock \showarticletitle{Making {Knowledge} in {Boundary} {Infrastructures}: {Inside} and {Beyond} a {Database} for {Rare} {Diseases}}.
\newblock \bibinfo{journal}{\emph{Science \& Technology Studies}} \bibinfo{volume}{29}, \bibinfo{number}{2} (\bibinfo{date}{May} \bibinfo{year}{2016}), \bibinfo{pages}{44--61}.
\newblock
\showISSN{2243-4690}
\urldef\tempurl%
\url{https://doi.org/10.23987/sts.55920}
\showDOI{\tempurl}
\newblock
\shownote{Number: 2.}


\bibitem[\protect\citeauthoryear{De~Liddo and Strube}{De~Liddo and Strube}{2021}]%
        {deliddoUnderstandingFailuresPotentials2021}
\bibfield{author}{\bibinfo{person}{Anna De~Liddo} {and} \bibinfo{person}{Rosa Strube}.} \bibinfo{year}{2021}\natexlab{}.
\newblock \showarticletitle{Understanding {Failures} and {Potentials} of {Argumentation} {Tools} for {Public} {Deliberation}}. In \bibinfo{booktitle}{\emph{Proceedings of the 10th {International} {Conference} on {Communities} \& {Technologies} - {Wicked} {Problems} in the {Age} of {Tech}}} \emph{(\bibinfo{series}{C\&amp;{T} '21})}. \bibinfo{publisher}{Association for Computing Machinery}, \bibinfo{address}{New York, NY, USA}, \bibinfo{pages}{75--88}.
\newblock
\showISBNx{978-1-4503-9056-9}
\urldef\tempurl%
\url{https://doi.org/10.1145/3461564.3461584}
\showDOI{\tempurl}


\bibitem[\protect\citeauthoryear{de~Ribaupierre and Falquet}{de~Ribaupierre and Falquet}{2017}]%
        {ribaupierreExtractingDiscourseElements2017}
\bibfield{author}{\bibinfo{person}{H\'el\`ene de Ribaupierre} {and} \bibinfo{person}{Gilles Falquet}.} \bibinfo{year}{2017}\natexlab{}.
\newblock \showarticletitle{Extracting Discourse Elements and Annotating Scientific Documents Using the {{SciAnnotDoc}} Model: A Use Case in Gender Documents}.
\newblock \bibinfo{journal}{\emph{International Journal on Digital Libraries}} (\bibinfo{date}{Aug.} \bibinfo{year}{2017}), \bibinfo{pages}{1--16}.
\newblock
\showISSN{1432-5012, 1432-1300}
\urldef\tempurl%
\url{https://doi.org/10.1007/s00799-017-0227-5}
\showDOI{\tempurl}


\bibitem[\protect\citeauthoryear{de~Waard}{de~Waard}{2010}]%
        {waardProteinsFairytalesDirections2010}
\bibfield{author}{\bibinfo{person}{Anita de Waard}.} \bibinfo{year}{2010}\natexlab{}.
\newblock \showarticletitle{From {Proteins} to {Fairytales}: {Directions} in {Semantic} {Publishing}}.
\newblock \bibinfo{journal}{\emph{IEEE Intelligent Systems}} (\bibinfo{year}{2010}).
\newblock


\bibitem[\protect\citeauthoryear{de~Waard, Shum, Carusi, Park, Samwald, and Sándor}{de~Waard et~al\mbox{.}}{2009}]%
        {dewaardHypothesesEvidenceRelationships2009}
\bibfield{author}{\bibinfo{person}{Anita de Waard}, \bibinfo{person}{Simon~Buckingham Shum}, \bibinfo{person}{Annamaria Carusi}, \bibinfo{person}{Jack Park}, \bibinfo{person}{Matthias Samwald}, {and} \bibinfo{person}{Agnes Sándor}.} \bibinfo{year}{2009}\natexlab{}.
\newblock \showarticletitle{Hypotheses, {Evidence} and {Relationships}: {The} {HypER} {Approach} for {Representing} {Scientific} {Knowledge} {Claims}}. In \bibinfo{booktitle}{\emph{Proceedings of the 8th {International} {Semantic} {Web} {Conference}, {Workshop} on {Semantic} {Web} {Applications} in {Scientific} {Discourse}}}. \bibinfo{pages}{12}.
\newblock


\bibitem[\protect\citeauthoryear{Edwards}{Edwards}{2010}]%
        {edwardsVastMachineComputer2010}
\bibfield{author}{\bibinfo{person}{Paul~N. Edwards}.} \bibinfo{year}{2010}\natexlab{}.
\newblock \bibinfo{booktitle}{\emph{A {Vast} {Machine}: {Computer} {Models}, {Climate} {Data}, and the {Politics} of {Global} {Warming}}}.
\newblock \bibinfo{publisher}{MIT Press}.
\newblock
\showISBNx{978-0-262-29071-5}
\newblock
\shownote{Google-Books-ID: K9\_LsJBCqWMC.}


\bibitem[\protect\citeauthoryear{Edwards, Bowker, Jackson, and Williams}{Edwards et~al\mbox{.}}{2009}]%
        {edwards2009introduction}
\bibfield{author}{\bibinfo{person}{Paul~N Edwards}, \bibinfo{person}{Geoffrey~C Bowker}, \bibinfo{person}{Steven~J Jackson}, {and} \bibinfo{person}{Robin Williams}.} \bibinfo{year}{2009}\natexlab{}.
\newblock \showarticletitle{Introduction: An Agenda for Infrastructure Studies}.
\newblock \bibinfo{journal}{\emph{Journal of the Association for Information Systems}} \bibinfo{volume}{10}, \bibinfo{number}{5} (\bibinfo{year}{2009}), \bibinfo{pages}{6}.
\newblock


\bibitem[\protect\citeauthoryear{Edwards, Jackson, Bowker, and Knobel}{Edwards et~al\mbox{.}}{2007}]%
        {edwards2007understanding}
\bibfield{author}{\bibinfo{person}{Paul~N Edwards}, \bibinfo{person}{Steven~J Jackson}, \bibinfo{person}{Geoffrey~C Bowker}, {and} \bibinfo{person}{Cory~P Knobel}.} \bibinfo{year}{2007}\natexlab{}.
\newblock \bibinfo{booktitle}{\emph{Understanding infrastructure: {Dynamics}, tensions, and design}}.
\newblock \bibinfo{type}{{T}echnical {R}eport}.
\newblock


\bibitem[\protect\citeauthoryear{Edwards, Mayernik, Batcheller, Bowker, and Borgman}{Edwards et~al\mbox{.}}{2011}]%
        {edwardsScienceFrictionData2011}
\bibfield{author}{\bibinfo{person}{Paul~N. Edwards}, \bibinfo{person}{Matthew~S. Mayernik}, \bibinfo{person}{Archer~L. Batcheller}, \bibinfo{person}{Geoffrey~C. Bowker}, {and} \bibinfo{person}{Christine~L. Borgman}.} \bibinfo{year}{2011}\natexlab{}.
\newblock \showarticletitle{Science friction: {Data}, metadata, and collaboration}.
\newblock \bibinfo{journal}{\emph{Social Studies of Science}} \bibinfo{volume}{41}, \bibinfo{number}{5} (\bibinfo{date}{Oct.} \bibinfo{year}{2011}), \bibinfo{pages}{667--690}.
\newblock
\showISSN{0306-3127}
\urldef\tempurl%
\url{https://doi.org/10.1177/0306312711413314}
\showDOI{\tempurl}
\newblock
\shownote{Publisher: SAGE Publications Ltd.}


\bibitem[\protect\citeauthoryear{Ellis}{Ellis}{1993}]%
        {ellisModelingInformationSeekingPatterns1993}
\bibfield{author}{\bibinfo{person}{David Ellis}.} \bibinfo{year}{1993}\natexlab{}.
\newblock \showarticletitle{Modeling the {Information}-{Seeking} {Patterns} of {Academic} {Researchers}: {A} {Grounded} {Theory} {Approach}}.
\newblock \bibinfo{journal}{\emph{The Library Quarterly}} \bibinfo{volume}{63}, \bibinfo{number}{4} (\bibinfo{date}{Oct.} \bibinfo{year}{1993}), \bibinfo{pages}{469--486}.
\newblock
\showISSN{0024-2519}
\urldef\tempurl%
\url{https://doi.org/10.1086/602622}
\showDOI{\tempurl}


\bibitem[\protect\citeauthoryear{Ervin}{Ervin}{2008}]%
        {ervinMotivatingAuthorsUpdate2008}
\bibfield{author}{\bibinfo{person}{Ann-Margret Ervin}.} \bibinfo{year}{2008}\natexlab{}.
\newblock \showarticletitle{Motivating authors to update systematic reviews: practical strategies from a behavioural science perspective}.
\newblock \bibinfo{journal}{\emph{Paediatric and perinatal epidemiology}} \bibinfo{volume}{22}, \bibinfo{number}{0 1} (\bibinfo{date}{Jan.} \bibinfo{year}{2008}), \bibinfo{pages}{33--37}.
\newblock
\showISSN{0269-5022}
\urldef\tempurl%
\url{https://doi.org/10.1111/j.1365-3016.2007.00910.x}
\showDOI{\tempurl}


\bibitem[\protect\citeauthoryear{Fischer}{Fischer}{1994}]%
        {fischerTurningBreakdownsOpportunities1994}
\bibfield{author}{\bibinfo{person}{Gerhard Fischer}.} \bibinfo{year}{1994}\natexlab{}.
\newblock \showarticletitle{Turning breakdowns into opportunities for creativity}.
\newblock \bibinfo{journal}{\emph{Knowledge-Based Systems}} \bibinfo{volume}{7}, \bibinfo{number}{4} (\bibinfo{date}{Dec.} \bibinfo{year}{1994}), \bibinfo{pages}{221--232}.
\newblock
\showISSN{0950-7051}
\urldef\tempurl%
\url{https://doi.org/10.1016/0950-7051(94)90033-7}
\showDOI{\tempurl}
\newblock
\shownote{00000.}


\bibitem[\protect\citeauthoryear{Fleming, Seehra, Polychronopoulou, Fedorowicz, and Pandis}{Fleming et~al\mbox{.}}{2013}]%
        {flemingCochraneNonCochraneSystematic2013}
\bibfield{author}{\bibinfo{person}{Padhraig~S. Fleming}, \bibinfo{person}{Jadbinder Seehra}, \bibinfo{person}{Argy Polychronopoulou}, \bibinfo{person}{Zbys Fedorowicz}, {and} \bibinfo{person}{Nikolaos Pandis}.} \bibinfo{year}{2013}\natexlab{}.
\newblock \showarticletitle{Cochrane and non-{Cochrane} systematic reviews in leading orthodontic journals: a quality paradigm?}
\newblock \bibinfo{journal}{\emph{European Journal of Orthodontics}} \bibinfo{volume}{35}, \bibinfo{number}{2} (\bibinfo{date}{April} \bibinfo{year}{2013}), \bibinfo{pages}{244--248}.
\newblock
\showISSN{0141-5387}
\urldef\tempurl%
\url{https://doi.org/10.1093/ejo/cjs016}
\showDOI{\tempurl}
\newblock
\shownote{Publisher: Oxford Academic.}


\bibitem[\protect\citeauthoryear{Gao, Xiong, Gao, Jia, Pan, Bi, Dai, Sun, Guo, Wang, and Wang}{Gao et~al\mbox{.}}{2024}]%
        {gaoRetrievalAugmentedGenerationLarge2024}
\bibfield{author}{\bibinfo{person}{Yunfan Gao}, \bibinfo{person}{Yun Xiong}, \bibinfo{person}{Xinyu Gao}, \bibinfo{person}{Kangxiang Jia}, \bibinfo{person}{Jinliu Pan}, \bibinfo{person}{Yuxi Bi}, \bibinfo{person}{Yi Dai}, \bibinfo{person}{Jiawei Sun}, \bibinfo{person}{Qianyu Guo}, \bibinfo{person}{Meng Wang}, {and} \bibinfo{person}{Haofen Wang}.} \bibinfo{year}{2024}\natexlab{}.
\newblock \bibinfo{title}{Retrieval-{Augmented} {Generation} for {Large} {Language} {Models}: {A} {Survey}}.
\newblock
\newblock
\urldef\tempurl%
\url{https://doi.org/10.48550/arXiv.2312.10997}
\showDOI{\tempurl}
\newblock
\shownote{arXiv:2312.10997 [cs].}


\bibitem[\protect\citeauthoryear{Gaver}{Gaver}{2014}]%
        {gaverScienceDesignImplications2014}
\bibfield{author}{\bibinfo{person}{William Gaver}.} \bibinfo{year}{2014}\natexlab{}.
\newblock \showarticletitle{Science and design: {The} implications of different forms of accountability}.
\newblock In \bibinfo{booktitle}{\emph{Ways of knowing in {HCI}}}, \bibfield{editor}{\bibinfo{person}{Judith~S. Olson} {and} \bibinfo{person}{Wendy~A. Kellogg}} (Eds.). \bibinfo{publisher}{Springer}, \bibinfo{address}{New York NY}, \bibinfo{pages}{143--165}.
\newblock


\bibitem[\protect\citeauthoryear{Granello}{Granello}{2001}]%
        {granelloPromotingCognitiveComplexity2001}
\bibfield{author}{\bibinfo{person}{Darcy~Haag Granello}.} \bibinfo{year}{2001}\natexlab{}.
\newblock \showarticletitle{Promoting {Cognitive} {Complexity} in {Graduate} {Written} {Work}: {Using} {Bloom}'s {Taxonomy} as a {Pedagogical} {Tool} to {Improve} {Literature} {Reviews}}.
\newblock \bibinfo{journal}{\emph{Counselor Education and Supervision}} \bibinfo{volume}{40}, \bibinfo{number}{4} (\bibinfo{date}{June} \bibinfo{year}{2001}), \bibinfo{pages}{292--307}.
\newblock
\showISSN{00110035}
\urldef\tempurl%
\url{https://doi.org/10.1002/j.1556-6978.2001.tb01261.x}
\showDOI{\tempurl}


\bibitem[\protect\citeauthoryear{Griffith, Spies, Krysiak, McMichael, Coffman, Danos, Ainscough, Ramirez, Rieke, Kujan, Barnell, Wagner, Skidmore, Wollam, Liu, Jones, Bilski, Lesurf, Feng, Shah, Bonakdar, Trani, Matlock, Ramu, Campbell, Spies, Graubert, Gangavarapu, Eldred, Larson, Walker, Good, Wu, Su, Dienstmann, Margolin, Tamborero, Lopez-Bigas, Jones, Bose, Spencer, Wartman, Wilson, Mardis, and Griffith}{Griffith et~al\mbox{.}}{2017}]%
        {griffithCIViCCommunityKnowledgebase2017}
\bibfield{author}{\bibinfo{person}{Malachi Griffith}, \bibinfo{person}{Nicholas~C. Spies}, \bibinfo{person}{Kilannin Krysiak}, \bibinfo{person}{Joshua~F. McMichael}, \bibinfo{person}{Adam~C. Coffman}, \bibinfo{person}{Arpad~M. Danos}, \bibinfo{person}{Benjamin~J. Ainscough}, \bibinfo{person}{Cody~A. Ramirez}, \bibinfo{person}{Damian~T. Rieke}, \bibinfo{person}{Lynzey Kujan}, \bibinfo{person}{Erica~K. Barnell}, \bibinfo{person}{Alex~H. Wagner}, \bibinfo{person}{Zachary~L. Skidmore}, \bibinfo{person}{Amber Wollam}, \bibinfo{person}{Connor~J. Liu}, \bibinfo{person}{Martin~R. Jones}, \bibinfo{person}{Rachel~L. Bilski}, \bibinfo{person}{Robert Lesurf}, \bibinfo{person}{Yan-Yang Feng}, \bibinfo{person}{Nakul~M. Shah}, \bibinfo{person}{Melika Bonakdar}, \bibinfo{person}{Lee Trani}, \bibinfo{person}{Matthew Matlock}, \bibinfo{person}{Avinash Ramu}, \bibinfo{person}{Katie~M. Campbell}, \bibinfo{person}{Gregory~C. Spies}, \bibinfo{person}{Aaron~P. Graubert}, \bibinfo{person}{Karthik Gangavarapu},
  \bibinfo{person}{James~M. Eldred}, \bibinfo{person}{David~E. Larson}, \bibinfo{person}{Jason~R. Walker}, \bibinfo{person}{Benjamin~M. Good}, \bibinfo{person}{Chunlei Wu}, \bibinfo{person}{Andrew~I. Su}, \bibinfo{person}{Rodrigo Dienstmann}, \bibinfo{person}{Adam~A. Margolin}, \bibinfo{person}{David Tamborero}, \bibinfo{person}{Nuria Lopez-Bigas}, \bibinfo{person}{Steven J.~M. Jones}, \bibinfo{person}{Ron Bose}, \bibinfo{person}{David~H. Spencer}, \bibinfo{person}{Lukas~D. Wartman}, \bibinfo{person}{Richard~K. Wilson}, \bibinfo{person}{Elaine~R. Mardis}, {and} \bibinfo{person}{Obi~L. Griffith}.} \bibinfo{year}{2017}\natexlab{}.
\newblock \showarticletitle{{CIViC} is a community knowledgebase for expert crowdsourcing the clinical interpretation of variants in cancer}.
\newblock \bibinfo{journal}{\emph{Nature Genetics}} \bibinfo{volume}{49}, \bibinfo{number}{2} (\bibinfo{date}{Feb.} \bibinfo{year}{2017}), \bibinfo{pages}{170--174}.
\newblock
\showISSN{1546-1718}
\urldef\tempurl%
\url{https://doi.org/10.1038/ng.3774}
\showDOI{\tempurl}
\newblock
\shownote{00239 Number: 2 Publisher: Nature Publishing Group.}


\bibitem[\protect\citeauthoryear{Groth, Gibson, and Velterop}{Groth et~al\mbox{.}}{2010}]%
        {grothAnatomyNanopublication2010}
\bibfield{author}{\bibinfo{person}{Paul Groth}, \bibinfo{person}{Andrew Gibson}, {and} \bibinfo{person}{Jan Velterop}.} \bibinfo{year}{2010}\natexlab{}.
\newblock \showarticletitle{The anatomy of a nanopublication}.
\newblock \bibinfo{journal}{\emph{Information Services \& Use}} \bibinfo{volume}{30}, \bibinfo{number}{1-2} (\bibinfo{date}{Jan.} \bibinfo{year}{2010}), \bibinfo{pages}{51--56}.
\newblock
\showISSN{0167-5265}
\urldef\tempurl%
\url{https://doi.org/10.3233/ISU-2010-0613}
\showDOI{\tempurl}


\bibitem[\protect\citeauthoryear{Groza, Möller, Handschuh, Trif, and Decker}{Groza et~al\mbox{.}}{2007}]%
        {grozaSALTWeavingClaim2007}
\bibfield{author}{\bibinfo{person}{Tudor Groza}, \bibinfo{person}{Knud Möller}, \bibinfo{person}{Siegfried Handschuh}, \bibinfo{person}{Diana Trif}, {and} \bibinfo{person}{Stefan Decker}.} \bibinfo{year}{2007}\natexlab{}.
\newblock \showarticletitle{{SALT}: {Weaving} the {Claim} {Web}}. In \bibinfo{booktitle}{\emph{The {Semantic} {Web}}} \emph{(\bibinfo{series}{Lecture {Notes} in {Computer} {Science}})}, \bibfield{editor}{\bibinfo{person}{Karl Aberer}, \bibinfo{person}{Key-Sun Choi}, \bibinfo{person}{Natasha Noy}, \bibinfo{person}{Dean Allemang}, \bibinfo{person}{Kyung-Il Lee}, \bibinfo{person}{Lyndon Nixon}, \bibinfo{person}{Jennifer Golbeck}, \bibinfo{person}{Peter Mika}, \bibinfo{person}{Diana Maynard}, \bibinfo{person}{Riichiro Mizoguchi}, \bibinfo{person}{Guus Schreiber}, {and} \bibinfo{person}{Philippe Cudré-Mauroux}} (Eds.). \bibinfo{publisher}{Springer Berlin Heidelberg}, \bibinfo{pages}{197--210}.
\newblock
\showISBNx{978-3-540-76298-0}


\bibitem[\protect\citeauthoryear{Hanseth and Braa}{Hanseth and Braa}{1998}]%
        {hansethTechnologyTraitorEmergent1998}
\bibfield{author}{\bibinfo{person}{Ole Hanseth} {and} \bibinfo{person}{Kristin Braa}.} \bibinfo{year}{1998}\natexlab{}.
\newblock \showarticletitle{Technology as {Traitor}: {Emergent} {SAP} {Infrastructure} in a {Global} {Organization}}.
\newblock \bibinfo{journal}{\emph{ICIS 1998 Proceedings}} (\bibinfo{date}{Dec.} \bibinfo{year}{1998}).
\newblock
\urldef\tempurl%
\url{https://aisel.aisnet.org/icis1998/17}
\showURL{%
\tempurl}


\bibitem[\protect\citeauthoryear{Heer}{Heer}{2019}]%
        {heerAgencyAutomationDesigning2019}
\bibfield{author}{\bibinfo{person}{Jeffrey Heer}.} \bibinfo{year}{2019}\natexlab{}.
\newblock \showarticletitle{Agency plus Automation: {{Designing}} Artificial Intelligence into Interactive Systems}.
\newblock \bibinfo{journal}{\emph{Proceedings of the National Academy of Sciences}} \bibinfo{volume}{116}, \bibinfo{number}{6} (\bibinfo{date}{Feb.} \bibinfo{year}{2019}), \bibinfo{pages}{1844--1850}.
\newblock
\showISSN{0027-8424, 1091-6490}
\urldef\tempurl%
\url{https://doi.org/10.1073/pnas.1807184115}
\showDOI{\tempurl}


\bibitem[\protect\citeauthoryear{Hoang and Schneider}{Hoang and Schneider}{2018}]%
        {hoangOpportunitiesComputerSupport2018}
\bibfield{author}{\bibinfo{person}{Linh Hoang} {and} \bibinfo{person}{Jodi Schneider}.} \bibinfo{year}{2018}\natexlab{}.
\newblock \showarticletitle{Opportunities for {Computer} {Support} for {Systematic} {Reviewing} - {A} {Gap} {Analysis}}. In \bibinfo{booktitle}{\emph{Transforming {Digital} {Worlds}}} \emph{(\bibinfo{series}{Lecture {Notes} in {Computer} {Science}})}, \bibfield{editor}{\bibinfo{person}{Gobinda Chowdhury}, \bibinfo{person}{Julie McLeod}, \bibinfo{person}{Val Gillet}, {and} \bibinfo{person}{Peter Willett}} (Eds.). \bibinfo{publisher}{Springer International Publishing}, \bibinfo{pages}{367--377}.
\newblock
\showISBNx{978-3-319-78105-1}


\bibitem[\protect\citeauthoryear{Holbrook, Bourke, Lovat, and Dally}{Holbrook et~al\mbox{.}}{2004}]%
        {holbrookInvestigatingPhDThesis2004}
\bibfield{author}{\bibinfo{person}{Allyson Holbrook}, \bibinfo{person}{Sid Bourke}, \bibinfo{person}{Terence Lovat}, {and} \bibinfo{person}{Kerry Dally}.} \bibinfo{year}{2004}\natexlab{}.
\newblock \showarticletitle{Investigating {PhD} thesis examination reports}.
\newblock \bibinfo{journal}{\emph{International Journal of Educational Research}} \bibinfo{volume}{41}, \bibinfo{number}{2} (\bibinfo{date}{Jan.} \bibinfo{year}{2004}), \bibinfo{pages}{98--120}.
\newblock
\showISSN{0883-0355}
\urldef\tempurl%
\url{https://doi.org/10.1016/j.ijer.2005.04.008}
\showDOI{\tempurl}


\bibitem[\protect\citeauthoryear{Horvitz}{Horvitz}{1999}]%
        {horvitzPrinciplesMixedinitiativeUser1999}
\bibfield{author}{\bibinfo{person}{Eric Horvitz}.} \bibinfo{year}{1999}\natexlab{}.
\newblock \showarticletitle{Principles of {{Mixed}}-Initiative {{User Interfaces}}}. In \bibinfo{booktitle}{\emph{Proceedings of the {{SIGCHI Conference}} on {{Human Factors}} in {{Computing Systems}}}} \emph{(\bibinfo{series}{{{CHI}} '99})}. \bibinfo{publisher}{{ACM}}, \bibinfo{address}{New York, NY, USA}, \bibinfo{pages}{159--166}.
\newblock
\showISBNx{978-0-201-48559-2}
\urldef\tempurl%
\url{https://doi.org/10.1145/302979.303030}
\showDOI{\tempurl}


\bibitem[\protect\citeauthoryear{Hughes}{Hughes}{1983}]%
        {hughesNetworksPowerElectrification1983}
\bibfield{author}{\bibinfo{person}{Thomas~Parke Hughes}.} \bibinfo{year}{1983}\natexlab{}.
\newblock \bibinfo{booktitle}{\emph{Networks of {Power}: {Electrification} in {Western} {Society}, 1880-1930}}.
\newblock \bibinfo{publisher}{JHU Press}.
\newblock
\showISBNx{978-0-8018-4614-4}
\newblock
\shownote{05700 Google-Books-ID: g07Q9M4agp4C.}


\bibitem[\protect\citeauthoryear{Jonnalagadda, Goyal, and Huffman}{Jonnalagadda et~al\mbox{.}}{2015}]%
        {jonnalagaddaAutomatingDataExtraction2015}
\bibfield{author}{\bibinfo{person}{Siddhartha~R. Jonnalagadda}, \bibinfo{person}{Pawan Goyal}, {and} \bibinfo{person}{Mark~D. Huffman}.} \bibinfo{year}{2015}\natexlab{}.
\newblock \showarticletitle{Automating data extraction in systematic reviews: a systematic review}.
\newblock \bibinfo{journal}{\emph{Systematic Reviews}} \bibinfo{volume}{4}, \bibinfo{number}{1} (\bibinfo{date}{June} \bibinfo{year}{2015}), \bibinfo{pages}{78}.
\newblock
\showISSN{2046-4053}
\urldef\tempurl%
\url{https://doi.org/10.1186/s13643-015-0066-7}
\showDOI{\tempurl}


\bibitem[\protect\citeauthoryear{Kilicoglu}{Kilicoglu}{2017}]%
        {kilicogluBiomedicalTextMining}
\bibfield{author}{\bibinfo{person}{Halil Kilicoglu}.} \bibinfo{year}{2017}\natexlab{}.
\newblock \showarticletitle{Biomedical Text Mining for Research Rigor and Integrity: Tasks, Challenges, Directions}.
\newblock \bibinfo{journal}{\emph{Briefings in Bioinformatics}} (\bibinfo{year}{2017}).
\newblock
\urldef\tempurl%
\url{https://doi.org/10.1093/bib/bbx057}
\showDOI{\tempurl}


\bibitem[\protect\citeauthoryear{Kim, Nguyen, Weir, Guo, Miller, and Gajos}{Kim et~al\mbox{.}}{2014}]%
        {kimCrowdsourcingStepbystepInformation2014}
\bibfield{author}{\bibinfo{person}{Juho Kim}, \bibinfo{person}{Phu~Tran Nguyen}, \bibinfo{person}{Sarah Weir}, \bibinfo{person}{Philip~J. Guo}, \bibinfo{person}{Robert~C. Miller}, {and} \bibinfo{person}{Krzysztof~Z. Gajos}.} \bibinfo{year}{2014}\natexlab{}.
\newblock \showarticletitle{Crowdsourcing {Step}-by-step {Information} {Extraction} to {Enhance} {Existing} {How}-to {Videos}}. In \bibinfo{booktitle}{\emph{Proceedings of the {SIGCHI} {Conference} on {Human} {Factors} in {Computing} {Systems}}} \emph{(\bibinfo{series}{{CHI} '14})}. \bibinfo{publisher}{ACM}, \bibinfo{address}{New York, NY, USA}, \bibinfo{pages}{4017--4026}.
\newblock
\showISBNx{978-1-4503-2473-1}
\urldef\tempurl%
\url{https://doi.org/10.1145/2556288.2556986}
\showDOI{\tempurl}


\bibitem[\protect\citeauthoryear{Kittur, Yu, Hope, Chan, {Lifshitz-Assaf}, Gilon, Ng, Kraut, and Shahaf}{Kittur et~al\mbox{.}}{2019}]%
        {kitturScalingAnalogicalInnovation2019}
\bibfield{author}{\bibinfo{person}{Aniket Kittur}, \bibinfo{person}{Lixiu Yu}, \bibinfo{person}{Tom Hope}, \bibinfo{person}{Joel Chan}, \bibinfo{person}{Hila {Lifshitz-Assaf}}, \bibinfo{person}{Karni Gilon}, \bibinfo{person}{Felicia Ng}, \bibinfo{person}{Robert~E. Kraut}, {and} \bibinfo{person}{Dafna Shahaf}.} \bibinfo{year}{2019}\natexlab{}.
\newblock \showarticletitle{Scaling up Analogical Innovation with Crowds and {{AI}}}.
\newblock \bibinfo{journal}{\emph{Proceedings of the National Academy of Sciences}} (\bibinfo{date}{Jan.} \bibinfo{year}{2019}), \bibinfo{pages}{201807185}.
\newblock
\showISSN{0027-8424, 1091-6490}
\urldef\tempurl%
\url{https://doi.org/10.1073/pnas.1807185116}
\showDOI{\tempurl}


\bibitem[\protect\citeauthoryear{Knight, Wilson, Brailsford, and Milic-Frayling}{Knight et~al\mbox{.}}{2019}]%
        {knightEnslavedTrappedData2019}
\bibfield{author}{\bibinfo{person}{Ian~A. Knight}, \bibinfo{person}{Max~L. Wilson}, \bibinfo{person}{David~F. Brailsford}, {and} \bibinfo{person}{Natasa Milic-Frayling}.} \bibinfo{year}{2019}\natexlab{}.
\newblock \showarticletitle{Enslaved to the {Trapped} {Data}: {A} {Cognitive} {Work} {Analysis} of {Medical} {Systematic} {Reviews}}. In \bibinfo{booktitle}{\emph{Proceedings of the 2019 {Conference} on {Human} {Information} {Interaction} and {Retrieval}}} \emph{(\bibinfo{series}{{CHIIR} '19})}. \bibinfo{publisher}{ACM}, \bibinfo{address}{New York, NY, USA}, \bibinfo{pages}{203--212}.
\newblock
\showISBNx{978-1-4503-6025-8}
\urldef\tempurl%
\url{https://doi.org/10.1145/3295750.3298937}
\showDOI{\tempurl}
\newblock
\shownote{event-place: Glasgow, Scotland UK.}


\bibitem[\protect\citeauthoryear{Kriplean, Morgan, Freelon, Borning, and Bennett}{Kriplean et~al\mbox{.}}{2012}]%
        {kripleanSupportingReflectivePublic2012}
\bibfield{author}{\bibinfo{person}{Travis Kriplean}, \bibinfo{person}{Jonathan Morgan}, \bibinfo{person}{Deen Freelon}, \bibinfo{person}{Alan Borning}, {and} \bibinfo{person}{Lance Bennett}.} \bibinfo{year}{2012}\natexlab{}.
\newblock \showarticletitle{Supporting {Reflective} {Public} {Thought} with {Considerit}}. In \bibinfo{booktitle}{\emph{Proceedings of the {ACM} 2012 {Conference} on {Computer} {Supported} {Cooperative} {Work}}} \emph{(\bibinfo{series}{{CSCW} '12})}. \bibinfo{publisher}{ACM}, \bibinfo{address}{New York, NY, USA}, \bibinfo{pages}{265--274}.
\newblock
\showISBNx{978-1-4503-1086-4}
\urldef\tempurl%
\url{https://doi.org/10.1145/2145204.2145249}
\showDOI{\tempurl}
\newblock
\shownote{00000.}


\bibitem[\protect\citeauthoryear{Kuhn}{Kuhn}{2009}]%
        {kuhnAceWikiNaturalExpressive2009}
\bibfield{author}{\bibinfo{person}{Tobias Kuhn}.} \bibinfo{year}{2009}\natexlab{}.
\newblock \showarticletitle{{AceWiki}: {A} {Natural} and {Expressive} {Semantic} {Wiki}}. In \bibinfo{booktitle}{\emph{Semantic {Web} {User} {Interaction} ({SWUI}) {CEUR} {Workshop}}}. \bibinfo{pages}{8}.
\newblock
\newblock
\shownote{00092.}


\bibitem[\protect\citeauthoryear{Kuhn, Barbano, Nagy, and Krauthammer}{Kuhn et~al\mbox{.}}{2013}]%
        {kuhnBroadeningScopeNanopublications2013}
\bibfield{author}{\bibinfo{person}{Tobias Kuhn}, \bibinfo{person}{Paolo~Emilio Barbano}, \bibinfo{person}{Mate~Levente Nagy}, {and} \bibinfo{person}{Michael Krauthammer}.} \bibinfo{year}{2013}\natexlab{}.
\newblock \showarticletitle{Broadening the {Scope} of {Nanopublications}}. In \bibinfo{booktitle}{\emph{The {Semantic} {Web}: {Semantics} and {Big} {Data}}} \emph{(\bibinfo{series}{Lecture {Notes} in {Computer} {Science}})}, \bibfield{editor}{\bibinfo{person}{Philipp Cimiano}, \bibinfo{person}{Oscar Corcho}, \bibinfo{person}{Valentina Presutti}, \bibinfo{person}{Laura Hollink}, {and} \bibinfo{person}{Sebastian Rudolph}} (Eds.). \bibinfo{publisher}{Springer Berlin Heidelberg}, \bibinfo{pages}{487--501}.
\newblock
\showISBNx{978-3-642-38288-8}


\bibitem[\protect\citeauthoryear{Kuhn and Dumontier}{Kuhn and Dumontier}{2017}]%
        {kuhnGenuineSemanticPublishing2017}
\bibfield{author}{\bibinfo{person}{Tobias Kuhn} {and} \bibinfo{person}{Michel Dumontier}.} \bibinfo{year}{2017}\natexlab{}.
\newblock \showarticletitle{Genuine semantic publishing}.
\newblock \bibinfo{journal}{\emph{Data Science}} \bibinfo{volume}{1}, \bibinfo{number}{1-2} (\bibinfo{date}{Jan.} \bibinfo{year}{2017}), \bibinfo{pages}{139--154}.
\newblock
\showISSN{2451-8484}
\urldef\tempurl%
\url{https://doi.org/10.3233/DS-170010}
\showDOI{\tempurl}


\bibitem[\protect\citeauthoryear{Kunz and Rittel}{Kunz and Rittel}{1970}]%
        {kunzIssuesElementsInformation1970}
\bibfield{author}{\bibinfo{person}{Werner Kunz} {and} \bibinfo{person}{Horst~WJ Rittel}.} \bibinfo{year}{1970}\natexlab{}.
\newblock \bibinfo{booktitle}{\emph{Issues as elements of information systems}}.
\newblock \bibinfo{type}{Working {Paper}} 131. \bibinfo{institution}{Institute of Urban and Regional Development, University of California Berkeley}, \bibinfo{address}{Berkeley, California, USA}.
\newblock


\bibitem[\protect\citeauthoryear{Lave and Wenger}{Lave and Wenger}{1991}]%
        {laveSituatedLearningLegitimate1991}
\bibfield{author}{\bibinfo{person}{Jean Lave} {and} \bibinfo{person}{Etienne Wenger}.} \bibinfo{year}{1991}\natexlab{}.
\newblock \bibinfo{booktitle}{\emph{Situated {Learning}: {Legitimate} {Peripheral} {Participation}}}.
\newblock \bibinfo{publisher}{Cambridge University Press}.
\newblock


\bibitem[\protect\citeauthoryear{Le~Dantec and DiSalvo}{Le~Dantec and DiSalvo}{2013}]%
        {ledantecInfrastructuringFormationPublics2013}
\bibfield{author}{\bibinfo{person}{Christopher~A Le~Dantec} {and} \bibinfo{person}{Carl DiSalvo}.} \bibinfo{year}{2013}\natexlab{}.
\newblock \showarticletitle{Infrastructuring and the formation of publics in participatory design}.
\newblock \bibinfo{journal}{\emph{Social Studies of Science}} \bibinfo{volume}{43}, \bibinfo{number}{2} (\bibinfo{date}{April} \bibinfo{year}{2013}), \bibinfo{pages}{241--264}.
\newblock
\showISSN{0306-3127, 1460-3659}
\urldef\tempurl%
\url{https://doi.org/10.1177/0306312712471581}
\showDOI{\tempurl}
\newblock
\shownote{00469.}


\bibitem[\protect\citeauthoryear{Lee, Dourish, and Mark}{Lee et~al\mbox{.}}{2006}]%
        {leeHumanInfrastructureCyberinfrastructure2006}
\bibfield{author}{\bibinfo{person}{Charlotte~P. Lee}, \bibinfo{person}{Paul Dourish}, {and} \bibinfo{person}{Gloria Mark}.} \bibinfo{year}{2006}\natexlab{}.
\newblock \showarticletitle{The {Human} {Infrastructure} of {Cyberinfrastructure}}. In \bibinfo{booktitle}{\emph{Proceedings of the 2006 20th {Anniversary} {Conference} on {Computer} {Supported} {Cooperative} {Work}}} \emph{(\bibinfo{series}{{CSCW} '06})}. \bibinfo{publisher}{ACM}, \bibinfo{address}{New York, NY, USA}, \bibinfo{pages}{483--492}.
\newblock
\showISBNx{978-1-59593-249-5}
\urldef\tempurl%
\url{https://doi.org/10.1145/1180875.1180950}
\showDOI{\tempurl}
\newblock
\shownote{event-place: Banff, Alberta, Canada.}


\bibitem[\protect\citeauthoryear{Leigh~Star}{Leigh~Star}{2010}]%
        {leighstarThisNotBoundary2010}
\bibfield{author}{\bibinfo{person}{Susan Leigh~Star}.} \bibinfo{year}{2010}\natexlab{}.
\newblock \showarticletitle{This Is {{Not}} a {{Boundary Object}}: {{Reflections}} on the {{Origin}} of a {{Concept}} , {{This}} Is {{Not}} a {{Boundary Object}}: {{Reflections}} on the {{Origin}} of a {{Concept}}}.
\newblock \bibinfo{journal}{\emph{Science, Technology, \& Human Values}} \bibinfo{volume}{35}, \bibinfo{number}{5} (\bibinfo{date}{Sept.} \bibinfo{year}{2010}), \bibinfo{pages}{601--617}.
\newblock
\showISSN{0162-2439}
\urldef\tempurl%
\url{https://doi.org/10.1177/0162243910377624}
\showDOI{\tempurl}


\bibitem[\protect\citeauthoryear{Liddo, S\'andor, and Shum}{Liddo et~al\mbox{.}}{2012}]%
        {liddoContestedCollectiveIntelligence2012a}
\bibfield{author}{\bibinfo{person}{Anna~De Liddo}, \bibinfo{person}{\'Agnes S\'andor}, {and} \bibinfo{person}{Simon~Buckingham Shum}.} \bibinfo{year}{2012}\natexlab{}.
\newblock \showarticletitle{Contested {{Collective Intelligence}}: {{Rationale}}, {{Technologies}}, and a {{Human}}-{{Machine Annotation Study}}}.
\newblock \bibinfo{journal}{\emph{Computer Supported Cooperative Work (CSCW)}} \bibinfo{volume}{21}, \bibinfo{number}{4-5} (\bibinfo{date}{Oct.} \bibinfo{year}{2012}), \bibinfo{pages}{417--448}.
\newblock
\showISSN{0925-9724, 1573-7551}
\urldef\tempurl%
\url{https://doi.org/10.1007/s10606-011-9155-x}
\showDOI{\tempurl}


\bibitem[\protect\citeauthoryear{Lovitts}{Lovitts}{2007}]%
        {lovittsMakingImplicitExplicit2007}
\bibfield{author}{\bibinfo{person}{Barbara~E. Lovitts}.} \bibinfo{year}{2007}\natexlab{}.
\newblock \bibinfo{booktitle}{\emph{Making the {Implicit} {Explicit}: {Creating} {Performance} {Expectations} for the {Dissertation}}}.
\newblock \bibinfo{publisher}{Stylus Publishing}, \bibinfo{address}{Sterling, Va}.
\newblock
\showISBNx{978-1-57922-181-2}


\bibitem[\protect\citeauthoryear{Marshall, Shipman, and Coombs}{Marshall et~al\mbox{.}}{1994}]%
        {marshallVIKISpatialHypertext1994}
\bibfield{author}{\bibinfo{person}{Catherine~C. Marshall}, \bibinfo{person}{Frank~M. Shipman}, {and} \bibinfo{person}{James~H. Coombs}.} \bibinfo{year}{1994}\natexlab{}.
\newblock \showarticletitle{{VIKI}: spatial hypertext supporting emergent structure}. In \bibinfo{booktitle}{\emph{Proceedings of the 1994 {ACM} {European} conference on {Hypermedia} technology}} \emph{(\bibinfo{series}{{ECHT} '94})}. \bibinfo{publisher}{Association for Computing Machinery}, \bibinfo{address}{New York, NY, USA}, \bibinfo{pages}{13--23}.
\newblock
\showISBNx{978-0-89791-640-0}
\urldef\tempurl%
\url{https://doi.org/10.1145/192757.192759}
\showDOI{\tempurl}


\bibitem[\protect\citeauthoryear{Marshall, Johnson, Wang, Rajasekaran, and Wallace}{Marshall et~al\mbox{.}}{2020}]%
        {marshallSemiAutomatedEvidenceSynthesis2020}
\bibfield{author}{\bibinfo{person}{Iain~J. Marshall}, \bibinfo{person}{Blair~T. Johnson}, \bibinfo{person}{Zigeng Wang}, \bibinfo{person}{Sanguthevar Rajasekaran}, {and} \bibinfo{person}{Byron~C. Wallace}.} \bibinfo{year}{2020}\natexlab{}.
\newblock \showarticletitle{Semi-{Automated} evidence synthesis in health psychology: current methods and future prospects}.
\newblock \bibinfo{journal}{\emph{Health Psychology Review}} \bibinfo{volume}{14}, \bibinfo{number}{1} (\bibinfo{date}{Jan.} \bibinfo{year}{2020}), \bibinfo{pages}{145--158}.
\newblock
\showISSN{1743-7199}
\urldef\tempurl%
\url{https://doi.org/10.1080/17437199.2020.1716198}
\showDOI{\tempurl}
\newblock
\shownote{Publisher: Routledge \_eprint: https://doi.org/10.1080/17437199.2020.1716198.}


\bibitem[\protect\citeauthoryear{McCrickard}{McCrickard}{2012}]%
        {mccrickardMakingClaimsKnowledge2012}
\bibfield{author}{\bibinfo{person}{D.~Scott McCrickard}.} \bibinfo{year}{2012}\natexlab{}.
\newblock \showarticletitle{Making {{Claims}}: {{Knowledge Design}}, {{Capture}}, and {{Sharing}} in {{HCI}}}.
\newblock \bibinfo{journal}{\emph{Synthesis Lectures on Human-Centered Informatics}} \bibinfo{volume}{5}, \bibinfo{number}{3} (\bibinfo{date}{June} \bibinfo{year}{2012}), \bibinfo{pages}{1--125}.
\newblock
\showISSN{1946-7680}
\urldef\tempurl%
\url{https://doi.org/10.2200/S00423ED1V01Y201205HCI015}
\showDOI{\tempurl}


\bibitem[\protect\citeauthoryear{McCrickard, Wahid, Branham, and Harrison}{McCrickard et~al\mbox{.}}{2013}]%
        {mccrickardAchievingBothCreativity2013}
\bibfield{author}{\bibinfo{person}{D.~Scott McCrickard}, \bibinfo{person}{Shahtab Wahid}, \bibinfo{person}{Stacy~M. Branham}, {and} \bibinfo{person}{Steve Harrison}.} \bibinfo{year}{2013}\natexlab{}.
\newblock \showarticletitle{Achieving {{Both Creativity}} and {{Rationale}}: {{Reuse}} in {{Design}} with {{Images}} and {{Claims}}}.
\newblock In \bibinfo{booktitle}{\emph{Creativity and {{Rationale}}}}, \bibfield{editor}{\bibinfo{person}{John~M. Carroll}} (Ed.). Number~20 in \bibinfo{series}{Human\textendash{{Computer Interaction Series}}}. \bibinfo{publisher}{{Springer London}}, \bibinfo{pages}{105--119}.
\newblock
\showISBNx{978-1-4471-4110-5 978-1-4471-4111-2}
\urldef\tempurl%
\url{https://doi.org/10.1007/978-1-4471-4111-2_6}
\showDOI{\tempurl}


\bibitem[\protect\citeauthoryear{McElreath and Smaldino}{McElreath and Smaldino}{2015}]%
        {mcelreathReplicationCommunicationPopulation2015}
\bibfield{author}{\bibinfo{person}{Richard McElreath} {and} \bibinfo{person}{Paul~E. Smaldino}.} \bibinfo{year}{2015}\natexlab{}.
\newblock \showarticletitle{Replication, {Communication}, and the {Population} {Dynamics} of {Scientific} {Discovery}}.
\newblock \bibinfo{journal}{\emph{PLOS ONE}} \bibinfo{volume}{10}, \bibinfo{number}{8} (\bibinfo{date}{Aug.} \bibinfo{year}{2015}), \bibinfo{pages}{e0136088}.
\newblock
\showISSN{1932-6203}
\urldef\tempurl%
\url{https://doi.org/10.1371/journal.pone.0136088}
\showDOI{\tempurl}
\newblock
\shownote{00107 Publisher: Public Library of Science.}


\bibitem[\protect\citeauthoryear{McPhetres, Albayrak-Aydemir, Mendes, Chow, Gonzalez-Marquez, Loukras, Maus, O'Mahony, Pomareda, Primbs, Sackman, Smithson, and Volodko}{McPhetres et~al\mbox{.}}{2020}]%
        {mcphetresDecadeTheoryReflected2020}
\bibfield{author}{\bibinfo{person}{Jonathon McPhetres}, \bibinfo{person}{Nihan Albayrak-Aydemir}, \bibinfo{person}{Ana~Barbosa Mendes}, \bibinfo{person}{Elvina~C. Chow}, \bibinfo{person}{Patricio Gonzalez-Marquez}, \bibinfo{person}{Erin Loukras}, \bibinfo{person}{Annika Maus}, \bibinfo{person}{Aoife O'Mahony}, \bibinfo{person}{Christina Pomareda}, \bibinfo{person}{Maximilian Primbs}, \bibinfo{person}{Shalaine Sackman}, \bibinfo{person}{Conor Smithson}, {and} \bibinfo{person}{Kirill Volodko}.} \bibinfo{year}{2020}\natexlab{}.
\newblock \bibinfo{booktitle}{\emph{A decade of theory as reflected in {Psychological} {Science} (2009-2019)}}.
\newblock \bibinfo{type}{{T}echnical {R}eport}. \bibinfo{institution}{PsyArXiv}.
\newblock
\urldef\tempurl%
\url{https://doi.org/10.31234/osf.io/hs5nx}
\showDOI{\tempurl}
\newblock
\shownote{00000 type: article.}


\bibitem[\protect\citeauthoryear{Michelson and Reuter}{Michelson and Reuter}{2019}]%
        {michelsonSignificantCostSystematic2019}
\bibfield{author}{\bibinfo{person}{Matthew Michelson} {and} \bibinfo{person}{Katja Reuter}.} \bibinfo{year}{2019}\natexlab{}.
\newblock \showarticletitle{The significant cost of systematic reviews and meta-analyses: {A} call for greater involvement of machine learning to assess the promise of clinical trials}.
\newblock \bibinfo{journal}{\emph{Contemporary Clinical Trials Communications}}  \bibinfo{volume}{16} (\bibinfo{date}{Dec.} \bibinfo{year}{2019}), \bibinfo{pages}{100443}.
\newblock
\showISSN{2451-8654}
\urldef\tempurl%
\url{https://doi.org/10.1016/j.conctc.2019.100443}
\showDOI{\tempurl}


\bibitem[\protect\citeauthoryear{Morabito and Chan}{Morabito and Chan}{2021}]%
        {morabitoManagingContextScholarly2021}
\bibfield{author}{\bibinfo{person}{John~S Morabito} {and} \bibinfo{person}{Joel Chan}.} \bibinfo{year}{2021}\natexlab{}.
\newblock \showarticletitle{Managing {Context} during {Scholarly} {Knowledge} {Synthesis}: {Process} {Patterns} and {System} {Mechanics}}. In \bibinfo{booktitle}{\emph{Creativity and {Cognition}}} \emph{(\bibinfo{series}{C\&amp;{C} '21})}. \bibinfo{publisher}{Association for Computing Machinery}, \bibinfo{address}{New York, NY, USA}, \bibinfo{pages}{1--5}.
\newblock
\showISBNx{978-1-4503-8376-9}
\urldef\tempurl%
\url{https://doi.org/10.1145/3450741.3465244}
\showDOI{\tempurl}
\newblock
\shownote{00000.}


\bibitem[\protect\citeauthoryear{Mosconi, Li, Randall, Karasti, Tolmie, Barutzky, Korn, and Pipek}{Mosconi et~al\mbox{.}}{2019}]%
        {mosconiThreeGapsOpening2019}
\bibfield{author}{\bibinfo{person}{Gaia Mosconi}, \bibinfo{person}{Qinyu Li}, \bibinfo{person}{Dave Randall}, \bibinfo{person}{Helena Karasti}, \bibinfo{person}{Peter Tolmie}, \bibinfo{person}{Jana Barutzky}, \bibinfo{person}{Matthias Korn}, {and} \bibinfo{person}{Volkmar Pipek}.} \bibinfo{year}{2019}\natexlab{}.
\newblock \showarticletitle{Three {Gaps} in {Opening} {Science}}.
\newblock \bibinfo{journal}{\emph{Computer Supported Cooperative Work (CSCW)}} \bibinfo{volume}{28}, \bibinfo{number}{3} (\bibinfo{date}{June} \bibinfo{year}{2019}), \bibinfo{pages}{749--789}.
\newblock
\showISSN{1573-7551}
\urldef\tempurl%
\url{https://doi.org/10.1007/s10606-019-09354-z}
\showDOI{\tempurl}


\bibitem[\protect\citeauthoryear{Myers, Ko, and Burnett}{Myers et~al\mbox{.}}{2006}]%
        {myersInvitedResearchOverview2006}
\bibfield{author}{\bibinfo{person}{Brad~A. Myers}, \bibinfo{person}{Amy~J. Ko}, {and} \bibinfo{person}{Margaret~M. Burnett}.} \bibinfo{year}{2006}\natexlab{}.
\newblock \showarticletitle{Invited research overview: end-user programming}. In \bibinfo{booktitle}{\emph{{CHI} '06 {Extended} {Abstracts} on {Human} {Factors} in {Computing} {Systems}}} \emph{(\bibinfo{series}{{CHI} {EA} '06})}. \bibinfo{publisher}{Association for Computing Machinery}, \bibinfo{address}{New York, NY, USA}, \bibinfo{pages}{75--80}.
\newblock
\showISBNx{978-1-59593-298-3}
\urldef\tempurl%
\url{https://doi.org/10.1145/1125451.1125472}
\showDOI{\tempurl}


\bibitem[\protect\citeauthoryear{Nelson}{Nelson}{1965}]%
        {nelsonComplexInformationProcessing1965}
\bibfield{author}{\bibinfo{person}{T.~H. Nelson}.} \bibinfo{year}{1965}\natexlab{}.
\newblock \showarticletitle{Complex information processing: a file structure for the complex, the changing and the indeterminate}. In \bibinfo{booktitle}{\emph{Proceedings of the 1965 20th national conference}} \emph{(\bibinfo{series}{{ACM} '65})}. \bibinfo{publisher}{Association for Computing Machinery}, \bibinfo{address}{Cleveland, Ohio, USA}, \bibinfo{pages}{84--100}.
\newblock
\showISBNx{978-1-4503-7495-8}
\urldef\tempurl%
\url{https://doi.org/10.1145/800197.806036}
\showDOI{\tempurl}


\bibitem[\protect\citeauthoryear{O'Hara, Smith, Newman, and Sellen}{O'Hara et~al\mbox{.}}{1998}]%
        {oharaStudentReadersUse1998}
\bibfield{author}{\bibinfo{person}{Kenton O'Hara}, \bibinfo{person}{Fiona Smith}, \bibinfo{person}{William Newman}, {and} \bibinfo{person}{Abigail Sellen}.} \bibinfo{year}{1998}\natexlab{}.
\newblock \showarticletitle{Student readers' use of library documents: implications for library technologies}. In \bibinfo{booktitle}{\emph{Proceedings of the {SIGCHI} {Conference} on {Human} {Factors} in {Computing} {Systems}}} \emph{(\bibinfo{series}{{CHI} '98})}. \bibinfo{publisher}{ACM Press/Addison-Wesley Publishing Co.}, \bibinfo{address}{Los Angeles, California, USA}, \bibinfo{pages}{233--240}.
\newblock
\showISBNx{978-0-201-30987-4}
\urldef\tempurl%
\url{https://doi.org/10.1145/274644.274678}
\showDOI{\tempurl}
\newblock
\shownote{00000.}


\bibitem[\protect\citeauthoryear{Palmer, Teffeau, and Pirmann}{Palmer et~al\mbox{.}}{2009}]%
        {palmerScholarlyInformationPractices2009}
\bibfield{author}{\bibinfo{person}{Carole~L. Palmer}, \bibinfo{person}{Lauren~C. Teffeau}, {and} \bibinfo{person}{Carrie~M. Pirmann}.} \bibinfo{year}{2009}\natexlab{}.
\newblock \bibinfo{booktitle}{\emph{Scholarly {Information} {Practices} in the {Online} {Environment}: {Themes} from the {Literature} and {Implications} for {Library} {Service} {Development}}}.
\newblock \bibinfo{type}{{T}echnical {R}eport}. \bibinfo{pages}{59} pages.
\newblock
\urldef\tempurl%
\url{http://www.oclc.org/programs/publications/reports/2009-02.pdf}
\showURL{%
\tempurl}


\bibitem[\protect\citeauthoryear{Petrosino}{Petrosino}{1999}]%
        {petrosino1999lead}
\bibfield{author}{\bibinfo{person}{Anthony Petrosino}.} \bibinfo{year}{1999}\natexlab{}.
\newblock \showarticletitle{Lead authors of cochrane reviews: {Survey} results}.
\newblock \bibinfo{journal}{\emph{Report to the Campbell Collaboration. Cambridge, MA: University of Pennsylvania}} (\bibinfo{year}{1999}).
\newblock


\bibitem[\protect\citeauthoryear{Pipek and Wulf}{Pipek and Wulf}{2009}]%
        {pipekInfrastructuringIntegratedPerspective2009}
\bibfield{author}{\bibinfo{person}{Volkmar Pipek} {and} \bibinfo{person}{Volker Wulf}.} \bibinfo{year}{2009}\natexlab{}.
\newblock \showarticletitle{Infrastructuring: {Toward} an {Integrated} {Perspective} on the {Design} and {Use} of {Information} {Technology}}.
\newblock \bibinfo{journal}{\emph{Journal of the Association for Information Systems}} \bibinfo{volume}{10}, \bibinfo{number}{5} (\bibinfo{date}{May} \bibinfo{year}{2009}), \bibinfo{pages}{447--473}.
\newblock
\showISSN{15369323}
\urldef\tempurl%
\url{https://doi.org/10.17705/1jais.00195}
\showDOI{\tempurl}


\bibitem[\protect\citeauthoryear{Qian, Fenlon, Lutters, and Chan}{Qian et~al\mbox{.}}{2020}]%
        {qianOpeningBlackBox2020}
\bibfield{author}{\bibinfo{person}{Xin Qian}, \bibinfo{person}{Katrina Fenlon}, \bibinfo{person}{Wayne~G. Lutters}, {and} \bibinfo{person}{Joel Chan}.} \bibinfo{year}{2020}\natexlab{}.
\newblock \showarticletitle{Opening {Up} the {Black} {Box} of {Scholarly} {Synthesis}: {Intermediate} {Products}, {Processes}, and {Tools}.}. In \bibinfo{booktitle}{\emph{Proceedings of {ASIST} 2020}}.
\newblock


\bibitem[\protect\citeauthoryear{Renear and Palmer}{Renear and Palmer}{2009a}]%
        {renearStrategicReadingOntologies2009}
\bibfield{author}{\bibinfo{person}{Allen~H. Renear} {and} \bibinfo{person}{Carole~L. Palmer}.} \bibinfo{year}{2009}\natexlab{a}.
\newblock \showarticletitle{Strategic {Reading}, {Ontologies}, and the {Future} of {Scientific} {Publishing}}.
\newblock \bibinfo{journal}{\emph{Science}} \bibinfo{volume}{325}, \bibinfo{number}{5942} (\bibinfo{date}{Aug.} \bibinfo{year}{2009}), \bibinfo{pages}{828--832}.
\newblock
\showISSN{0036-8075, 1095-9203}
\urldef\tempurl%
\url{https://doi.org/10.1126/science.1157784}
\showDOI{\tempurl}
\newblock
\shownote{00000.}


\bibitem[\protect\citeauthoryear{Renear and Palmer}{Renear and Palmer}{2009b}]%
        {renear_strategic_2009}
\bibfield{author}{\bibinfo{person}{A.~H. Renear} {and} \bibinfo{person}{C.~L. Palmer}.} \bibinfo{year}{2009}\natexlab{b}.
\newblock \showarticletitle{Strategic reading, ontologies, and the future of scientific publishing}.
\newblock \bibinfo{journal}{\emph{Science}} \bibinfo{volume}{325}, \bibinfo{number}{5942} (\bibinfo{year}{2009}), \bibinfo{pages}{828}.
\newblock


\bibitem[\protect\citeauthoryear{Ribes and Lee}{Ribes and Lee}{2010}]%
        {ribesSociotechnicalStudiesCyberinfrastructure2010}
\bibfield{author}{\bibinfo{person}{David Ribes} {and} \bibinfo{person}{Charlotte~P. Lee}.} \bibinfo{year}{2010}\natexlab{}.
\newblock \showarticletitle{Sociotechnical {Studies} of {Cyberinfrastructure} and e-{Research}: {Current} {Themes} and {Future} {Trajectories}}.
\newblock \bibinfo{journal}{\emph{Computer Supported Cooperative Work (CSCW)}} \bibinfo{volume}{19}, \bibinfo{number}{3} (\bibinfo{date}{Aug.} \bibinfo{year}{2010}), \bibinfo{pages}{231--244}.
\newblock
\showISSN{1573-7551}
\urldef\tempurl%
\url{https://doi.org/10.1007/s10606-010-9120-0}
\showDOI{\tempurl}


\bibitem[\protect\citeauthoryear{Rolland and Lee}{Rolland and Lee}{2013}]%
        {rollandTrustReliabilityReusing2013}
\bibfield{author}{\bibinfo{person}{Betsy Rolland} {and} \bibinfo{person}{Charlotte~P. Lee}.} \bibinfo{year}{2013}\natexlab{}.
\newblock \showarticletitle{Beyond trust and reliability: reusing data in collaborative cancer epidemiology research}. In \bibinfo{booktitle}{\emph{Proceedings of the 2013 conference on {Computer} supported cooperative work}} \emph{(\bibinfo{series}{{CSCW} '13})}. \bibinfo{publisher}{Association for Computing Machinery}, \bibinfo{address}{New York, NY, USA}, \bibinfo{pages}{435--444}.
\newblock
\showISBNx{978-1-4503-1331-5}
\urldef\tempurl%
\url{https://doi.org/10.1145/2441776.2441826}
\showDOI{\tempurl}


\bibitem[\protect\citeauthoryear{Russell, Stefik, Pirolli, and Card}{Russell et~al\mbox{.}}{1993}]%
        {russellCostStructureSensemaking1993}
\bibfield{author}{\bibinfo{person}{Daniel~M. Russell}, \bibinfo{person}{Mark~J. Stefik}, \bibinfo{person}{Peter Pirolli}, {and} \bibinfo{person}{Stuart~K. Card}.} \bibinfo{year}{1993}\natexlab{}.
\newblock \showarticletitle{The {{Cost Structure}} of {{Sensemaking}}}. In \bibinfo{booktitle}{\emph{Proceedings of the {{INTERACT}} '93 and {{CHI}} '93 {{Conference}} on {{Human Factors}} in {{Computing Systems}}}} \emph{(\bibinfo{series}{{{CHI}} '93})}. \bibinfo{publisher}{{ACM}}, \bibinfo{address}{New York, NY, USA}, \bibinfo{pages}{269--276}.
\newblock
\showISBNx{0-89791-575-5}
\urldef\tempurl%
\url{https://doi.org/10.1145/169059.169209}
\showDOI{\tempurl}


\bibitem[\protect\citeauthoryear{Sanders and Stappers}{Sanders and Stappers}{2008}]%
        {sandersCocreationNewLandscapes2008}
\bibfield{author}{\bibinfo{person}{Elizabeth B.-N. Sanders} {and} \bibinfo{person}{Pieter~Jan Stappers}.} \bibinfo{year}{2008}\natexlab{}.
\newblock \showarticletitle{Co-creation and the new landscapes of design}.
\newblock \bibinfo{journal}{\emph{CoDesign}} \bibinfo{volume}{4}, \bibinfo{number}{1} (\bibinfo{date}{March} \bibinfo{year}{2008}), \bibinfo{pages}{5--18}.
\newblock
\showISSN{1571-0882, 1745-3755}
\urldef\tempurl%
\url{https://doi.org/10.1080/15710880701875068}
\showDOI{\tempurl}


\bibitem[\protect\citeauthoryear{Sawyer, Kaziunas, and \O{}esterlund}{Sawyer et~al\mbox{.}}{2012}]%
        {sawyerSocialScientistsCyberinfrastructure2012}
\bibfield{author}{\bibinfo{person}{Steve Sawyer}, \bibinfo{person}{Elizabeth Kaziunas}, {and} \bibinfo{person}{Carsten \O{}esterlund}.} \bibinfo{year}{2012}\natexlab{}.
\newblock \showarticletitle{Social {{Scientists}} and {{Cyberinfrastructure}}: {{Insights}} from a {{Document Perspective}}}. In \bibinfo{booktitle}{\emph{Proceedings of the {{ACM}} 2012 {{Conference}} on {{Computer Supported Cooperative Work}}}} \emph{(\bibinfo{series}{{{CSCW}} '12})}. \bibinfo{publisher}{{ACM}}, \bibinfo{address}{New York, NY, USA}, \bibinfo{pages}{931--934}.
\newblock
\showISBNx{978-1-4503-1086-4}
\urldef\tempurl%
\url{https://doi.org/10.1145/2145204.2145342}
\showDOI{\tempurl}


\bibitem[\protect\citeauthoryear{Scheel, Tiokhin, Isager, and Lakens}{Scheel et~al\mbox{.}}{2020}]%
        {scheelWhyHypothesisTesters2020}
\bibfield{author}{\bibinfo{person}{Anne~M. Scheel}, \bibinfo{person}{Leonid Tiokhin}, \bibinfo{person}{Peder~M. Isager}, {and} \bibinfo{person}{Daniël Lakens}.} \bibinfo{year}{2020}\natexlab{}.
\newblock \showarticletitle{Why {Hypothesis} {Testers} {Should} {Spend} {Less} {Time} {Testing} {Hypotheses}}.
\newblock \bibinfo{journal}{\emph{Perspectives on Psychological Science: A Journal of the Association for Psychological Science}} (\bibinfo{date}{Dec.} \bibinfo{year}{2020}), \bibinfo{pages}{1745691620966795}.
\newblock
\showISSN{1745-6924}
\urldef\tempurl%
\url{https://doi.org/10.1177/1745691620966795}
\showDOI{\tempurl}
\newblock
\shownote{00008.}


\bibitem[\protect\citeauthoryear{Schneider, Groza, and Passant}{Schneider et~al\mbox{.}}{2013}]%
        {schneiderReviewArgumentationSocial2013}
\bibfield{author}{\bibinfo{person}{Jodi Schneider}, \bibinfo{person}{Tudor Groza}, {and} \bibinfo{person}{Alexandre Passant}.} \bibinfo{year}{2013}\natexlab{}.
\newblock \showarticletitle{A review of argumentation for the social semantic web}.
\newblock \bibinfo{journal}{\emph{Semantic Web}} \bibinfo{volume}{4}, \bibinfo{number}{2} (\bibinfo{year}{2013}), \bibinfo{pages}{159--218}.
\newblock
\showISSN{1570-0844}
\newblock
\shownote{Publisher: IOS Press.}


\bibitem[\protect\citeauthoryear{Schwandt}{Schwandt}{1994}]%
        {schwandtConstructivistInterpretivistApproaches1994}
\bibfield{author}{\bibinfo{person}{Thomas Schwandt}.} \bibinfo{year}{1994}\natexlab{}.
\newblock \showarticletitle{Constructivist, {Interpretivist} {Approaches} to {Human} {Inquiry}}.
\newblock In \bibinfo{booktitle}{\emph{Handbook of {Qualitative} {Research}}}. \bibinfo{publisher}{SAGE}, \bibinfo{address}{Thousand Oaks, California}.
\newblock


\bibitem[\protect\citeauthoryear{Schön}{Schön}{1983}]%
        {schonReflectivePractitionerHow1983}
\bibfield{author}{\bibinfo{person}{Donald~A. Schön}.} \bibinfo{year}{1983}\natexlab{}.
\newblock \bibinfo{booktitle}{\emph{The reflective practitioner: {How} professionals think in action}}.
\newblock \bibinfo{publisher}{Basic books}.
\newblock


\bibitem[\protect\citeauthoryear{Seering, Kraut, and Dabbish}{Seering et~al\mbox{.}}{2017}]%
        {seeringShapingProAntiSocial2017}
\bibfield{author}{\bibinfo{person}{Joseph Seering}, \bibinfo{person}{Robert Kraut}, {and} \bibinfo{person}{Laura Dabbish}.} \bibinfo{year}{2017}\natexlab{}.
\newblock \showarticletitle{Shaping {Pro} and {Anti}-{Social} {Behavior} on {Twitch} {Through} {Moderation} and {Example}-{Setting}}. In \bibinfo{booktitle}{\emph{Proceedings of the 2017 {ACM} {Conference} on {Computer} {Supported} {Cooperative} {Work} and {Social} {Computing}}} \emph{(\bibinfo{series}{{CSCW} '17})}. \bibinfo{publisher}{ACM}, \bibinfo{address}{New York, NY, USA}, \bibinfo{pages}{111--125}.
\newblock
\showISBNx{978-1-4503-4335-0}
\urldef\tempurl%
\url{https://doi.org/10.1145/2998181.2998277}
\showDOI{\tempurl}


\bibitem[\protect\citeauthoryear{Shipman and Marshall}{Shipman and Marshall}{1999}]%
        {shipmanFormalityConsideredHarmful1999}
\bibfield{author}{\bibinfo{person}{Frank~M. Shipman} {and} \bibinfo{person}{Catherine~C. Marshall}.} \bibinfo{year}{1999}\natexlab{}.
\newblock \showarticletitle{Formality {{Considered Harmful}}: {{Experiences}}, {{Emerging Themes}}, and {{Directions}} on the {{Use}} of {{Formal Representations}} in {{Interactive Systems}}}.
\newblock \bibinfo{journal}{\emph{Computer Supported Cooperative Work (CSCW)}} \bibinfo{volume}{8}, \bibinfo{number}{4} (\bibinfo{date}{Dec.} \bibinfo{year}{1999}), \bibinfo{pages}{333--352}.
\newblock
\showISSN{0925-9724, 1573-7551}
\urldef\tempurl%
\url{https://doi.org/10.1023/A:1008716330212}
\showDOI{\tempurl}


\bibitem[\protect\citeauthoryear{Shipman and McCall}{Shipman and McCall}{1994}]%
        {shipmanSupportingKnowledgebaseEvolution1994}
\bibfield{author}{\bibinfo{person}{Frank~M. Shipman} {and} \bibinfo{person}{Raymond McCall}.} \bibinfo{year}{1994}\natexlab{}.
\newblock \showarticletitle{Supporting knowledge-base evolution with incremental formalization}. In \bibinfo{booktitle}{\emph{Proceedings of the {SIGCHI} {Conference} on {Human} {Factors} in {Computing} {Systems}}} \emph{(\bibinfo{series}{{CHI} '94})}. \bibinfo{publisher}{Association for Computing Machinery}, \bibinfo{address}{Boston, Massachusetts, USA}, \bibinfo{pages}{285--291}.
\newblock
\showISBNx{978-0-89791-650-9}
\urldef\tempurl%
\url{https://doi.org/10.1145/191666.191768}
\showDOI{\tempurl}
\newblock
\shownote{00000.}


\bibitem[\protect\citeauthoryear{Shojania, Sampson, Ansari, Ji, Doucette, and Moher}{Shojania et~al\mbox{.}}{2007}]%
        {shojaniaHowQuicklySystematic2007}
\bibfield{author}{\bibinfo{person}{Kaveh~G. Shojania}, \bibinfo{person}{Margaret Sampson}, \bibinfo{person}{Mohammed~T. Ansari}, \bibinfo{person}{Jun Ji}, \bibinfo{person}{Steve Doucette}, {and} \bibinfo{person}{David Moher}.} \bibinfo{year}{2007}\natexlab{}.
\newblock \showarticletitle{How {{Quickly Do Systematic Reviews Go Out}} of {{Date}}? {{A Survival Analysis}}}.
\newblock \bibinfo{journal}{\emph{Annals of Internal Medicine}} \bibinfo{volume}{147}, \bibinfo{number}{4} (\bibinfo{date}{Aug.} \bibinfo{year}{2007}), \bibinfo{pages}{224}.
\newblock
\showISSN{0003-4819}
\urldef\tempurl%
\url{https://doi.org/10.7326/0003-4819-147-4-200708210-00179}
\showDOI{\tempurl}


\bibitem[\protect\citeauthoryear{Shum, Motta, and Domingue}{Shum et~al\mbox{.}}{2000}]%
        {shumScholOntoOntologybasedDigital2000}
\bibfield{author}{\bibinfo{person}{Simon~Buckingham Shum}, \bibinfo{person}{Enrico Motta}, {and} \bibinfo{person}{John Domingue}.} \bibinfo{year}{2000}\natexlab{}.
\newblock \showarticletitle{{{ScholOnto}}: An Ontology-Based Digital Library Server for Research Documents and Discourse}.
\newblock \bibinfo{journal}{\emph{International Journal on Digital Libraries}} \bibinfo{volume}{3}, \bibinfo{number}{3} (\bibinfo{date}{Oct.} \bibinfo{year}{2000}), \bibinfo{pages}{237--248}.
\newblock
\showISSN{1432-5012}
\urldef\tempurl%
\url{https://doi.org/10.1007/s007990000034}
\showDOI{\tempurl}


\bibitem[\protect\citeauthoryear{Shum, Uren, Li, Sereno, and Mancini}{Shum et~al\mbox{.}}{2006}]%
        {shumModelingNaturalisticArgumentation2006}
\bibfield{author}{\bibinfo{person}{Simon J.~Buckingham Shum}, \bibinfo{person}{Victoria Uren}, \bibinfo{person}{Gangmin Li}, \bibinfo{person}{Bertrand Sereno}, {and} \bibinfo{person}{Clara Mancini}.} \bibinfo{year}{2006}\natexlab{}.
\newblock \showarticletitle{Modeling Naturalistic Argumentation in Research Literatures: {{Representation}} and Interaction Design Issues}.
\newblock \bibinfo{journal}{\emph{International Journal of Intelligent Systems}} \bibinfo{volume}{22}, \bibinfo{number}{1} (\bibinfo{date}{Nov.} \bibinfo{year}{2006}), \bibinfo{pages}{17--47}.
\newblock
\showISSN{1098-111X}
\urldef\tempurl%
\url{https://doi.org/10.1002/int.20188}
\showDOI{\tempurl}


\bibitem[\protect\citeauthoryear{Siangliulue, Chan, Dow, and Gajos}{Siangliulue et~al\mbox{.}}{2016}]%
        {siangliulueIdeaHoundImprovingLargescale2016}
\bibfield{author}{\bibinfo{person}{Pao Siangliulue}, \bibinfo{person}{Joel Chan}, \bibinfo{person}{Steven~P. Dow}, {and} \bibinfo{person}{Krzysztof~Z. Gajos}.} \bibinfo{year}{2016}\natexlab{}.
\newblock \showarticletitle{{{IdeaHound}}: {{Improving Large}}-Scale {{Collaborative Ideation}} with {{Crowd}}-{{Powered Real}}-Time {{Semantic Modeling}}}. In \bibinfo{booktitle}{\emph{Proceedings of the 29th {{Annual Symposium}} on {{User Interface Software}} and {{Technology}}}} \emph{(\bibinfo{series}{{{UIST}} '16})}. \bibinfo{publisher}{{ACM}}, \bibinfo{address}{New York, NY, USA}, \bibinfo{pages}{609--624}.
\newblock
\showISBNx{978-1-4503-4189-9}
\urldef\tempurl%
\url{https://doi.org/10.1145/2984511.2984578}
\showDOI{\tempurl}


\bibitem[\protect\citeauthoryear{Silver}{Silver}{2020}]%
        {silverCruxLiteratureReviewing2020}
\bibfield{author}{\bibinfo{person}{Christina Silver}.} \bibinfo{year}{2020}\natexlab{}.
\newblock \bibinfo{title}{The crux of literature reviewing: structuring critical appraisals and using {CAQDAS}-packages}.
\newblock
\newblock
\urldef\tempurl%
\url{https://www.qdaservices.co.uk/post/the-crux-of-literature-reviewing-structuring-critical-appraisals-and-using-caqdas-packages}
\showURL{%
\tempurl}
\newblock
\shownote{Library Catalog: www.qdaservices.co.uk.}


\bibitem[\protect\citeauthoryear{Star and Griesemer}{Star and Griesemer}{1989}]%
        {starInstitutionalEcologyTranslations1989}
\bibfield{author}{\bibinfo{person}{Susan~Leigh Star} {and} \bibinfo{person}{James~R. Griesemer}.} \bibinfo{year}{1989}\natexlab{}.
\newblock \showarticletitle{Institutional {{Ecology}}, `{{Translations}}' and {{Boundary Objects}}: {{Amateurs}} and {{Professionals}} in {{Berkeley}}'s {{Museum}} of {{Vertebrate Zoology}}, 1907-39}.
\newblock \bibinfo{journal}{\emph{Social Studies of Science}} \bibinfo{volume}{19}, \bibinfo{number}{3} (\bibinfo{date}{Aug.} \bibinfo{year}{1989}), \bibinfo{pages}{387--420}.
\newblock
\showISSN{0306-3127}
\urldef\tempurl%
\url{https://doi.org/10.1177/030631289019003001}
\showDOI{\tempurl}


\bibitem[\protect\citeauthoryear{Star and Ruhleder}{Star and Ruhleder}{1996}]%
        {starStepsEcologyInfrastructure1996}
\bibfield{author}{\bibinfo{person}{Susan~Leigh Star} {and} \bibinfo{person}{Karen Ruhleder}.} \bibinfo{year}{1996}\natexlab{}.
\newblock \showarticletitle{Steps toward an ecology of infrastructure: {Design} and access for large information spaces}.
\newblock \bibinfo{journal}{\emph{Information systems research}} \bibinfo{volume}{7}, \bibinfo{number}{1} (\bibinfo{year}{1996}), \bibinfo{pages}{111--134}.
\newblock
\urldef\tempurl%
\url{https://pubsonline.informs.org/doi/abs/10.1287/isre.7.1.111}
\showURL{%
\tempurl}
\newblock
\shownote{03070 tex.publisher: INFORMS.}


\bibitem[\protect\citeauthoryear{Strike and Posner}{Strike and Posner}{1983}]%
        {strikeTypesSynthesisTheir1983}
\bibfield{author}{\bibinfo{person}{Kenneth Strike} {and} \bibinfo{person}{George Posner}.} \bibinfo{year}{1983}\natexlab{}.
\newblock \showarticletitle{Types of synthesis and their criteria}.
\newblock In \bibinfo{booktitle}{\emph{Knowledge {Structure} and {Use}}}, \bibfield{editor}{\bibinfo{person}{S~Ward} {and} \bibinfo{person}{L~Reed}} (Eds.). \bibinfo{publisher}{Temple University Press}, \bibinfo{address}{Philadelphia, PA}.
\newblock


\bibitem[\protect\citeauthoryear{Thomas, {Noel-Storr}, Marshall, Wallace, McDonald, Mavergames, Glasziou, Shemilt, Synnot, Turner, Elliott, Agoritsas, Hilton, Perron, Akl, Hodder, Pestridge, Albrecht, Horsley, Platt, Armstrong, Nguyen, Plovnick, Arno, Ivers, Quinn, Au, Johnston, Rada, Bagg, Jones, Ravaud, Boden, Kahale, Richter, Boisvert, Keshavarz, Ryan, Brandt, {Kolakowsky-Hayner}, Salama, Brazinova, Nagraj, Salanti, Buchbinder, Lasserson, Santaguida, Champion, Lawrence, Santesso, Chandler, Les, Sch\"unemann, Charidimou, Leucht, Shemilt, Chou, Low, Sherifali, Churchill, Maas, Siemieniuk, Cnossen, MacLehose, Simmonds, Cossi, Macleod, Skoetz, Counotte, Marshall, {Soares-Weiser}, Craigie, Marshall, Srikanth, Dahm, Martin, Sullivan, Danilkewich, Garc\'ia, Synnot, Danko, Mavergames, Taylor, Donoghue, Maxwell, Thayer, Dressler, McAuley, Thomas, Egan, McDonald, Tritton, Elliott, McKenzie, Tsafnat, Elliott, Meerpohl, Tugwell, Etxeandia, Merner, Turgeon, Featherstone, Mondello, Turner, Foxlee, Morley, van
  Valkenhoef, Garner, Munafo, Vandvik, Gerrity, Munn, Wallace, Glasziou, Murano, Wallace, Green, Newman, Watts, Grimshaw, Nieuwlaat, Weeks, Gurusamy, Nikolakopoulou, Weigl, Haddaway, {Noel-Storr}, Wells, Hartling, O'Connor, Wiercioch, Hayden, Page, Wolfenden, Helfand, Pahwa, Nu\~nez, Higgins, Pardo, Yost, Hill, and Pearson}{Thomas et~al\mbox{.}}{2017}]%
        {thomasLivingSystematicReviews2017}
\bibfield{author}{\bibinfo{person}{James Thomas}, \bibinfo{person}{Anna {Noel-Storr}}, \bibinfo{person}{Iain Marshall}, \bibinfo{person}{Byron Wallace}, \bibinfo{person}{Steven McDonald}, \bibinfo{person}{Chris Mavergames}, \bibinfo{person}{Paul Glasziou}, \bibinfo{person}{Ian Shemilt}, \bibinfo{person}{Anneliese Synnot}, \bibinfo{person}{Tari Turner}, \bibinfo{person}{Julian Elliott}, \bibinfo{person}{Thomas Agoritsas}, \bibinfo{person}{John Hilton}, \bibinfo{person}{Caroline Perron}, \bibinfo{person}{Elie Akl}, \bibinfo{person}{Rebecca Hodder}, \bibinfo{person}{Charlotte Pestridge}, \bibinfo{person}{Lauren Albrecht}, \bibinfo{person}{Tanya Horsley}, \bibinfo{person}{Joanne Platt}, \bibinfo{person}{Rebecca Armstrong}, \bibinfo{person}{Phi~Hung Nguyen}, \bibinfo{person}{Robert Plovnick}, \bibinfo{person}{Anneliese Arno}, \bibinfo{person}{Noah Ivers}, \bibinfo{person}{Gail Quinn}, \bibinfo{person}{Agnes Au}, \bibinfo{person}{Renea Johnston}, \bibinfo{person}{Gabriel Rada}, \bibinfo{person}{Matthew Bagg},
  \bibinfo{person}{Arwel Jones}, \bibinfo{person}{Philippe Ravaud}, \bibinfo{person}{Catherine Boden}, \bibinfo{person}{Lara Kahale}, \bibinfo{person}{Bernt Richter}, \bibinfo{person}{Isabelle Boisvert}, \bibinfo{person}{Homa Keshavarz}, \bibinfo{person}{Rebecca Ryan}, \bibinfo{person}{Linn Brandt}, \bibinfo{person}{Stephanie~A. {Kolakowsky-Hayner}}, \bibinfo{person}{Dina Salama}, \bibinfo{person}{Alexandra Brazinova}, \bibinfo{person}{Sumanth~Kumbargere Nagraj}, \bibinfo{person}{Georgia Salanti}, \bibinfo{person}{Rachelle Buchbinder}, \bibinfo{person}{Toby Lasserson}, \bibinfo{person}{Lina Santaguida}, \bibinfo{person}{Chris Champion}, \bibinfo{person}{Rebecca Lawrence}, \bibinfo{person}{Nancy Santesso}, \bibinfo{person}{Jackie Chandler}, \bibinfo{person}{Zbigniew Les}, \bibinfo{person}{Holger~J. Sch\"unemann}, \bibinfo{person}{Andreas Charidimou}, \bibinfo{person}{Stefan Leucht}, \bibinfo{person}{Ian Shemilt}, \bibinfo{person}{Roger Chou}, \bibinfo{person}{Nicola Low}, \bibinfo{person}{Diana Sherifali},
  \bibinfo{person}{Rachel Churchill}, \bibinfo{person}{Andrew Maas}, \bibinfo{person}{Reed Siemieniuk}, \bibinfo{person}{Maryse~C. Cnossen}, \bibinfo{person}{Harriet MacLehose}, \bibinfo{person}{Mark Simmonds}, \bibinfo{person}{Marie-Joelle Cossi}, \bibinfo{person}{Malcolm Macleod}, \bibinfo{person}{Nicole Skoetz}, \bibinfo{person}{Michel Counotte}, \bibinfo{person}{Iain Marshall}, \bibinfo{person}{Karla {Soares-Weiser}}, \bibinfo{person}{Samantha Craigie}, \bibinfo{person}{Rachel Marshall}, \bibinfo{person}{Velandai Srikanth}, \bibinfo{person}{Philipp Dahm}, \bibinfo{person}{Nicole Martin}, \bibinfo{person}{Katrina Sullivan}, \bibinfo{person}{Alanna Danilkewich}, \bibinfo{person}{Laura~Mart\'inez Garc\'ia}, \bibinfo{person}{Anneliese Synnot}, \bibinfo{person}{Kristen Danko}, \bibinfo{person}{Chris Mavergames}, \bibinfo{person}{Mark Taylor}, \bibinfo{person}{Emma Donoghue}, \bibinfo{person}{Lara~J. Maxwell}, \bibinfo{person}{Kris Thayer}, \bibinfo{person}{Corinna Dressler}, \bibinfo{person}{James McAuley},
  \bibinfo{person}{James Thomas}, \bibinfo{person}{Cathy Egan}, \bibinfo{person}{Steve McDonald}, \bibinfo{person}{Roger Tritton}, \bibinfo{person}{Julian Elliott}, \bibinfo{person}{Joanne McKenzie}, \bibinfo{person}{Guy Tsafnat}, \bibinfo{person}{Sarah~A. Elliott}, \bibinfo{person}{Joerg Meerpohl}, \bibinfo{person}{Peter Tugwell}, \bibinfo{person}{Itziar Etxeandia}, \bibinfo{person}{Bronwen Merner}, \bibinfo{person}{Alexis Turgeon}, \bibinfo{person}{Robin Featherstone}, \bibinfo{person}{Stefania Mondello}, \bibinfo{person}{Tari Turner}, \bibinfo{person}{Ruth Foxlee}, \bibinfo{person}{Richard Morley}, \bibinfo{person}{Gert van Valkenhoef}, \bibinfo{person}{Paul Garner}, \bibinfo{person}{Marcus Munafo}, \bibinfo{person}{Per Vandvik}, \bibinfo{person}{Martha Gerrity}, \bibinfo{person}{Zachary Munn}, \bibinfo{person}{Byron Wallace}, \bibinfo{person}{Paul Glasziou}, \bibinfo{person}{Melissa Murano}, \bibinfo{person}{Sheila~A. Wallace}, \bibinfo{person}{Sally Green}, \bibinfo{person}{Kristine Newman},
  \bibinfo{person}{Chris Watts}, \bibinfo{person}{Jeremy Grimshaw}, \bibinfo{person}{Robby Nieuwlaat}, \bibinfo{person}{Laura Weeks}, \bibinfo{person}{Kurinchi Gurusamy}, \bibinfo{person}{Adriani Nikolakopoulou}, \bibinfo{person}{Aaron Weigl}, \bibinfo{person}{Neal Haddaway}, \bibinfo{person}{Anna {Noel-Storr}}, \bibinfo{person}{George Wells}, \bibinfo{person}{Lisa Hartling}, \bibinfo{person}{Annette O'Connor}, \bibinfo{person}{Wojtek Wiercioch}, \bibinfo{person}{Jill Hayden}, \bibinfo{person}{Matthew Page}, \bibinfo{person}{Luke Wolfenden}, \bibinfo{person}{Mark Helfand}, \bibinfo{person}{Manisha Pahwa}, \bibinfo{person}{Juan Jos\'e~Yepes Nu\~nez}, \bibinfo{person}{Julian Higgins}, \bibinfo{person}{Jordi~Pardo Pardo}, \bibinfo{person}{Jennifer Yost}, \bibinfo{person}{Sophie Hill}, {and} \bibinfo{person}{Leslea Pearson}.} \bibinfo{year}{2017}\natexlab{}.
\newblock \showarticletitle{Living Systematic Reviews: 2. {{Combining}} Human and Machine Effort}.
\newblock \bibinfo{journal}{\emph{Journal of Clinical Epidemiology}}  \bibinfo{volume}{91} (\bibinfo{date}{Nov.} \bibinfo{year}{2017}), \bibinfo{pages}{31--37}.
\newblock
\showISSN{0895-4356, 1878-5921}
\urldef\tempurl%
\url{https://doi.org/10.1016/j.jclinepi.2017.08.011}
\showDOI{\tempurl}


\bibitem[\protect\citeauthoryear{Toulmin}{Toulmin}{2003}]%
        {toulminUsesArgument2003}
\bibfield{author}{\bibinfo{person}{Stephen~E. Toulmin}.} \bibinfo{year}{2003}\natexlab{}.
\newblock \bibinfo{booktitle}{\emph{The {Uses} of {Argument}}}.
\newblock \bibinfo{publisher}{Cambridge University Press}.
\newblock
\showISBNx{978-0-521-53483-3}
\newblock
\shownote{00064 Google-Books-ID: 8UYgegaB1S0C.}


\bibitem[\protect\citeauthoryear{Uren, Buckingham~Shum, Bachler, and Li}{Uren et~al\mbox{.}}{2006}]%
        {urenSensemakingToolsUnderstanding2006}
\bibfield{author}{\bibinfo{person}{Victoria Uren}, \bibinfo{person}{Simon Buckingham~Shum}, \bibinfo{person}{Michelle Bachler}, {and} \bibinfo{person}{Gangmin Li}.} \bibinfo{year}{2006}\natexlab{}.
\newblock \showarticletitle{Sensemaking Tools for Understanding Research Literatures: {{Design}}, Implementation and User Evaluation}.
\newblock \bibinfo{journal}{\emph{International Journal of Human-Computer Studies}} \bibinfo{volume}{64}, \bibinfo{number}{5} (\bibinfo{date}{May} \bibinfo{year}{2006}), \bibinfo{pages}{420--445}.
\newblock
\showISSN{1071-5819}
\urldef\tempurl%
\url{https://doi.org/10.1016/j.ijhcs.2005.09.004}
\showDOI{\tempurl}


\bibitem[\protect\citeauthoryear{Van~Kleek, Bernstein, Karger, and {schraefel}}{Van~Kleek et~al\mbox{.}}{2007}]%
        {vankleekGuiPhooeyCase2007}
\bibfield{author}{\bibinfo{person}{Max Van~Kleek}, \bibinfo{person}{Michael Bernstein}, \bibinfo{person}{David~R. Karger}, {and} \bibinfo{person}{mc {schraefel}}.} \bibinfo{year}{2007}\natexlab{}.
\newblock \showarticletitle{Gui \textemdash{} {{Phooey}}!: {{The Case}} for {{Text Input}}}. In \bibinfo{booktitle}{\emph{Proceedings of the 20th {{Annual ACM Symposium}} on {{User Interface Software}} and {{Technology}}}} \emph{(\bibinfo{series}{{{UIST}} '07})}. \bibinfo{publisher}{{ACM}}, \bibinfo{address}{New York, NY, USA}, \bibinfo{pages}{193--202}.
\newblock
\showISBNx{978-1-59593-679-0}
\urldef\tempurl%
\url{https://doi.org/10.1145/1294211.1294247}
\showDOI{\tempurl}


\bibitem[\protect\citeauthoryear{van Rooij and Baggio}{van Rooij and Baggio}{2021}]%
        {vanrooijTheoryTestHow2021}
\bibfield{author}{\bibinfo{person}{Iris van Rooij} {and} \bibinfo{person}{Giosuè Baggio}.} \bibinfo{year}{2021}\natexlab{}.
\newblock \showarticletitle{Theory {Before} the {Test}: {How} to {Build} {High}-{Verisimilitude} {Explanatory} {Theories} in {Psychological} {Science}}.
\newblock \bibinfo{journal}{\emph{Perspectives on Psychological Science}} (\bibinfo{date}{Jan.} \bibinfo{year}{2021}), \bibinfo{pages}{1745691620970604}.
\newblock
\showISSN{1745-6916}
\urldef\tempurl%
\url{https://doi.org/10.1177/1745691620970604}
\showDOI{\tempurl}
\newblock
\shownote{00029 Publisher: SAGE Publications Inc.}


\bibitem[\protect\citeauthoryear{Vaughn and Jacquez}{Vaughn and Jacquez}{2020}]%
        {vaughnParticipatoryResearchMethods2020}
\bibfield{author}{\bibinfo{person}{Lisa~M. Vaughn} {and} \bibinfo{person}{Farrah Jacquez}.} \bibinfo{year}{2020}\natexlab{}.
\newblock \showarticletitle{Participatory {Research} {Methods} – {Choice} {Points} in the {Research} {Process}}.
\newblock \bibinfo{journal}{\emph{Journal of Participatory Research Methods}} \bibinfo{volume}{1}, \bibinfo{number}{1} (\bibinfo{date}{July} \bibinfo{year}{2020}).
\newblock
\urldef\tempurl%
\url{https://doi.org/10.35844/001c.13244}
\showDOI{\tempurl}


\bibitem[\protect\citeauthoryear{Weir, Kim, Gajos, and Miller}{Weir et~al\mbox{.}}{2015}]%
        {weirLearnersourcingSubgoalLabels2015}
\bibfield{author}{\bibinfo{person}{Sarah Weir}, \bibinfo{person}{Juho Kim}, \bibinfo{person}{Krzysztof~Z. Gajos}, {and} \bibinfo{person}{Robert~C. Miller}.} \bibinfo{year}{2015}\natexlab{}.
\newblock \showarticletitle{Learnersourcing {Subgoal} {Labels} for {How}-to {Videos}}. In \bibinfo{booktitle}{\emph{Proceedings of the 18th {ACM} {Conference} on {Computer} {Supported} {Cooperative} {Work} \& {Social} {Computing}}} \emph{(\bibinfo{series}{{CSCW} '15})}. \bibinfo{publisher}{ACM}, \bibinfo{address}{New York, NY, USA}, \bibinfo{pages}{405--416}.
\newblock
\showISBNx{978-1-4503-2922-4}
\urldef\tempurl%
\url{https://doi.org/10.1145/2675133.2675219}
\showDOI{\tempurl}


\bibitem[\protect\citeauthoryear{Willis, Sharma, Snyder, Brown, \O{}sterlund, and Sawyer}{Willis et~al\mbox{.}}{2014}]%
        {willisDocumentsDistributedScientific2014}
\bibfield{author}{\bibinfo{person}{Matt Willis}, \bibinfo{person}{Sarika Sharma}, \bibinfo{person}{Jaime Snyder}, \bibinfo{person}{Michelle Brown}, \bibinfo{person}{Carsten \O{}sterlund}, {and} \bibinfo{person}{Steve Sawyer}.} \bibinfo{year}{2014}\natexlab{}.
\newblock \showarticletitle{Documents and {{Distributed Scientific Collaboration}}}. In \bibinfo{booktitle}{\emph{Proceedings of the {{Companion Publication}} of the 17th {{ACM Conference}} on {{Computer Supported Cooperative Work}} \& {{Social Computing}}}} \emph{(\bibinfo{series}{{{CSCW Companion}} '14})}. \bibinfo{publisher}{{ACM}}, \bibinfo{address}{New York, NY, USA}, \bibinfo{pages}{257--260}.
\newblock
\showISBNx{978-1-4503-2541-7}
\urldef\tempurl%
\url{https://doi.org/10.1145/2556420.2556491}
\showDOI{\tempurl}


\bibitem[\protect\citeauthoryear{Wolfswinkel, Furtmueller, and Wilderom}{Wolfswinkel et~al\mbox{.}}{2013}]%
        {wolfswinkelUsingGroundedTheory2013}
\bibfield{author}{\bibinfo{person}{Joost~F Wolfswinkel}, \bibinfo{person}{Elfi Furtmueller}, {and} \bibinfo{person}{Celeste P~M Wilderom}.} \bibinfo{year}{2013}\natexlab{}.
\newblock \showarticletitle{Using grounded theory as a method for rigorously reviewing literature}.
\newblock \bibinfo{journal}{\emph{European Journal of Information Systems}} \bibinfo{volume}{22}, \bibinfo{number}{1} (\bibinfo{date}{Jan.} \bibinfo{year}{2013}), \bibinfo{pages}{45--55}.
\newblock
\showISSN{0960-085X}
\urldef\tempurl%
\url{https://doi.org/10.1057/ejis.2011.51}
\showDOI{\tempurl}
\newblock
\shownote{Publisher: Taylor \& Francis.}


\bibitem[\protect\citeauthoryear{Wright, Wadden, Lo, Kuehl, Cohan, Augenstein, and Wang}{Wright et~al\mbox{.}}{2022}]%
        {wrightGeneratingScientificClaims2022}
\bibfield{author}{\bibinfo{person}{Dustin Wright}, \bibinfo{person}{David Wadden}, \bibinfo{person}{Kyle Lo}, \bibinfo{person}{Bailey Kuehl}, \bibinfo{person}{Arman Cohan}, \bibinfo{person}{Isabelle Augenstein}, {and} \bibinfo{person}{Lucy~Lu Wang}.} \bibinfo{year}{2022}\natexlab{}.
\newblock \showarticletitle{Generating {Scientific} {Claims} for {Zero}-{Shot} {Scientific} {Fact} {Checking}}.
\newblock \bibinfo{journal}{\emph{arXiv:2203.12990 [cs]}} (\bibinfo{date}{March} \bibinfo{year}{2022}).
\newblock
\urldef\tempurl%
\url{http://arxiv.org/abs/2203.12990}
\showURL{%
\tempurl}
\newblock
\shownote{arXiv: 2203.12990.}


\bibitem[\protect\citeauthoryear{Yamamoto, Takada, Gross, and Nakakoji}{Yamamoto et~al\mbox{.}}{1998}]%
        {yamamotoRepresentationalTalkbackApproach1998}
\bibfield{author}{\bibinfo{person}{Y. Yamamoto}, \bibinfo{person}{S. Takada}, \bibinfo{person}{M. Gross}, {and} \bibinfo{person}{K. Nakakoji}.} \bibinfo{year}{1998}\natexlab{}.
\newblock \showarticletitle{Representational talkback: an approach to support writing as design}. In \bibinfo{booktitle}{\emph{Proceedings. 3rd {Asia} {Pacific} {Computer} {Human} {Interaction} ({Cat}. {No}.{98EX110})}}. \bibinfo{pages}{125--131}.
\newblock
\urldef\tempurl%
\url{https://doi.org/10.1109/APCHI.1998.704176}
\showDOI{\tempurl}


\bibitem[\protect\citeauthoryear{Young and Lutters}{Young and Lutters}{2017}]%
        {youngInfrastructuringCrossDisciplinarySynthetic2017}
\bibfield{author}{\bibinfo{person}{Alyson~L. Young} {and} \bibinfo{person}{Wayne~G. Lutters}.} \bibinfo{year}{2017}\natexlab{}.
\newblock \showarticletitle{Infrastructuring for {Cross}-{Disciplinary} {Synthetic} {Science}: {Meta}-{Study} {Research} in {Land} {System} {Science}}.
\newblock \bibinfo{journal}{\emph{Computer Supported Cooperative Work (CSCW)}} \bibinfo{volume}{26}, \bibinfo{number}{1} (\bibinfo{date}{April} \bibinfo{year}{2017}), \bibinfo{pages}{165--203}.
\newblock
\showISSN{1573-7551}
\urldef\tempurl%
\url{https://doi.org/10.1007/s10606-017-9267-z}
\showDOI{\tempurl}


\bibitem[\protect\citeauthoryear{Zimmerman and Forlizzi}{Zimmerman and Forlizzi}{2014}]%
        {zimmermanResearchDesignHCI2014}
\bibfield{author}{\bibinfo{person}{John Zimmerman} {and} \bibinfo{person}{Jodi Forlizzi}.} \bibinfo{year}{2014}\natexlab{}.
\newblock \showarticletitle{Research {Through} {Design} in {HCI}}.
\newblock In \bibinfo{booktitle}{\emph{Ways of {Knowing} in {HCI}}}, \bibfield{editor}{\bibinfo{person}{Judith~S. Olson} {and} \bibinfo{person}{Wendy~A. Kellogg}} (Eds.). \bibinfo{publisher}{Springer New York}, \bibinfo{address}{New York, NY}, \bibinfo{pages}{167--189}.
\newblock
\showISBNx{978-1-4939-0377-1 978-1-4939-0378-8}
\urldef\tempurl%
\url{https://doi.org/10.1007/978-1-4939-0378-8_8}
\showDOI{\tempurl}


\end{thebibliography}

\appendix

\section{Usage survey questions}
\label{ap:usage-survey-items}

\begin{enumerate}
    \item \textbf{What is your occupation?} \textit{Options:} 
    \begin{itemize}
        \item Research - Academic
        \item Research - Independent / applied
        \item Other
    \end{itemize}
    \item \textbf{Where are you in the world?} Please share city/state and country as you feel comfortable. \textit{(Free text)}
    \item \textbf{What sorts of things do you work on?} If you're a researcher, you can briefly describe your general research area(s) and some examples of problems you work on. If you're not doing research, briefly describe what you do. Think about the length of a Twitter bio or so. \textit{(Free text)}
    \item If Occupation = ``Research - Academic'':
    \begin{enumerate}
        \item \textbf{What is the nature of your academic research position?} \textit{Options:}
        \begin{itemize}
            \item Undergraduate Student Researcher
            \item Graduate Student Researcher - Masters
            \item Graduate Student Researcher - PhD
            \item Postdoctoral researcher
            \item Tenure-track faculty - pre-tenure
            \item Tenure-track faculty - post-tenure
            \item Professional / teaching faculty - junior
            \item Professional / teaching faculty - senior
            \item Other full-time position (e.g., research scientist)
            \item Other
        \end{itemize}
    \end{enumerate}
    \item \textbf{Are you a current user of the extension?} \textit{Options:}
    \begin{itemize}
        \item Yes
        \item Maybe - still exploring if will fit
        \item No - tried but didn't stick
    \end{itemize}
    \item If user = ``Yes''
    \begin{enumerate}
        \item \textbf{How did you learn about and start using the extension?} Common sources include Roam courses, Twitter, and Slack/Discord. We want to understand the channels through which the extension is spreading: what communities or sub-communities seem to be open to or need something like this. We also want to understand what sorts of support is needed to learn to use the extension. \textit{(Free text)}
        \item \textbf{Why did you start using the extension? What value is it providing to you now?} We want to understand the value proposition of this extension and its concepts for other users like yourself. It would be especially helpful if you are able to share specific examples of the extension's functionality delivering value to you, perhaps in contrast to how things used to be before you started using the extension.
        \item \textbf{Have you have you shared any part of your discourse graph with others? Why and how? What went well / could go better?} We want to understand the value proposition of this extension and its concepts for improving collaborative/distributed synthesis. 
        \item \textbf{Which of these features do you use?} If you're unsure of the list of features, you might learn more here: https://oasis-lab.gitbook.io/roamresearch-discourse-graph-extension/ \textit{(Options for each feature: ``Didn't know about this'', ``Never used``, ``Use on occasion'', ``Core part of my workflow'')}
        \begin{itemize}
            \item Node menu (for creating nodes)
            \item Discourse context
            \item Query drawer (ad-hoc)
            \item Discourse overlay
            \item Grammar editor (for creating/modifying nodes and relation patterns)
            \item Export
            \item Playground
            \item Node discourse attributes
            \item Node index
            \item Node templates
            \item Query drawer (saved to pages)
        \end{itemize}
        \item \textbf{Is there a screenshot (or set of screenshots) or video walkthrough you are willing to share to illustrate your usage of the extension and what value it provides to you?} \textit{(File upload)}
        \item \textbf{Alternatively is there a link to a recording (Loom/Youtube) you are willing to share to illustrate your usage of the extension and what value it provides to you?}
        \item \textbf{If you shared a screenshot/video, would you be ok with me sharing this with other potential users and in research reports to help others understand the extension?} If you say no, I will use it only for confidential analysis. \textit{(Options: ``Yes'', ``No''}
        \item \textbf{Would you be willing to share your csv export?} We are interested in understanding how people are using and modifying the grammar. We would like to analyze how many nodes and edges have been created, and how interconnected these are in your graph. If possible, we would also like to understand how the core question/claim/evidence nodes are used and written. If you are concerned about privacy, please feel free to remove the title and/or author names from the nodes csv before uploading here. If you wish to do so, and are not sure how: open the command palette (CMD/CTRL + P) and select "Export Discourse Graph" and then select the Neo4j export type. You will be prompted to save a .zip file, from which you can extract two csv files (<yourgraphname>\_query-results\_<datetime>\_nodes.csv and <yourgraphname>\_query-results\_<datetime>\_relations.csv). Feel free to attach the zip file, or the modified csv files separately. \textit{(File upload)}
        \item \textbf{What are your biggest pain points or wishlist items at the moment with the extension?} We are interested in understanding the main blockers to something like this extension being used by users like yourself, and what we might be able to do to remove those blockers. \textit{(Free text)}
    \end{enumerate}
    \item If user = ``Maybe''
    \begin{enumerate}
        \item \textbf{How did you learn about and start exploring the extension?} Common sources include Roam courses, Twitter, and Slack/Discord. We want to understand the channels through which the extension is spreading: what communities or sub-communities seem to be open to or need something like this. We also want to understand what sorts of support is needed to learn to use the extension. \textit{(Free text)}
        \item \textbf{Why did you start exploring the extension? What value do you hope it could provide to you?} We want to understand the potential value proposition of this extension and its concepts for other users like yourself. It would be especially helpful if you are able to share specific examples of how you hope the extension's functionality would deliver value to you, perhaps in contrast to how things are right now in your workflows and work.
        \item \textbf{Which of these features have you explored?} If you're unsure of the list of features, you might learn more here: https://oasis-lab.gitbook.io/roamresearch-discourse-graph-extension/ \textit{(Options for each feature: ``Didn't know about this'', ``Never used``, ``Use on occasion'', ``Core part of my workflow'')}
        \begin{itemize}
            \item Node menu (for creating nodes)
            \item Discourse context
            \item Query drawer (ad-hoc)
            \item Discourse overlay
            \item Grammar editor (for creating/modifying nodes and relation patterns)
            \item Export
            \item Playground
            \item Node discourse attributes
            \item Node index
            \item Node templates
            \item Query drawer (saved to pages)
        \end{itemize}
        \item \textbf{What are your biggest pain points or wishlist items at the moment with the extension?} We are interested in understanding the main blockers to something like this extension being used by users like yourself, and what we might be able to do to remove those blockers. \textit{(Free text)}
    \end{enumerate}
    \item If user = ``No''
    \begin{enumerate}
        \item \textbf{How did you learn about and start exploring the extension?} Common sources include Roam courses, Twitter, and Slack/Discord. We want to understand the channels through which the extension is spreading: what communities or sub-communities seem to be open to or need something like this. We also want to understand what sorts of support is needed to learn to use the extension. \textit{(Free text)}
        \item \textbf{Why did you start exploring the extension? What value did you hope it could provide to you?} We want to understand the potential value proposition of this extension and its concepts for other users like yourself. It would be especially helpful if you are able to share specific examples of how you hoped the extension's functionality would have delivered value to you, perhaps in contrast to how things are right now in your workflows and work.
        \item \textbf{Which of these features did you explore?} If you're unsure of the list of features, you might learn more here: https://oasis-lab.gitbook.io/roamresearch-discourse-graph-extension/ \textit{(Options for each feature: ``Didn't know about this'', ``Never used``, ``Use on occasion'', ``Core part of my workflow'')}
        \begin{itemize}
            \item Node menu (for creating nodes)
            \item Discourse context
            \item Query drawer (ad-hoc)
            \item Discourse overlay
            \item Grammar editor (for creating/modifying nodes and relation patterns)
            \item Export
            \item Playground
            \item Node discourse attributes
            \item Node index
            \item Node templates
            \item Query drawer (saved to pages)
        \end{itemize}
        \item \textbf{What were the main reasons you decided not to use the extension?} We are interested in understanding the main blockers to something like this extension being used by users like yourself, and what we might be able to do to remove those blockers. \textit{(Free text)}
    \end{enumerate}
        
\end{enumerate}

\section{Technical details and key design changes}
\label{ap:designs}
\begin{figure}
    \centering
    \includegraphics[width=\linewidth]{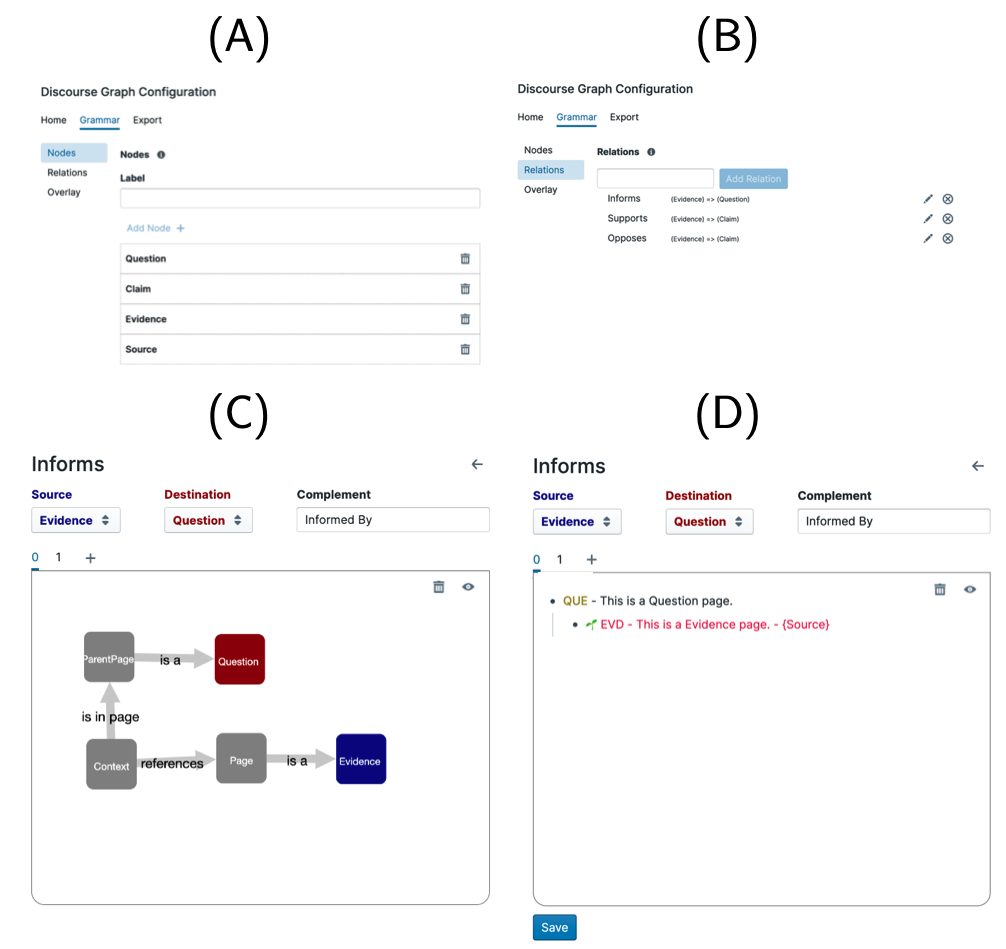}
    \caption{Snapshots from the grammar editor that shipped with the extension. Users could add new discourse nodes (A), add new discourse relation patterns (B), and use a graphical layout to specify the underlying Datalog query in the Roam database that mapped to writing patterns that would enable the discourse graph extension to recognize the discourse relations (C) and also preview the writing patterns associated with the discourse relation pattern (D)}
    \label{fig:sys-base-grammar-editor}
\end{figure}

\begin{figure}
    \centering
    \includegraphics[width=\linewidth]{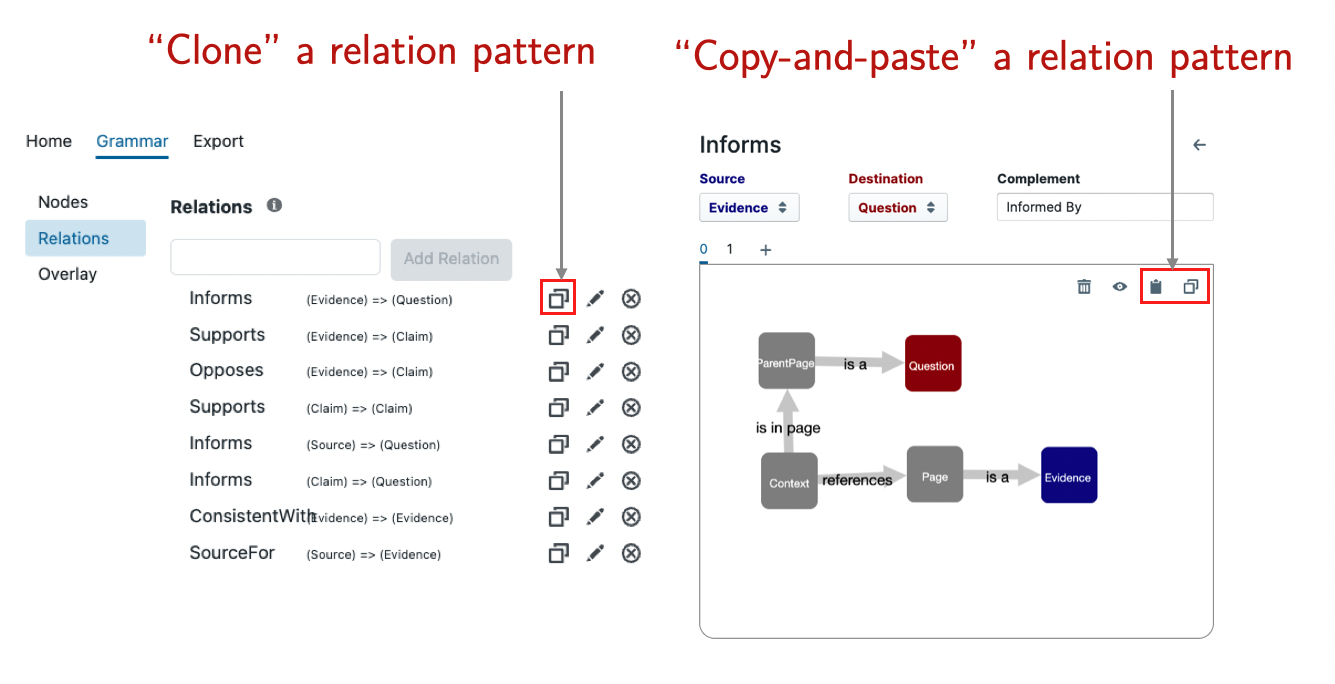}
    \caption{Key design changes to the graph grammar editor --- in response to user requests during the deploy period --- to enable users to to efficiently extend their discourse relation grammars by 1) "cloning" existing duplicate relation patterns, and 2) "copy-pasting" existing relation patterns.}
    \label{fig:grammar-editor-changes}
\end{figure}

Figure \ref{fig:sys-base-grammar-editor} shows the grammar editor that shipped with the extension, which allowed users to add new discourse nodes (A), add new discourse relation patterns (B), and use a graphical layout to specify the underlying Datalog query in the Roam database that mapped to writing patterns that would enable the discourse graph extension to recognize the discourse relations (C) and also preview the writing patterns associated with the discourse relation pattern (D). As described in Section \ref{sec:extending-grammar}, user behavior led to significant changes that enabled more efficient editing of the grammar (Figure \ref{fig:grammar-editor-changes})

\section{Supplementary data}
\label{ap:supplemental-data}
\begin{figure}
    \centering
    \includegraphics[width=\linewidth]{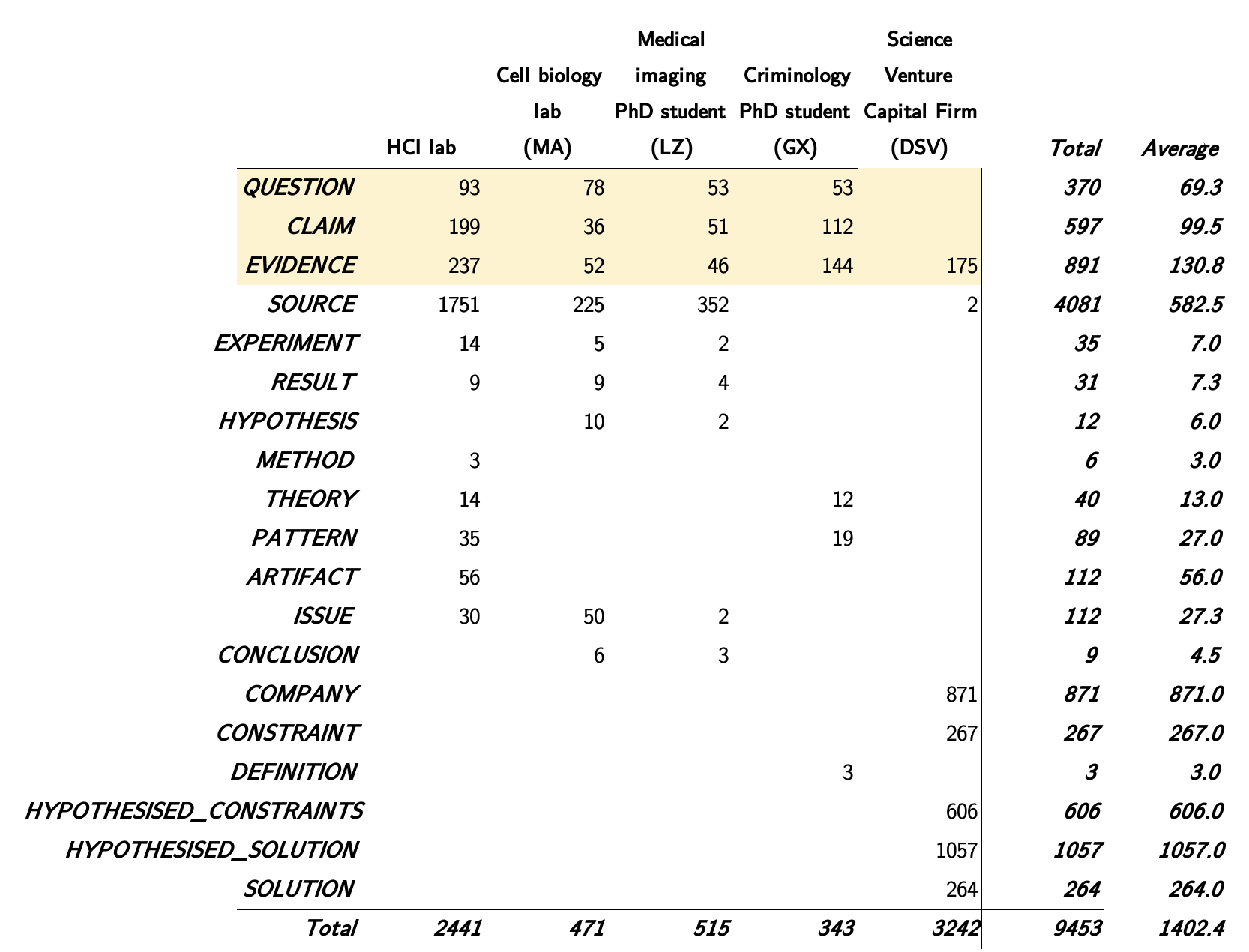}
    \caption{Frequencies of each type of node created across 5 users who shared their discourse graph CSV export in the usage survey as of Fall 2022, after about 1 year of usage.}
    \label{fig:node-frequencies}
\end{figure}

Figure \ref{fig:node-frequencies} shows the frequencies of each type of node created across 5 users who shared their discourse graph CSV export in the usage survey as of Fall 2022, after about 1 year of usage.

\end{document}